\documentclass[12pt]{article}
\usepackage{natbib}
\usepackage{amsfonts,amsbsy,amsmath,amssymb}
\numberwithin{equation}{section} 
\usepackage{indentfirst}
\usepackage{subfigure} 
\usepackage{graphicx}
\usepackage{float}
\usepackage{caption}
\usepackage[top=1in, bottom=1in, left=1in, right=1in]{geometry}
\usepackage[justification=centering]{caption}
\usepackage[normalem]{ulem}
\usepackage{appendix}
\usepackage{multirow}
\usepackage{algorithm,algorithmic}
\usepackage{longtable}

\usepackage{booktabs}
\usepackage{mathtools}
\usepackage{rotating}
\usepackage[colorlinks=true,linkcolor=blue,urlcolor=black,bookmarks=false]{hyperref}
\usepackage{bookmark}
\usepackage{soul}    
\usepackage{setspace}
\usepackage{numprint}
\usepackage{bm}
\usepackage{ulem}
\usepackage{tablefootnote}
\usepackage{threeparttable}
\usepackage{amssymb}
\usepackage{subfigure}
\usepackage{bbm}

\npdecimalsign{.}
\nprounddigits{3}
\hypersetup{
	colorlinks,
	citecolor=[RGB]{0,0,0},
	linkcolor=[RGB]{0,0,0},
	urlcolor=black}
\makeatletter

\newcommand{\Rmnum}[1]{\expandafter\@slowromancap\romannumeral #1@}
\makeatother
\usepackage{amsthm}
\makeatletter

\def\thmheadbrackets#1#2#3{%
	\thmname{#1}\thmnumber{\@ifnotempty{#1}{ }\@upn{#2}}%
	\thmnote{ {\the\thm@notefont[#3]}}}
\makeatother
\newtheoremstyle{brackets}
{}
{}
{\itshape}
{}
{\bfseries}
{.}
{ }
{\thmheadbrackets{#1}{#2}{#3}}
\theoremstyle{brackets}
\newtheoremstyle{break}
{\topsep}{\topsep}%
{\itshape}{}%
{\bfseries}{}%
{\newline}{}%

\theoremstyle{plain}
\newtheorem{definition}{Definition}
\newtheorem{theorem}{Theorem}
\newtheorem{lemma}{Lemma}
\newtheorem{proposition}{Proposition}

\makeatother
\usepackage{latexsym}
\newcommand{\bD}{{\bf D}}
\newcommand{\bR}{{\bf R}}
\newcommand{\bI}{{\bf I}}

\newcommand{\bY}{{\bf Y}}

\newcommand{\bB}{{\bf B}}

\newcommand{\bx}{{\bf x}}
\newcommand{\ba}{{\bf a}}
\newcommand{\bg}{{\bf g}}
\newcommand{\bw}{{\bf w}}

\newcommand{\by}{{\bf y}}
\newcommand{\bz}{{\bf z}}

\newcommand{\be}{{\bf e}}
\newcommand{\bmm}{{\bf m}}

\newcommand{\cG}{\mathcal{G}}
\newcommand{\cN}{\mathcal{N}}
\newcommand{\cS}{\mathcal{S}}

\newcommand{\cP}{\mathcal{P}}
\newcommand{\cR}{\mathcal{R}}

\newcommand{\wG}{\widehat{G}}

\newcommand{\mbE}{\mathbb{E}}

\newcommand{\mbR}{\mathbb{R}}

\newcommand{\mbP}{\mathbb{P}}

\newcommand{\bbeta}{\boldsymbol{\beta}}
\newcommand{\aalpha}{\boldsymbol{\alpha}}

\newcommand{\ttheta}{\boldsymbol{\theta}}

\newcommand{\ppsi}{\boldsymbol{\psi}}

\newcommand{\OOmega}{\boldsymbol{\Omega}}

\newcommand{\SSigma}{\boldsymbol{\Sigma}}

\newcommand{\XXi}{\boldsymbol{\Xi}}

\newcommand{\trans}{^{\mbox{\tiny{T}}}}

\newcommand{\ini}{\text{ini}}
\newcommand{\Naive}{\text{Naive}}
\newcommand{\dir}{\text{dir}}
\newcommand{\Cluster}{\text{Cluster}}
\newcommand{\less}{\text{less}}

\newcommand{\tbx}{\tilde{{\bf x}}}
\newcommand{\tby}{{\tilde{\bf y}}}
\newcommand{\tbY}{{\tilde{\bf Y}}}
\newcommand{\ty}{\tilde{y}}
\newcommand{\tu}{\tilde{u}}

\begin{document}

\title{\textbf{Conditional Selective Inference for the Selected Groups in Panel Data}}

\author{Chuang Wan\thanks{Department of Statistics and Data Science, School of Economics, Jinan University, Guangzhou, China }
\and Jiajun Sun\thanks{Department of Statistics and Data Science, School of Economics, Xiamen University, Xiamen, China}
\and Xingbai Xu\thanks{Corresponding author: The Wang Yanan Institute for Studies in Economics, Department of Statistics and Data Science,
School of Economics, Xiamen University, Xiamen, China. Email: xuxingbai@xmu.edu.cn.}}
	\maketitle
    
\begin{abstract}
\doublespacing
This paper addresses the problem of testing differences in group-specific slopes between clusters identified via $k$-means clustering in panel data. In this setting, classical Wald-type tests are inherently invalid, as they yield severely inflated Type I error probabilities. This failure stems from "double dipping''—using the same dataset for both group identification and subsequent inference. To resolve this, we propose a valid selective inference approach that conditions on the selection event to account for data-driven clustering. We derive an efficient algorithm to compute exact p-values for $k$-means outcomes and further develop a less-conditioned p-value that avoids unnecessary conditioning on the direction of the selected contrast, thereby improving power while preserving selective validity. Our framework is notably flexible: it requires only the asymptotic normality of initial estimators and extends naturally to GMM estimation, making it applicable to dynamic panel models and panel data models with endogenous variables. Simulation studies demonstrate robust finite-sample performance and show that the less-conditioned procedure improves power relative to the direction-conditioned benchmark. Applying our method to the global relationship between economic growth and CO$_2$ emissions, we uncover novel evidence of cross-country heterogeneity. We also provide the {\bf R} package, \texttt{TestHomoPanel}, to facilitate implementation.
\end{abstract}

\bigskip
	\textbf{JEL classification:} C12, C33, C38 
	\\
	\textbf{Keywords:}  Post-selection inference; $k$-means clustering; Hypothesis testing; Latent groups;  Panel data models.
	\doublespacing

\section{Introduction}\label{s1}

Panel  data have attracted long-standing and continuing interest in literature \citep{hsiao2022analysis}, and have a wide spectrum of applications in various fields, ranging from social sciences to medical studies. Such data typically consist of repeated observations from diverse units such as workers, firms, or nations, which possess diverse characteristics, leading to
inherent heterogeneity, such as variation in educational investment preferences or differential responses to economic policies.
A central challenge in panel data analysis is the parsimonious and effective modeling of such individual-level heterogeneity. A common strategy is the use of fixed effects to absorb such heterogeneity. Beyond this, two contrasting approaches are frequently adopted for modeling slope heterogeneity. One approach imposes homogeneous coefficients across all individuals, which simplifies model structure and facilitates inference.
However,  homogeneity assumptions are frequently rejected in many practical applications, as empirically evidenced by \cite{phillips2007transition,chen2019forensic} and \cite{huang2023detecting}. Alternatively, another strand of literature permits full heterogeneity in coefficients across units (e.g., \citealp{hsiao2008random,baltagi2008pool}). While this avoids restrictive assumptions, it often results in prohibitively large parameter spaces and inefficient estimation, thereby sacrificing the fundamental pooling advantage of panel data.

Consequently, an increasing number of researchers have embraced an intermediate paradigm that bridges these two extremes: panel group structure models. This approach posits that individuals are partitioned into a finite number of latent groups, where units within the same group share common slope parameters, while heterogeneity manifests across different groups.
Research in this area is very abundant, and most related works fall into two main categories based on their objectives. The first strand of literature aims to develop various detection algorithms such as the Classifier-LASSO (C-LASSO) method \citep{su2016identifying}, the $k$-means clustering \citep{liu2020identification} and the  sequential binary segmentation algorithm (SBSA) \citep{wang2021identifying} to improve the accuracy of identifying unknown group structures. Applications of these detection algorithms have been extended to various types of panel models, see, e.g., \cite{su2018identifying}, \cite{wang2018homogeneity}, \cite{su2019sieve}, \cite{huang2020identifying}, \cite{mehrabani2023estimation}, \cite{su2023identifying}, \cite{wang2024panel} and \cite{loyo2025grouped}. This research field is very active, and this review is surely incomplete.
The second strand aims to conduct hypothesis tests for the homogeneity of individuals, i.e., whether the slopes across individuals are statistically equivalent, see, e.g., \cite{pesaran2008testing}, \cite{su2013testing,breitung2016lagrange} and \cite{campello2019testing}.

We address a third and largely unaddressed issue: testing slope differences between identified groups after estimating latent structures.
Addressing this problem is challenging as it falls within the framework called
 post-selection inference, a novel research area that has drawn tremendous  attention in recent years. In this paper, we mainly focus on the problem of testing for a difference in slopes between two identified clusters by applying the $k$-means clustering algorithm \citep{liu2020identification}.
Suppose we observe a panel dataset $\{(y_{it},\bx_{it}):i\in[N];t\in[T]\}$ collected from $N$ individuals with $T$ time periods, where $[a]=\{1,\cdots,a\}$ for any positive integer $a$, $y_{it}$ is a real-valued response variable, and $\bx_{it}$ is the associated $p$-dimensional covariate.
Consider the following linear panel model
\begin{equation}\label{eq:1}
y_{it}=\bx_{it}\trans\bbeta_i^0+\eta_{i}+u_{it},~~~i\in[N],~~t\in[T],
\end{equation}
where $\bbeta_i^0$ is a $p\times1$ vector of slope parameters, $\eta_{i}$ is the individual fixed effect and $u_{it}$ is the idiosyncratic error term with zero mean. To mitigate model complexity, we assume that $\bbeta_i$ admits the following grouping structure of the form
\begin{equation}\label{eq:11}
\bbeta_i^0=\sum_{k=1}^{K^0}\aalpha_{G_k^0}^0\cdot{\bf1}\{i\in G_k^0\},~~G_k^0\cap G_{k'}^0=\emptyset~~\forall k\neq k',~~\text{and}~~\cup_{k=1}^{K^0}G_k^0=[N],
\end{equation}
where $\aalpha_k^0$ is the common group-specific slope satisfying  $\aalpha_{G_k^0}^0\neq\aalpha_{G_{k'}^0}^0$ for any $k\neq k'$, $G_k^0$ is the set of memberships for the $k$th group, $\cG^0=\{G_1^0,\cdots,G_{K^0}^0\}$ forms a partition of the set $[N]$, and $K^0$ is the true number of groups. These quantities  are all unknown to us.
Throughout this paper, a superscript zero on any quantity refers to the true quantity.

\subsection{Why is post-selection inference necessary?}
To identify the latent structure in \eqref{eq:1}, many notable clustering algorithms have been utilized to implement  homogeneous individuals clustering. All these methods perform well under certain circumstances and could achieve selection consistency theoretically as $N$ and $T$ go to infinity. However, it is well known that in unsupervised clustering analysis, clustering can easily lead to spurious discoveries.
 Specifically, individuals belonging to a single homogeneous group may be erroneously partitioned into multiple distinct clusters \citep{chen2023selective,gao2024selective}. To illustrate this phenomenon, let us examine a stylized example.
 Consider the same data generating process (DGP) as in model \eqref{eq:1},
where $\bx_{it}=(0.2\eta_i+e_{it1},0.2\eta_i+e_{it2})\trans$ is a $2\times1$ vector of exogenous variables, and $e_{it1},e_{it2}$,  $u_{it}$ and $\eta_i$ are all independently generated from the standard normal distribution $\cN(0,1)$. This basic setup has been frequently employed in the simulation studies such as DGP 1 of \cite{su2016identifying} and \cite{liu2020identification}.
We consider a heterogenous panel model with  $K^0=3$ groups, where the group proportions are $0.3:0.3:0.4$, and the true coefficients are (0.4, 1.6), (1, 1) and (1.6, 0.4), respectively. Three error distributions for $u_{it}=\sigma\varepsilon_{it}$ are considered: (1) $\varepsilon_{it}\stackrel{i.i.d}{\sim} \cN(0,1)$; (2) $\varepsilon_{it}\stackrel{i.i.d}{\sim}t(3)/\sqrt{3}$ and (3) $\varepsilon_{it}\stackrel{i.i.d}{\sim}(\chi_3^2-3)/\sqrt{6}$.
The noise scale $\sigma$ is adjusted to  maintain a signal-to-noise ratio of 1.  We use sample sizes $N=100,200$ and time spans $T=15,25,50$.

\begin{figure}[!ht]
	\centering
    \includegraphics[width=\textwidth]{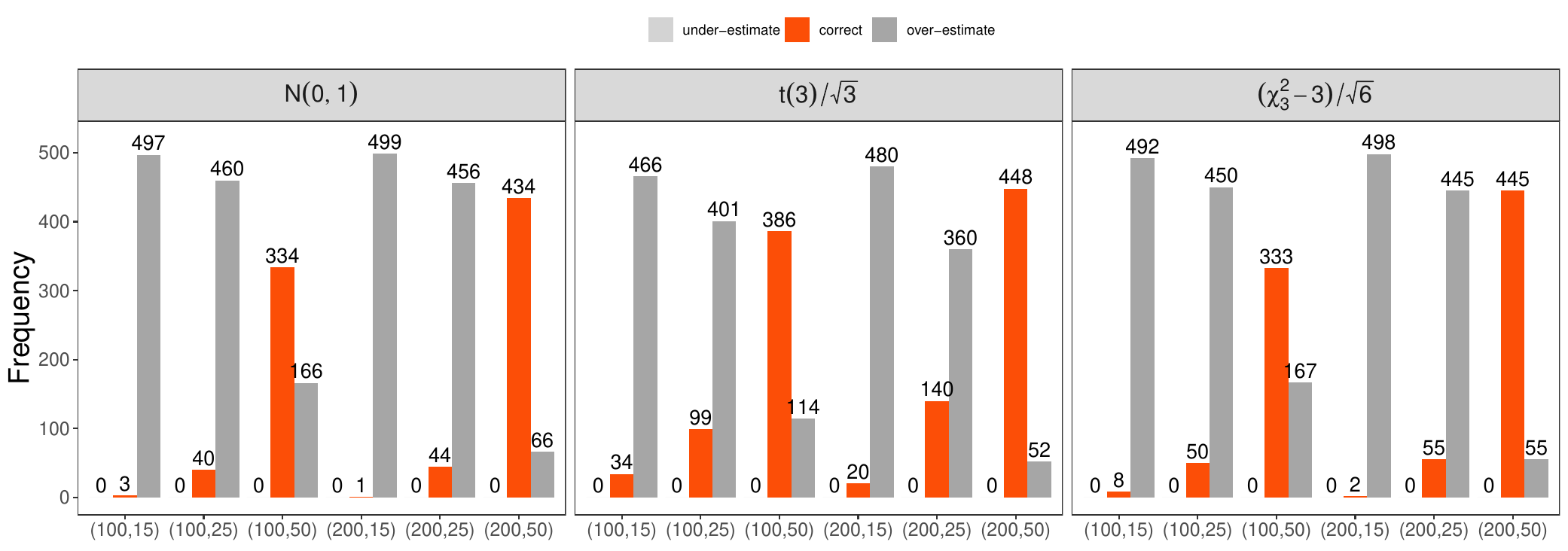}
	\caption{Frequencies of the estimated number of groups across 500 repetitions obtained by  $k$-means algorithm  from model \eqref{eq:1} with $N=(100,200)$, $T=(15,25,50)$ and $K^0=3$. }
	\label{fig:toy1}
\end{figure}

Figure \ref{fig:toy1} reports the histogram  of the estimated number of groups ($\widehat{K}$) over 500 replications
 using the $k$-means algorithm \citep{liu2020identification}. In each scenario, a noticeable tendency to overestimate $K^0$ (i.e., $\widehat{K}>K^0$) is observed, especially when $N$ or $T$ is small. For instance, when $(N,T)=(100,15)$, the $k$-means algorithm overestimates the number of groups in nearly all repetitions, with frequencies of 497, 466, and 492 for the three error types, respectively. Although increasing $T$ to 50 alleviates the overestimation to some extent, the probability remains non-negligible. A plausible  explanation is that the penalty term in the information criterion proposed by \cite{liu2020identification} is insufficiently stringent in this setting. This example clearly demonstrates that the risk of overestimating the number of groups warrants careful consideration, even when latent group structures are present.

\subsection{Why does naive approach fail?}
Given the potential for misclassification in identifying latent groups for panel data, it is meaningful to verify whether  the selected groups are actually from a common group. That is to say, we need to test whether the estimated group-specific estimates  are  statistically different after the group structure has been determined by the data. In this paper, we mainly apply the $k$-means algorithm \citep{liu2020identification}  to cluster the individuals, and leave the study of other algorithms to future research.
Suppose we have estimated group structures obtained by the $k$-means clustering, denoted as $\widehat{\cG}=(\widehat{G}_1,\cdots,\widehat{G}_K)$. Then we consider testing the null hypothesis that the group-specific slopes are the same between two estimated clusters,
\begin{equation}\label{eq:t}
H_0^{(\widehat{G}_k,\widehat{G}_{k'})}:{\aalpha}_{\wG_k}={\aalpha}_{\wG_{k'}},~~~\text{versus}~~~
H_1^{(\widehat{G}_k,\widehat{G}_{k'})}:{\aalpha}_{\wG_k}\neq{\aalpha}_{\wG_{k'}},
\end{equation}
where $\widehat{G}_k,\widehat{G}_{k'}\in\widehat{\cG}$, and ${\aalpha}_{\wG_k}$ and ${\aalpha}_{\wG_{k'}}$ are their corresponding group-specific slope estimates. To test hypothesis \eqref{eq:t}, it is tempting to simply apply the classical Wald-type test statistic based on the asymptotic theory for $\widehat{\aalpha}_{\widehat{\cG}}=(\widehat{\aalpha}_{\wG_1}\trans,\cdots,\widehat{\aalpha}_{\wG_K}\trans)\trans$, which we refer to as the naive approach. As shown by
\cite{liu2020identification}, when $K=K^0$, the estimator $\widehat{\aalpha}_{\widehat{\cG}}$ admits the following limiting distribution
$$
\sqrt{NT}\left(\widehat{\aalpha}_{\widehat{\cG}}-\aalpha_{\cG^0}^0\right)\stackrel{d}{\rightarrow}\cN\left({\bf0},
\SSigma_{{\cG^0}}^0\right),~~~\text{as $(N,T)\rightarrow\infty$},
$$
under some mild conditions, where $\aalpha_{\cG^0}^0=(\aalpha_{G_1^0}^{0\mbox{\tiny{T}}},\cdots,\aalpha_{G_{K^0}^0}^{0\mbox{\tiny{T}}})\trans$ is the true parameter, and $\SSigma_{{\cG^0}}^0$ is the $K^0p\times K^0p$ covariance matrix, which can be estimated by the plug-in method. The plug-in estimator for $\SSigma_{{\cG^0}}^0$ takes a block diagonal matrix form, i.e., $\widehat{\SSigma}_{\widehat{\cG}}=\text{bdiag}(\widehat{\SSigma}_{\widehat{G}_k},k=1,\cdots,K)$, where
$\widehat{\SSigma}_{\widehat{G}_k}$ is the estimated covariance matrix of $\widehat{\aalpha}_k$. One can refer to Theorem 3  of \cite{liu2020identification} for detailed expressions. The Wald test statistic for a given pair of clusters is
\begin{equation}
\widehat{W}_{k,k'}=\left(\bR\widehat{\aalpha}_{\widehat{\cG}}\right)\trans\left(\bR\widehat{\SSigma}_{\widehat{\cG}}
\bR\trans\right)^{-1}\bR\widehat{\aalpha}_{\widehat{\cG}},
\end{equation}
where $\bR$ is a $p\times Kp$ index matrix. For example, when $\widehat{G}_k=\widehat{G}_1$ and $\widehat{G}_{k'}=\widehat{G}_2$,
$$
\bR=\left(
\begin{array}{ccccc}
\bI_{p\times p} & -\bI_{p\times p} & {\bf0}_{p\times p} & \cdots & {\bf0}_{p\times p}
\end{array}
\right),
$$
where $\bI_{p\times p}$ denotes the $p\times p$ identity matrix, and ${\bf0}_{p\times p}$ denotes the $p\times p$ zero matrix.

At first glance, the p-value for $\widehat{W}_{k,k'}$ is given by
\begin{equation}\label{eq:n}
P_{\Naive}=\mbP_{H_0^{(\widehat{G}_k,\widehat{G}_{k'})}}\left(
\chi_p^2\geq \widehat{W}_{k,k'}
\right),
\end{equation}
where $\mbP_{H_0^{(\widehat{G}_k,\widehat{G}_{k'})}}(\cdot)$ is the probability measure under $H_0^{(\widehat{G}_k,\widehat{G}_{k'})}$.
However, this ``naive" p-value is {\bf invalid} due to the fact that $H_0^{(\widehat{G}_k,\widehat{G}_{k'})}$ is chosen based on the data, because  $\widehat{G}_k,\widehat{G}_{k'}$ are estimated clusters in $\widehat{\cG}$. This would lead to a misleading conclusion that
substantial differences between $\{\widehat{\aalpha}_{\wG_k}\}_{k=1}^K$ are observed even if there is no group pattern in the data.

\begin{figure}[!h]
	\centering
	\includegraphics[width=\textwidth]{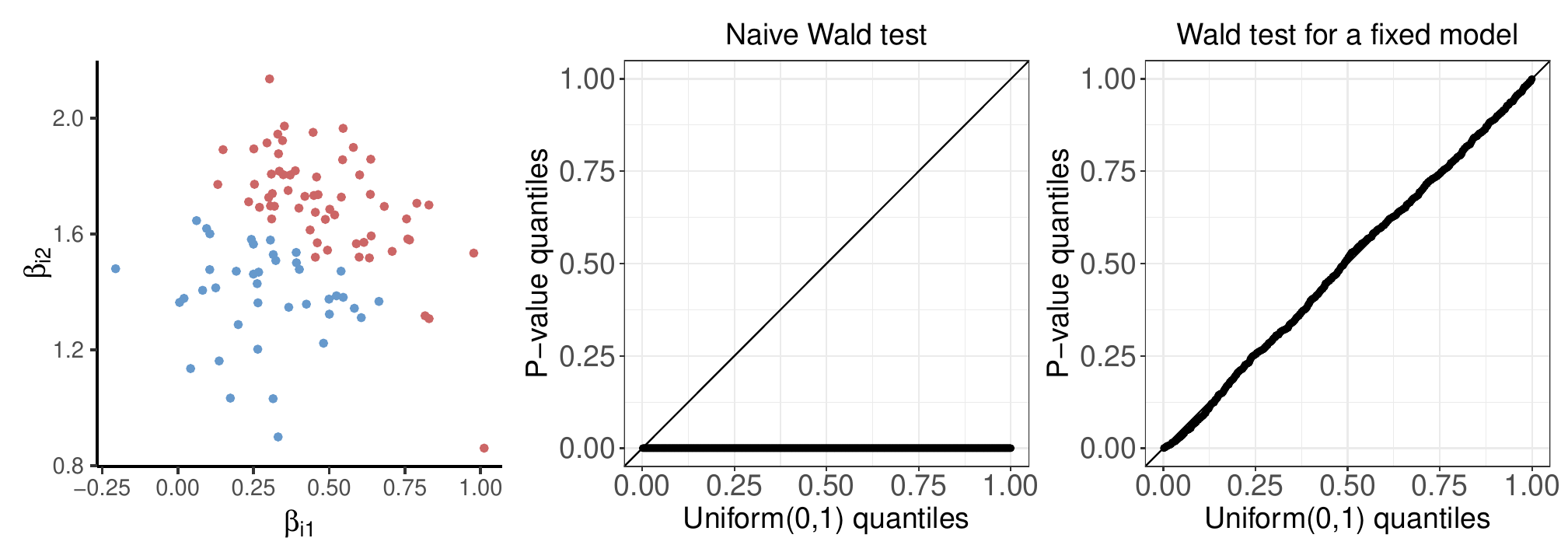}
	\caption{\emph{Left}: We apply the $k$-means clustering to one certain simulated data set, which yields two clusters.
\emph{Center:} Q-Q plot of the naive p-values, aggregated over 1000 simulated datasets. \emph{Right}: Q-Q plot of the p-values  based on a fixed group structure.}
	\label{fig:toy2}
\end{figure}

To better illustrate this, we  take the homogeneous linear panel model \eqref{eq:1} with $p=2$ regressors and $K^0=1$ as an example.
 Suppose we mistakenly classify the 100 individuals into two groups using the $k$-means algorithm, i.e., $K=2$,  and we need to examine
 the differences between the two selected groups, that is $H_0^{(\widehat{G}_1,\widehat{G}_2)}:\widehat{\aalpha}_{\widehat{G}_1}=\widehat{\aalpha}_{\wG_2}$
 versus $H_0^{(\widehat{G}_1,\widehat{G}_2)}:\widehat{\aalpha}_{\wG_1}\neq\widehat{\aalpha}_{\wG_2}$.
We generate 1000 simulated datasets and calculate their p-values.
The middle panel of Figure \ref{fig:toy2} shows the Quantile-Quantile (Q-Q) plots of the naive p-values (defined in \eqref{eq:n}).
We see that the obtained p-values are all near zero, which implies that  the naive Wald test statistic is extremely anti-conservative and cannot control the type I error. Hence, we would draw a misleading conclusion that there exists  a significant difference between the selected groups even if the true model is homogeneous.
The left column of Figure \ref{fig:toy2} displays the scatter plot of the preliminary individual slope estimates.   It is evident that for the $k$-means algorithms, all individuals are erroneously partitioned into two distinct groups, and  individuals with closer initial estimates are more likely to be grouped together.  This artificial separation creates a spurious gap between the group-specific estimates, leading to the breakdown of classical inference.

The underlying reason why the naive Wald statistic $\widehat{W}_{k,k'}$ fails is that
classical statistical theory implicitly assumes that the group structure  is fixed or predetermined
before the data are observed.  However, in our problem, we first select the group structures and then pose the hypothesis
based on the selected result. That means we use the observed dataset to guide the choice of the statistical question in the sense that $H_0^{(\widehat{G}_k,\widehat{G}_{k'})}$ is a function of the observed
dataset,
and we use the same dataset once again to conduct inference. This creates dependence between the selection and inference stages, which violates
 the assumption of traditional statistical theory.
This phenomenon is also known as ``double dipping" or ``data snooping". To verify our argument, we conduct another simulation study based on the setting where the group pattern is predetermined. Specifically,  we treat the first 40 individuals as Group 1 and the remaining 60  as Group 2, i.e., $G_1^f=\{1,\cdots,40\}$ and $G_2^f=\{41,\cdots,100\}$. Now, $G_1^f$ and $G_2^f$ are predefined and not functions of the observed dataset. Let the group structures be fixed across all simulations.  We then employ the
classical Wald statistic to test $H_0^{(G_1^f,G_2^f)}$, and calculate the p-values. The results are displayed in the right panel of Figure \ref{fig:toy2}.
Now, the Q-Q plot for the p-values perfectly aligns with the 45-degree line, which confirms that
the test statistic based on a fixed model is valid and can control the type I error rate reasonably well.

\subsection{Contributions and organization}

In classical statistical inference, hypotheses are typically predefined prior to data observation. However, in many contemporary supervised and unsupervised learning applications, including regression and clustering, the hypotheses of interest are frequently data-contingent, which violates the classical statistical theory. The significance of selective inference has been thoroughly discussed by \cite{taylor2015statistical} and \cite{benjamini2020selective}, where the latter illustrates its importance from the perspective of replicability. Most existing studies conduct selective inference in supervised settings, such as the recent literature on selective inference for model selection and changepoint detection \citep{lee2016exact,tian2018selective,hyun2021post,jewell2022testing}. In unsupervised clustering, \cite{gao2024selective} proposes an elegant selective inference procedure for post-clustering datasets based on the hierarchical clustering algorithm, inspiring a series of studies on this topic \citep{chen2023selective,yun2023selective,chen2025testing}. Our work is closely related to the contemporaneous and independent study by \cite{akgun2025robust}. While both papers address selective inference in panel data, our approach differs in three significant aspects. First, whereas \cite{akgun2025robust} rely on the exact normality of error terms, our framework requires only the asymptotic normality of the initial estimators, a property widely satisfied in the panel data literature, thus offering broader applicability. Second, we extend the selective inference framework to the GMM estimation of dynamic panel data models, while their focus remains on static settings. Finally, our underlying projection mechanism is distinct; by using a more intuitive oblique projection strategy, we provide a more transparent mathematical framework that avoids the complex constrained-estimator algebra required in their study.



In this work, we develop a \emph{conditional selective inference} framework for panel group structure models designed to test for  differences in a pair of group-specific slopes after clustering. Unlike existing work on post-selection inference for clustering, which clusters the raw observational data directly, we apply the $k$-means clustering to the initial  coefficient estimates of all individuals, a process that introduces additional challenges.
The core idea of our approach hinges on the fact that we test the hypothesis $H_0^{(\wG_k,\wG_{k'})}:{\aalpha}_{\wG_k}={\aalpha}_{\wG_{k'}}$ because the clusters $\wG_k$ and $\wG_{k'}$ were estimated from the panel dataset. To properly control for this data-driven selection, we construct a p-value conditioned on  $\{\wG_k,\wG_{k'}\in\widehat{\cG}\}$ and other events, thereby ensuring valid size control. We further propose a less-conditioned selective p-value by integrating over the direction of the selected contrast rather than conditioning on its observed value. This refinement preserves selective validity and improves power by avoiding an unnecessary conditioning event. We also find that the construction of test statistics and the computation of the p-value developed in this paper can be directly extended to the post-selection inference for evaluating a difference in coefficients of a single regressor after clustering. Furthermore, with minor modifications, the methodology can also be adapted to the GMM estimation framework, making it applicable to dynamic panel models  and addressing endogeneity.
As part of the contribution of this work, we provide a user-friendly {\bf R} package called \texttt{TestHomoPanel} that implements the selective inference procedure for identified group structures in  panel data. 

 The rest of our paper is organized as follows. In Section \ref{sec:2}, we briefly review the $k$-means clustering algorithm in the context of linear panel models. In Section \ref{sec:3}, we develop a framework to test for differences in slopes after  clustering and provide an efficient approach to compute the selective p-values. Section \ref{sec:direction-average-panel} introduces the less-conditioned p-value for improving power. In Section \ref{sec:4}, we extend the main idea of our proposal to test for differences due to a single feature, and the GMM framework.
 We evaluate the proposed methodologies through simulation studies in Section \ref{sec:5} and apply our methods to a real dataset  in Section \ref{sec:6}.
 Section \ref{sec:7} concludes. To save space, all proofs and additional illustrations of our extensions are relegated to the  Appendix.

\section{Review of the $k$ -means Clustering for Panel Data}\label{sec:2}
In this section, we first review the $k$-means clustering algorithm developed by \cite{liu2020identification} to identify the latent group structures for panel model \eqref{eq:1}. This algorithm is popular and efficient in capturing the group structures of slopes in panel models. \cite{bonhomme2015grouped} proposed a similar algorithm to identify the group patterns of the fixed effects in panel models. Since the individual effects, $\eta_i$, are not the primary interest, we eliminate them by taking the average within-subject, which yields the following equation
\begin{equation}\label{eq:2}
\tilde{y}_{it}=\tilde{\bx}_{it}\trans\bbeta_i^0+\tu_{it},
\end{equation}
where $\tilde{\bx}_{it}=\bx_{it}-T^{-1}\sum_{t=1}\trans\bx_{it}$, $\tu_{it}$ and $\ty_{it}$ are analogously defined.  Let $g_i\in[K^0]$ denote the group membership. Obviously, if individuals $i$ and $j$ belong to the same group, i.e., $i,j\in G_k^0$, then $g_i=g_j=k$ and $\bbeta_{i}^0=\bbeta_{j}^0=\aalpha_k^0=\aalpha_{G_k^0}^0$. Let $\underline{\aalpha}=(\aalpha_1,\cdots,\aalpha_K)$ denote the tuple of $K$ group-specific parameters, and $\bg_N=(g_1,\cdots,g_N)\in[K]^N$ denote the collection of  label memberships.
The identification of the latent group structure  for a given $K$ can be implemented through the following optimization problem
\begin{equation}\label{eq:21}
\min_{\underline{\aalpha},\bg_N}\sum_{i=1}^N\sum_{t=1}\trans(\ty_{it}-\tilde{\bx}_{it}\trans\aalpha_{g_i})^2.
\end{equation}

However, due to the complex group structures, it is challenging to directly minimize the least square (LS) function in \eqref{eq:21}. To overcome this limitation, \cite{liu2020identification} proposed an efficient iterative algorithm based on  $k$-means clustering.
Note that the LS objective function of the $i$th individual is
\begin{eqnarray*}
\sum_{t=1}\trans(\tilde{y}_{it}-\tilde{\bx}_{it}\trans\bbeta)^2
&=&(\tilde{\by}_i-\tilde{\bx}_i\trans\bbeta)\trans(\tilde{\by}_i-\tilde{\bx}_i\trans\bbeta)\\
&=&(\tilde{\by}_i-\tilde{\bx}_i\trans\widehat{\bbeta}_i^{\ini}+
\tilde{\bx}_i\trans\widehat{\bbeta}_i^{\ini}-\tilde{\bx}_i\trans\bbeta)\trans(\tilde{\by}_i-\tilde{\bx}_i\trans\widehat{\bbeta}_i^{\ini}+
\tilde{\bx}_i\trans\widehat{\bbeta}_i^{\ini}-\tilde{\bx}_i\trans\bbeta)\\
&\equiv&(\widehat{\bbeta}_i^{\ini}-\bbeta)\trans\tilde{\bx}_i\trans\tilde{\bx}_i(\widehat{\bbeta}_i^{\ini}-\bbeta)+\text{constant},
\end{eqnarray*}
where $\widehat{\bbeta}_i^{\ini}=(\tilde{\bx}_i\trans\tilde{\bx}_i)^{-1}\tilde{\bx}_i\trans\tilde{\by}_i$ is the initial LS estimate of individual $i$,  $\tilde{\by}_i=(\tilde{y}_{i1},\cdots,\tilde{y}_{iT})\trans$ is a $T\times1$ vector of response, and $\tilde{\bx}_i=(\tbx_{i1},\cdots,\tbx_{iT})\trans$ is a $T\times p$ matrix. The pooled objective function in \eqref{eq:21} is equivalent to
$\sum_{i=1}^N(\widehat{\bbeta}_i^{\ini}-\aalpha_{g_i})\trans\tilde{\bx}_i\trans\tilde{\bx}_i(\widehat{\bbeta}_i^{\ini}-\aalpha_{g_i})$.
 If the partition set $\cG=\{G_1,\cdots,G_K\}$ is given, then the group-specific parameter is given by
\begin{eqnarray*}
\widehat{\aalpha}_{G_k}&=&\left(\sum_{i\in G_k}\tilde{\bx}_i\trans\tilde{\bx}_i\right)^{-1}\sum_{i\in G_k}\tilde{\bx}_i\tilde{\by}_i
=\sum_{i\in G_k}\left\{
\left(\sum_{i'\in G_k}\tilde{\bx}_{i'}\trans\tilde{\bx}_{i'}\right)^{-1}\tilde{\bx}_i\trans\tilde{\by}_i
\right\}\\
&=&
\sum_{i\in G_k}\left\{
\left(\sum_{i'\in G_k}\tilde{\bx}_{i'}\trans\tilde{\bx}_{i'}\right)^{-1}(\tilde{\bx}_i\trans\tilde{\bx}_i)
(\tilde{\bx}_i\trans\tilde{\bx}_i)^{-1}\tilde{\bx}_i\trans\tilde{\by}_i
\right\}\\
&=&\sum_{i:g_i=k}\bw_i(k)\widehat{\bbeta}_i^{\ini},
\end{eqnarray*}
where $\bw_i(k)=\left(\sum_{i':g_{i'}=k}\tilde{\bx}_{i'}\trans\tilde{\bx}_{i'}\right)^{-1}(\tilde{\bx}_i\trans\tilde{\bx}_i)$ is a $p\times p$ matrix.
That is to say, $\widehat{\aalpha}_{G_k}$ is a linear combination of initial estimate $\widehat{\bbeta}_i^{\ini}$ for any $i\in G_k$.
The $k$-means clustering partitions the $N$ individuals into $K$ disjoint subsets $\widehat{G}_1,\cdots,\widehat{G}_K$ by solving the following problem
\begin{eqnarray*}
&&\underset{G_1,\cdots,G_K}{\text{minimize}}
\left\{
\sum_{k=1}^K\sum_{i\in G_k}\left(\widehat{\bbeta}_i^{\ini}-\sum_{i\in G_k}\bw_i(k)\widehat{\bbeta}_i^{\ini}\right)\trans\tilde{\bx}_i\trans\tilde{\bx}_i\left(
\widehat{\bbeta}_i^{\ini}-\sum_{i\in G_k}\bw_i(k)\widehat{\bbeta}_i^{\ini}\right)
\right\},\\
&&~~~~\text{subject to}~~\cup_{k=1}^KG_k=[N],~~G_k\cap G_{k'}=\emptyset,~~\forall\text{$k\neq k'$.}
\end{eqnarray*}
We summarize the main procedure in Algorithm \ref{alg:1}, which can be viewed as an adaptation of the $k$-means algorithm proposed by \cite{liu2020identification} for linear panel models. \footnote{ Algorithm \ref{alg:1} differs from the $k$-means approach in \cite{liu2020identification} primarily in handling the individual effects $\eta_i$. We concentrate them out beforehand, whereas they treat $\eta_i$ as unknown parameters to be estimated. } 

\begin{algorithm*}[h]
\caption{$k$-means clustering algorithm for linear panel model \eqref{eq:2}.}
\label{alg:1}
\begin{algorithmic}[1]
\STATE Initialize the centroids $(\widehat{\bmm}_1^{(0)},\cdots,\widehat{\bmm}_K^{(0)})$ by sampling $K$ individuals from
$\widehat{\bbeta}_1^{\ini},\cdots,\widehat{\bbeta}_N^{\ini}$ without replacement using a random seed.
\STATE Compute assignments $\hat{g}_i^{(0)}\leftarrow\arg\min_{k\in[K]}
(\widehat{\bbeta}_i^{\ini}-\widehat{\bmm}_k^{(0)})\trans\tilde{\bx}_i\trans\tilde{\bx}_i(\widehat{\bbeta}_i^{\ini}-\widehat{\bmm}_k^{(0)})$;
\STATE Initialize $s=0$:
\WHILE{$s\leq S$}
\STATE Define $\bw_i^{(s)}(k)=\left(\sum_{i':g_{i'}^{(s)}=k}\tilde{\bx}_{i'}\trans\tilde{\bx}_{i'}\right)^{-1}(\tilde{\bx}_i\trans\tilde{\bx}_i)$;
\STATE Update centroids: $\widehat{\bmm}_{k}^{(s+1)}=\sum_{i:\hat{g}_i^{(s)}=k}\bw_i^{{(s)}}(k)\widehat{\bbeta}_i^{\ini}$;
\STATE Update assignment: $\hat{g}_i^{(s+1)}\leftarrow\arg\min_{k\in[K]}
(\widehat{\bbeta}_i^{\ini}-\widehat{\bmm}_{k}^{(s+1)})\trans\tilde{\bx}_i\trans\tilde{\bx}_i(\widehat{\bbeta}_i^{
\ini}-\widehat{\bmm}_{k}^{(s+1)})$;
\IF{$\hat{g}_i^{(s+1)}=\hat{g}_i^{(s)}$ for all $1\leq i\leq N$}
\STATE break
\ELSE
\STATE $s\leftarrow s+1$
\ENDIF
\ENDWHILE
\RETURN $\{\widehat{\aalpha}_{\wG_k}=\widehat{\bmm}_k^{(S)}\}_{k=1}^K$ and $\wG_k=\{i:\hat{g}_i=k\}$ with $\hat{g}_i=\hat{g}_i^{(S)}$.
\end{algorithmic}
\end{algorithm*}

\section{Selective Inference for the Selected Groups}\label{sec:3}

\subsection{Selective type I error and selective p-value }\label{sec:31}
We consider a setting where the quantitative response vector $\tilde{\bY}=(\tilde{\bY}_1\trans,\cdots,\tilde{\bY}_N\trans)\trans\in\mbR^{NT}$ is random, while the predictor matrix $\tilde{\bx}=(\tilde{\bx}_1\trans,\cdots,\tilde{\bx}_N\trans)\trans\in\mbR^{NT\times p}$ is fixed. Let
\(\tilde{\by}=(\tilde{\by}_1\trans,\cdots,\tilde{\by}_N\trans)\trans\) denote the observed realization of
\(\tilde{\bY}\).
This setting is commonly used in the literature of post-selection inference for regression analysis, as seen in \cite{berk2013valid,taylor2015statistical,tibshirani2016exact} and \cite{kuchibhotla2020valid}. Based on this setting,
 we set
$\bB=\left(({\bbeta}_1^{\ini})\trans,\cdots,({\bbeta}_N^{\ini})\trans\right)\trans$ as the $Np\times1$-vector of initial random coefficients where $\bbeta_i^{\ini}=(\tbx_i\trans\tbx_i)^{-1}\tbx_i\trans\tbY_i$. Its observed realization, obtained by replacing $\tilde{\bY}$ with $\tilde{\by}$, is denoted by
$\widehat{\bB}=\left((\widehat{\bbeta}_1^{\ini})\trans,\cdots,
(\widehat{\bbeta}_N^{\ini})\trans\right)\trans$, where
$\widehat{\bbeta}_i^{\ini}=(\tbx_i\trans\tbx_i)^{-1}\tbx_i\trans\tilde{\by}_i$.
From  the $k$-means iterative process in  Algorithm \ref{alg:1}, we know that the final group-specific estimator is determined by a function of
initial individual-specific estimators.
Then, the difference in slopes between two selected groups $\wG_k$ and $\wG_{k'}$ is $\aalpha_{\wG_k}-\aalpha_{\wG_{k'}}=\bR\aalpha_{\widehat{\cG}}=\ttheta\trans\bB$, and the corresponding observed difference is
$\widehat{\aalpha}_{\wG_k}-\widehat{\aalpha}_{\wG_{k'}}=\bR\widehat{\aalpha}_{\widehat{\cG}}=\ttheta\trans\widehat{\bB}$, where
\begin{equation}\label{eq:nu}
\ttheta=\left(v_1\bw_1(\hat{g}_1),\cdots,v_N\bw_N(\hat{g}_N)\right)\trans
\end{equation}
is an $Np\times p$ matrix  with $v_i=1\{i\in\wG_k\}-1\{i\in\wG_{k'}\}$ taking values 1, $-1$ or 0. Here, $\ttheta$ is constructed from the observed selected groups and is treated as fixed when evaluating the conditional distribution of the random counterpart $\bB$.

%

With a slight abuse of notation, we use $\text{Cluster}(\cdot)$ to denote the estimated groups obtained by applying $k$-means clustering to the preliminary individual estimates. Hence, $\widehat{\cG}=\text{Cluster}(\widehat{\bB})$.
Since
 $H_0^{(\wG_k,\wG_{k'})}$ is chosen because $\{\wG_k,\wG_{k'}\in\text{Cluster}(\widehat{\bB})\}$, it is natural to define a conditional
 version of \eqref{eq:n}, i.e.,
\begin{equation}\label{eq:cp0}
\mbP_{H_0^{(\wG_k,\wG_{k'})}}\left(
W_{k,k'}\geq \widehat{W}_{k,k'}~\Big|~ \wG_k,\wG_{k'}\in\text{Cluster}({\bB})
\right),
\end{equation}
where $W_{k,k'}=\left(\bR{\aalpha}_{\widehat{\cG}}\right)\trans\left(\bR{\SSigma}_{\widehat{\cG}}
\bR\trans\right)^{-1}\bR{\aalpha}_{\widehat{\cG}}\stackrel{d}{\rightarrow}\chi_p^2$ under $H_0^{(\hat{G}_k,\hat{G}_{k'})}$.
We can understand \eqref{eq:cp0} as follows: ``Suppose we observe a new realization of $\tbY$, say $\tby^*$, in
place of the actual data $\tby$. We then apply the $k$-means algorithm to the new response vector $\tby^*$ and obtain  the same
clustering results, i.e., $\text{Cluster}(\widehat{\bB})=\text{Cluster}(\bB^*)$, where $\bB^*=
\left((\bbeta_1^{*\ini})\trans,
\cdots,(\bbeta_N^{*\ini})\trans\right)\trans$ with $\bbeta_i^{*\ini}=(\tbx_i\trans\tbx_i)^{-1}\tbx_i\trans\tby_i^*$.
What proportion of such $\tby^*$ yields a difference in slopes between $\widehat{G}_k$ and $\widehat{G}_{k'}$ that is at least as large as the difference  observed in the original dataset, when $H_0^{(\wG_k,\wG_{k'})}$ holds?"
Therefore, a valid testing procedure that rejects $H_0^{(\wG_k,\wG_{k'})}$ when \eqref{eq:cp0} is less than $\alpha$ should properly
control the  selective type I error rate \citep{gao2024selective} at level $\alpha$. A formal definition is provided below.


\begin{definition}[Selective type I error for panel clustering]\label{def1}
For any pair of non-overlapping groups $\{{G}_k,{G}_{k'}\in\text{Cluster}(\bB)\}$ selected via  $k$-means clustering, let $H_0^{({\wG}_k,\wG_{k'})}$ denote the null hypothesis that $\aalpha_{\wG_k}=\aalpha_{\wG_{k'}}$. A test of $H_0^{({\wG}_k,\wG_{k'})}$ based on $\bB$ is said to control the selective type I error rate at level $\alpha$ if
\begin{equation}\label{eq:dfp}
\mbP_{H_0^{({\wG}_k,{\wG}_{k'})}}\left(
\text{reject $H_0^{({\wG}_k,{\wG}_{k'})}$ at level $\alpha$ $\vert$ $\wG_k,\wG_{k'}\in\text{Cluster}(\bB)$ }
\right)\leq\alpha,~~~\forall~ 0\leq\alpha\leq1.
\end{equation}
\end{definition}

Definition \ref{def1} implies that if $H_0^{({\wG}_k,{\wG}_{k'})}$ is true,  the conditional probability of rejecting $H_0^{({\wG}_k,{\wG}_{k'})}$
 at level $\alpha$, given that $\wG_k$ and $\wG_{k'}$ are clusters in $\text{Cluster}(\bB)$, is bounded by $\alpha$.
However, \eqref{eq:cp0} cannot be calculated directly,
 due to the data-dependent selection of clusters.
 Consequently,  $\|\ttheta\trans\bB\|_2$ does not follow a re-scaled $\sqrt{\chi_p^2}$ distribution. To address this, we decompose $\bB$ into the following two components \citep{yun2023selective}:
$$
\bB=\mathcal{P}_{\ttheta}\bB+\mathcal{P}_{\ttheta}^{\perp}\bB=\frac{\XXi\ttheta(\ttheta\trans\XXi\ttheta)^{-1}\ttheta\trans\bB}{
\|\XXi\ttheta(\ttheta\trans\XXi\ttheta)^{-1}\ttheta\trans\bB\|_2}\cdot\|\XXi\ttheta(\ttheta\trans\XXi
\ttheta)^{-1}\ttheta\trans\bB\|_2+\mathcal{P}_{\ttheta}^{\perp}\bB,
$$
where $\XXi$ is the $Np\times Np$ variance-covariance matrix of $\bB$ taking the form $\XXi=\text{bdiag}(\XXi_i,i=1,\cdots,N)$,  $\mathcal{P}_{\ttheta}=\XXi\ttheta(\ttheta\trans\XXi\ttheta)^{-1}\ttheta\trans$, and
$\mathcal{P}_{\ttheta}^{\perp}=\bI_{Np}-\XXi\ttheta(\ttheta\trans\XXi\ttheta)^{-1}\ttheta\trans$ is the orthogonal complement of $\mathcal{P}_{\ttheta}$. From this decomposition, the final test statistic is $\|\cP_{\ttheta}{\bB}\|_2$, and a valid p-value can be constructed as
\begin{equation}\label{eq:ccp}
\mbP_{H_0^{(\wG_k,\wG_{k'})}}\left(
W_{k,k'}\geq\widehat{W}_{k,k'}~\Big|~ \text{Cluster}({\bB})=\Cluster(\widehat{\bB})
\right).
\end{equation}
Here, we condition on the cluster assignments of all individuals $\left\{\text{Cluster}({\bB})=\Cluster(\widehat{\bB})\right\}$ rather than just on $\{\wG_k,\wG_{k'}\in\Cluster(\widehat{\bB})\}$,  because the $k$-means algorithm \ref{alg:1} partitions all $N$ individuals.


As suggested by
\cite{gao2024selective}, to handle the data-dependent cluster selection, we need to condition on the normalized
vector $\frac{\cP_{\ttheta}\bB}{
\|\cP_{\ttheta}\bB\|_2}$ and the orthogonal projection matrix $\mathcal{P}_{\ttheta}^{\perp}\bB$, so that only the
test statistic $\|\cP_{\ttheta}{\bB}\|_2$ remains unknown, and moreover to condition on the range of values of
$\|\cP_{\ttheta}\bB\|_2$ such that the same clustering selection is made.
We thus need to condition on  two additional events:
\begin{eqnarray}\nonumber
P(\widehat{\bB};\wG_{k},\wG_{k'})&=&
\mbP_{H_0^{(\wG_k,\wG_{k'})}}\Big(
W_{k,k'}\geq\widehat{W}_{k,k'}~\Big|~
 \wG_k,\wG_{k'}\in\text{Cluster}(\bB),\\\label{eq:pbb}
&&~~~~~~~\mathcal{P}_{\ttheta}^{\perp}\bB=\mathcal{P}_{\ttheta}^{\perp}\widehat{\bB},
\dir(\cP_{\ttheta}\bB)=
\dir(\cP_{\ttheta}\widehat{\bB})
\Big),
\end{eqnarray}
where $\dir({\bf a})=\frac{{\bf a}}{\|{\bf a}\|_2}1\{{\bf a}\neq0\}$ is the direction of ${\bf a}$.
To find such $\bB$ that satisfies the conditions in \eqref{eq:pbb}, define
$\bB(\phi)=(\bbeta_1^{\ini}(\phi)\trans\cdots,\bbeta_N^{\ini}(\phi)\trans)\trans$, where
\begin{equation}\label{eq:bB}
{\bB}(\phi)=\dir(\cP_{\ttheta}\widehat{\bB})\cdot\phi+\mathcal{P}_{\ttheta}^{\perp}\widehat{\bB}.
\end{equation}
Here, $\phi$ is a random scalar. Note that if $\phi=\|\cP_{\ttheta}\widehat{\bB}\|_2$, then $\bB(\phi)=\widehat{\bB}$. Additionally,
it is straightforward  to verify that $\mathcal{P}_{\ttheta}^{\perp}\widehat{\bB}=\mathcal{P}_{\ttheta}^{\perp}\bB(\phi)$
and $\dir(\cP_{\ttheta}\widehat{\bB})=\dir(\cP_{\ttheta}{\bB}(\phi))$ for any $\phi\geq0$.
It remains to condition on the event $\Cluster(\widehat{\bB})=\Cluster(\bB(\phi))$. Define the one-dimensional set for $\phi$ as
\begin{equation}\label{eq:phi}
\mathcal{S}=\left\{\phi\geq0:\text{Cluster}(\widehat{\bB})=\text{Cluster}({\bB}(\phi))\right\}.
\end{equation}
That is, $\mathcal{S}$ characterizes all possible $\phi$ for which the clustering outcome $\widehat{\cG}$ remains unchanged when $\phi$ replaces the observed test statistic  $\|\cP_{\ttheta}\widehat{\bB}\|_2$.
The following result shows that, conditional on these additional events, the selective p-value in \eqref{eq:pbb} is
computationally tractable and controls the selective type I error rate. The formal validity statement is established under a fixed-\(N\) but \(T\to\infty\) asymptotic framework. Under this framework, \(\bB\in\mathbb R^{Np}\) is a fixed-dimensional random vector, and the selection event is characterized by finitely many quadratic inequalities.
This framework should be interpreted as a conditional selective inference theory for the realized selected groups, rather than as the consistency theory for recovering the true group structure in \cite{liu2020identification}. The latter objective typically requires joint asymptotics with both \(N\) and \(T\) diverging, whereas the validity result below concerns the conditional distribution after the clustering outcome has been selected.

\begin{theorem}\label{thm1}
Suppose that $N$ is fixed while \(T\to\infty\). Conditional on the fixed design matrix \(\tilde{\bx}\), assume that
\[
\sqrt{NT}\,(\bB-\bB^0)\overset{d}{\longrightarrow} \cN(0,\XXi),
\]
where \(\XXi\) is positive definite. Let $\bB$ be an initial estimate generated from the panel model \eqref{eq:1}. Then, conditional on $\text{Cluster}(\widehat{\bB})$,
$\dir(\cP_{\ttheta}\widehat{\bB})$, and $\mathcal{P}_{\ttheta}^{\perp}\widehat{\bB}$, and  for any $\wG_k,\wG_{k'}\in\text{Cluster}(\widehat{\bB})$, under the null hypothesis $H_0^{(\wG_k,\wG_{k'})}$, the test statistic $W_{k,k'}$ converges to a truncated $\chi_p^2$ distribution over the set $\cR$, where
$$
\cR=\left\{\omega:\omega=\phi^2\left(\ttheta\trans\dir(\cP_{\ttheta}\widehat{\bB})\right)\trans
\left(\bR\widehat{\SSigma}_{\widehat{\cG}}\bR\right)^{-1}\left(\ttheta\trans\dir(\cP_{\ttheta}\widehat{\bB})\right),\phi\in\cS\right\}.
 $$
 In particular, the p-value is given by
\begin{equation}\label{eq:t1}
\lim_{T\rightarrow\infty}P(\widehat{\bB};G_{k},G_{k'})=1-\mathbb{F}_{\chi_p^2}\left(
\widehat{W}_{k,k'};\mathcal{R}
\right),
\end{equation}
 where $\mathbb{F}_{\chi_p^2}(\cdot;\cR)$ denotes the cumulative distribution function (CDF) of the $\chi_p^2$ random variable truncated to the set $\cR$.
 Furthermore, under $H_0^{(\wG_k,\wG_{k'})}$,
\begin{equation}\label{eq:tt2}
 \lim_{T\rightarrow\infty}\mbP_{H_0^{(\wG_k,\wG_{k'})}}\left(P({\bB};\wG_{k},\wG_{k'})\leq\alpha\Big\vert \wG_k,\wG_{k'}\in\text{Cluster}(\bB)\right)=\alpha,~~\forall~0\leq\alpha\leq1.
 \end{equation}
\end{theorem}
Equation \eqref{eq:t1} in Theorem \ref{thm1} states that the selective p-value $P(\widehat{\bB};G_{k},G_{k'})$ can be expressed as the survival function of the truncated $\chi_p^2$ random variable, truncated to the set $\cR$ defined in \eqref{eq:phi}. In Section \ref{sec:p},
we provide a detailed guide on
computing this selective p-value in \eqref{eq:t1}.
From \eqref{eq:tt2},  the rejection rule whenever $P(\widehat{\bB};G_{k},G_{k'})$ is below $\alpha$  controls the selective type I
error (Definition \ref{def1}) at level $\alpha$. Therefore, $P(\widehat{\bB};G_{k},G_{k'})$ is a valid  selective p-value.

\subsection{Computation of the selective p-value}\label{sec:p}
As shown by Theorem \ref{thm1},  computing the p-value  in \eqref{eq:t1} requires  characterizing the set $\cS$ defined in \eqref{eq:phi}. Based on the $k$-means clustering in Algorithm \ref{alg:1}, $\cS$ is equivalent to
\begin{equation}\label{eq:cs}
\cS=\left\{
\phi\geq0:\bigcap_{s=0}^S\bigcap_{i=1}^N
\left\{
g_i^{(s)}\left(\bB(\phi)\right)=
g_i^{(s)}(\widehat{\bB})
\right\}
\right\},
\end{equation} 
where $g_i^{(s)}\left(\bB(\phi)\right)$ denotes the group assigned to individual $i$ at the $s$th iteration.
We now outline how to characterize $\cS$. First, we reformulate $\cS$ as
$\left\{\phi\geq0:\cap_{i=1}^N\left\{g_i^{(0)}(\bB(\phi))=g_i^{(0)}(\widehat{\bB})\right\}\right\}\cap
\left\{\phi\geq0:\cap_{s=1}^S\cap_{i=1}^N\left\{g_i^{(s)}(\bB(\phi))=g_i^{(s)}(\widehat{\bB})\right\}\right\}$.
 We analyze the first term of the intersection: according to Step 2 of Algorithm \ref{alg:1}, for $i=1,\cdots,N$,
$g_i^{(0)}(\bB(\phi))=g_i^{(0)}(\widehat{\bB})$ holds if and only if the initially randomly-sampled centroid closest to
$\bbeta_i^{\ini}(\phi)$ is the same as that closest  to $\widehat{\bbeta}_i^{\ini}$.
 This condition can be mathematically expressed using $K-1$ inequalities for each individual. Similarly, the second term follows the same logic, but with centroids that depend on the cluster assignments from the previous iteration.
 We formalize this intuition in Theorem \ref{thm:2}.

\begin{theorem}\label{thm:2}
Suppose that we apply the $k$-means clustering (Algorithm \ref{alg:1}) to the coefficient vector $\widehat{\bB}$ obtained from the linear panel model \eqref{eq:2}, yielding $K$ clusters in at most $S$ iterations. Define
\begin{equation}\label{eq:ws}
\bw_i^{(s)}(k)=\left(\sum_{\{i':\hat{g}_{i'}^{(s)}(\widehat{\bB})=k\}}\tbx_{i'}\trans\tbx_{i'}\right)^{-1}(\tbx_i\trans\tbx_i)\cdot1\left
\{\hat{g}_i^{(s)}(\widehat{\bB})=k\right\}.
\end{equation}
Then, for $\cS$ defined in \eqref{eq:cs}, we have
\begin{align}\label{eq:s1}
\cS=&\left(
\bigcap_{i=1}^N\bigcap_{k=1}^K
\left\{
\phi:\left\|{\bbeta}_i^{\ini}(\phi)-\widehat{\bmm}_{\hat{g}_i^{(0)}(\widehat{\bB})}^{(0)}\left(\bB(\phi)\right)\right\|_{\tbx_i}^2
\leq
\left\|{\bbeta}_i^{\ini}(\phi)-\widehat{\bmm}_{k}^{(0)}\left(\bB(\phi)\right)\right\|_{\tbx_i}^2
\right\}
\right)\\\nonumber
&\cap\left(
\bigcap_{s=1}^S\bigcap_{i=1}^N\bigcap_{k=1}^K
\left\{
\phi:\left\|{\bbeta}_i^{\ini}(\phi)-\sum_{i'=1}^N\bw_{i'}^{(s-1)}\left(
\hat{g}_i^{(s)}(\widehat{\bB})
\right)\bbeta_{i'}^{\ini}(\phi)\right\|_{\tbx_i}^2\right.\right.\\\label{eq:s2}
&~~~~~~~~~~~~~~~~~~~~~~~~~~~~~~~~~~~~~~~~\left.\left.\leq
\left\|{\bbeta}_i^{\ini}(\phi)-\sum_{i'=1}^N\bw_{i'}^{(s-1)}\left(
k\right)\bbeta_{i'}^{\ini}(\phi)\right\|_{\tbx_i}^2
\right\}
\right),
\end{align}
where $\|\ba\|_{\tbx_i}^2=\ba\trans(\tbx_i\trans\tbx_i)\ba$ for any $p\times1$-vector $\ba$.
\end{theorem}

Recall that $\widehat{\bmm}_k^{(0)}(\widehat{\bB})$ denotes the $k$th centroid sampled from the initial estimates $\{\widehat{\bbeta}_i^{\ini}\}_{i=1}^N$ in Step 2 of Algorithm \ref{alg:1}, and
$\hat{g}_i^{(s)}(\widehat{\bB})$ denotes the group index assigned to the $i$th individual at the $s$th iteration (Step 6).
 In summary, Theorem \ref{thm:2} establishes  that
$\cS$ can be formulated as the intersection of ${O}(NKS)$ constituent  sets. Thus, characterizing
the sets in \eqref{eq:s1} and \eqref{eq:s2} is sufficient for our purpose. We now introduce two key lemmas.

\begin{lemma}\label{lem1}
For $\bB(\phi)$ defined in \eqref{eq:bB}, we have that $\|{\bbeta}_i^{\ini}(\phi)-\bbeta_{i'}^{\ini}(\phi)\|_{\tbx_i}^2=a\phi^2+b\phi+\gamma$, where
\begin{eqnarray*}
a&=&\left\|\frac{\left\{v_i\XXi_i\bw_i(\hat{g}_i)^{\mbox{\tiny{T}}}-
v_{i'}\XXi_{i'}\bw_{i'}(\hat{g}_{i'})^{\mbox{\tiny{T}}}\right\}
(\ttheta\trans\XXi\ttheta)^{-1}\bR\widehat{\aalpha}_{\widehat{\cG}}}
{\|\cP_{\ttheta}\widehat{\bB}\|_2}\right\|_{\tbx_i}^2,\\
b&=&2\frac{\left[\left\{v_i\XXi_i\bw_i(\hat{g}_i)^{\mbox{\tiny{T}}}-
v_{i'}\XXi_{i'}\bw_{i'}(\hat{g}_{i'})^{\mbox{\tiny{T}}}\right\}
(\ttheta\trans\XXi\ttheta)^{-1}\bR\widehat{\aalpha}_{\widehat{\cG}}\right]\trans
}
{\|\cP_{\ttheta}\widehat{\bB}\|_2}(\tbx_i\trans\tbx_i)\\
&&\times \left[
\widehat{\bbeta}_i^{\ini}-\widehat{\bbeta}_{i'}^{\ini}-\left\{v_i\XXi_i\bw_i(\hat{g}_i)^{\mbox{\tiny{T}}}-
v_{i'}\XXi_{i'}\bw_{i'}(\hat{g}_{i'})^{\mbox{\tiny{T}}}\right\}
(\ttheta\trans\XXi\ttheta)^{-1}\bR\widehat{\aalpha}_{\widehat{\cG}}
\right],\\
\gamma&=&\left\|\widehat{\bbeta}_i^{\ini}-\widehat{\bbeta}_{i'}^{\ini}-\left\{v_i\XXi_i\bw_i(\hat{g}_i)^{\mbox{\tiny{T}}}-
v_{i'}\XXi_{i'}\bw_{i'}(\hat{g}_{i'})^{\mbox{\tiny{T}}}\right\}
(\ttheta\trans\XXi\ttheta)^{-1}\bR\widehat{\aalpha}_{\widehat{\cG}}\right\|_{\tbx_i}^2.
\end{eqnarray*}
\end{lemma}

\begin{lemma}\label{lem2}
For $\bB(\phi)$ defined in \eqref{eq:bB} and $\bw_i^{(s)}(k)$ in \eqref{eq:ws}, we have that $\|\bbeta_i^{\ini}(\phi)-
\sum_{i'=1}^N\bw_{i'}^{(s-1)}(k)\bbeta_{i'}^{\ini}(\phi)\|_{\tbx_i}^2=\tilde{a}\phi^2+\tilde{b}\phi+\tilde{\gamma}$, where
\begin{eqnarray*}
\tilde{a}&=&\left\|\frac{
\left\{v_i\XXi_i\bw_i(\hat{g}_i)^{\mbox{\tiny{T}}}-\sum_{i'=1}^Nv_{i'}\bw_{i'}^{(s-1)}(k)\XXi_{i'}\bw_{i'}(\hat{g}_{i'})^{\mbox{\tiny{T}}} \right\}(\ttheta\trans\XXi\ttheta)^{-1}\bR\widehat{\aalpha}_{\widehat{\cG}}}
{\|\cP_{\ttheta}\widehat{\bB}\|_2}\right\|_{\tbx_i}^2,\\
\tilde{b}&=&2\frac{\left[
\left\{v_i\XXi_i\bw_i(\hat{g}_i)^{\mbox{\tiny{T}}}-\sum_{i'=1}^Nv_{i'}\bw_{i'}^{(s-1)}(k)
\XXi_{i'}\bw_{i'}(\hat{g}_{i'})^{\mbox{\tiny{T}}}
\right\}(\ttheta\trans\XXi\ttheta)^{-1}\bR\widehat{\aalpha}_{\widehat{\cG}}\right]\trans
}
{\|\cP_{\ttheta}\widehat{\bB}\|_2}(\tbx_i\trans\tbx_i)\times\\
&& \left[
\widehat{\bbeta}_i^{\ini}-\sum_{i'=1}^N\bw_{i'}^{(s-1)}(k)\widehat{\bbeta}_{i'}^{\ini}-
\left\{v_i\XXi_i\bw_i(\hat{g}_i)^{\mbox{\tiny{T}}}-\sum_{i'=1}^Nv_{i'}\bw_{i'}^{(s-1)}(k)
\XXi_{i'}\bw_{i'}(\hat{g}_{i'})^{\mbox{\tiny{T}}}
\right\}(\ttheta\trans\XXi\ttheta)^{-1}\bR\widehat{\aalpha}_{\widehat{\cG}}
\right],\\
\tilde{\gamma}&=&\left\|
\widehat{\bbeta}_i^{\ini}-\sum_{i'=1}^N\bw_{i'}^{(s-1)}(k)\widehat{\bbeta}_{i'}^{\ini}-
\left\{v_i\XXi_i\bw_i(\hat{g}_i)^{\mbox{\tiny{T}}}-\sum_{i'=1}^Nv_{i'}\bw_{i'}^{(s-1)}(k)
\XXi_{i'}\bw_{i'}(\hat{g}_{i'})^{\mbox{\tiny{T}}}
\right\}(\ttheta\trans\XXi\ttheta)^{-1}\bR\widehat{\aalpha}_{\widehat{\cG}}
\right\|_{\tbx_i}^2.
\end{eqnarray*}
\end{lemma}
It follows from Lemmas \ref{lem1} and \ref{lem2} that all inequalities in \eqref{eq:s1} and \eqref{eq:s2} are  quadratic in $\phi$, and
their coefficients can be derived analytically.  Hence, characterizing  the set $\cS$ reduces to solving $\mathcal{O}(NKS)$ quadratic inequalities in $\phi$.
Analogous to Proposition 3 of \cite{chen2023selective}, for fixed
dimension \(p\), the additional computational cost of evaluating the selective
p-value is therefore of order \(\mathcal{O}(NKS)\), apart from the preliminary
cost of obtaining the initial estimators and running the \(k\)-means algorithm.

\section{Improve Power by Conditioning on Less}\label{sec:direction-average-panel}
\subsection{Direction-averaged selective p-value}
The selective p-value in \eqref{eq:pbb} is valid because it conditions on the
selection event, the orthogonal projection $\mathcal{P}_{\ttheta}^{\perp}\bB$,
and the direction $\dir(\cP_{\ttheta}\bB)$.  The last conditioning event is
introduced for computational tractability: after fixing the direction, only the
length $\|\cP_{\ttheta}\bB\|_2$ remains random, and the selection event can be
characterized through a one-dimensional perturbation $\bB(\phi)$.  However, this
direction contains information that is not needed to define the selected null
hypothesis.
 Following the conditioning-on-less principle in post-selection
inference \citep{fithian2014optimal,liu2018more,jewell2022testing,carrington2025improving}, we can potentially improve power by integrating over this direction
instead of conditioning on its observed value.
The reduced-conditioning  p-value is
\begin{equation}
\label{eq:panel-less-p}
    P_{\mathrm{less}}(\widehat{\bB};\wG_{k},\wG_{k'})
    =
    \mathbb P_{{H_0^{(\wG_k,\wG_{k'})}}}
    \left(
    W_{k,k'}\geq\widehat{W}_{k,k'}~\Big|~
    \Cluster(\bB)=\Cluster(\widehat{\bB}),
    \mathcal P_{\ttheta}^{\perp}\bB
    =
    \mathcal P_{\ttheta}^{\perp}\widehat{\bB}
    \right).
\end{equation}


Let $\bD=(\ttheta\trans\XXi\ttheta)^{-1/2}\ttheta\trans\bB$. Then, under the null hypothesis $H_0^{(\wG_k,\wG_{k'})}$, $\bD\stackrel{d}{\rightarrow}\cN(0,\bI_p)$.
Write the polar decomposition:
$$
\bD=\ppsi \xi,~~~~\xi=\|\bD\|_2,~~~\ppsi=\dir(\bD),
$$
so that $\xi\sim\chi_p$, $\ppsi$ is uniformly distributed on the unit sphere
$\mathbb S^{p-1}$, and $\xi$ is independent of $\ppsi$.
Given the information we are conditioning on, as we vary $\ppsi$ and $\xi$ we get data
\begin{equation}
\label{eq:Bu-panel}
    \bB(\ppsi,\xi)
    =
    \mathcal{P}_{\ttheta}^{\perp}\widehat{\bB}
    +
    \XXi\ttheta(\ttheta\trans\XXi\ttheta)^{-1/2}\ppsi \xi.
\end{equation}
It is obvious that $\cP_{\ttheta}^\perp\bB(\ppsi,\xi)=\cP_{\ttheta}^{\perp}\widehat{\bB}$, and $\bB({\ppsi},{\xi})=\widehat{\bB}$
when ${\ppsi}=\dir(\widehat{\bD})$ and
${\xi}=\|\widehat{\bD}\|_2$ with $\widehat{\bD}=(\ttheta\trans\XXi\ttheta)^{-1/2}\ttheta\trans\widehat{\bB}$.  An equivalent formula for the  post-selection p-value \eqref{eq:panel-less-p} is
\begin{equation}\label{eq:redu}
\mbP_{{H_0^{(\wG_k,\wG_{k'})}}}\left(
W_{k,k'}\geq\widehat{W}_{k,k'}~\Big|~\text{Cluster}(\bB(\ppsi,\xi))=\Cluster(\widehat{\bB})
\right).
\end{equation}

In the Appendix \ref{sec:ap4}, we demonstrate that the
less-conditioned selective p-value leads to weakly higher attainable power
than the direction-conditioned selective p-value in \eqref{eq:pbb}, since it avoids
conditioning on the random direction \(\ppsi\), which may carry useful
information under the alternative.


\subsection{Monte Carlo approximation with sampling}

Unfortunately it is not possible to analytically calculate the  p-value \eqref{eq:redu}.  Instead we will resort to using Monte Carlo sampling to estimate it, under the fact that we have a method for calculating the null distribution of $\xi$ given $\ppsi$. Let
$$
\cR_{\less}=\left\{(\ppsi,\xi):\wG_k,\wG_{k'}\in\text{Cluster}(\bB(\ppsi,\xi))\right\}
$$
so the conditioning event for (\ref{eq:redu}) corresponds to $(\ppsi,\xi)$. Furthermore, for a fixed direction $\ppsi$,  define
$$
\cR_{\ppsi}=\left\{
\xi:\wG_k,\wG_{k'}\in\text{Cluster}(\bB(\ppsi,\xi))
\right\}
$$
the one-dimensional  set  of $\xi$ values corresponding to data where we estimate $(\wG_k,\wG_{k'})$ as a pair of clusters. Now we have $p$ additional parameters in $\ppsi$, the truncation region $\cR_{\less}$ becomes much  more complicated to calculate explicitly, as the values of $\xi$ that yields $\wG_k,\wG_{k'}\in\text{Cluster}(\bB(\ppsi,\xi))$ depend on $\ppsi$. However, for a given $\ppsi^*$, by replacing $\bB$ with
$\bB(\ppsi^*,\xi)= \mathcal{P}_{\ttheta}^{\perp}\widehat{\bB}+\XXi\ttheta(\ttheta\trans\XXi\ttheta)^{-1/2}\ppsi^*\xi$, we can calculate
 $\cR_{\ppsi^*}$ using the method in Section \ref{sec:3}. We can then calculate a p-value conditional on $\ppsi^*$:
 $$
 P_{\ppsi^*}(\widehat{\bB};\wG_k,\wG_{k'}) =\frac{\mathbb{P}\left(W_{k,k'}\geq \widehat{W}_{k,k'}\cap \xi\in\cR_{\ppsi^*}\right)}
 {\mathbb{P}\left(\xi\in\cR_{\ppsi^*}\right)}.
 $$

To estimate the overall p-value, we take $M$ samples, $\{\ppsi^{(1)},\cdots,\ppsi^{(M)}\}$, and calculate $\cR_{{\ppsi}^{(m)}}$
for each $\ppsi^{(m)}$. We then estimate the p-value as
\begin{equation}
\frac{\mathbb{P}\left(W_{k,k'}\geq \widehat{W}_{k,k'}\cap \xi\in\cR_{\less}\right)}
 {\mathbb{P}\left(\xi\in\cR_{\less}\right)}
 \approx
 \frac{\frac{1}{M}\sum_{m=1}^M\mathbb{P}\left(W_{k,k'}\geq \widehat{W}_{k,k'}\cap \xi\in\cR_{\ppsi^{(m)}}\right)}
 {\frac{1}{M}\sum_{m=1}^M\mathbb{P}\left(\xi\in\cR_{\ppsi^{(m)}}\right)}.
\end{equation}
This can also be written as a weighted average of individual p-value estimates
\begin{equation}\label{6}
\widetilde{P}_{\less}(\widehat{\bB};\wG_k,\wG_{k'})=\frac{1}{\sum_{m=1}^Mw_m}\sum_{m=1}^Mw_m P_{\ppsi^{(m)}}(\widehat{\bB};\wG_k,\wG_{k'}),
\end{equation}
where $w_m=\mbP(\xi\in\cR_{\ppsi^{(m)}})$. Since each $\cR_{\ppsi^{(m)}}$ consists of a union of intervals, and $\xi\sim\chi_p$, it is straightforward
to calculate $w_m$ and $ P_{\ppsi^{(m)}}(\widehat{\bB};\wG_k,\wG_{k'})$.

As $M\to \infty$, this Monte Carlo approximation converges to the post-selection p-value (\ref{eq:panel-less-p}). For finite $M$, however, it is not guaranteed to be a proper p-value, since there is no assurance that its distribution will be uniform on \([0,1]\) when the null holds and we condition on opting to test the null hypothesis.
As enlightened by \cite{carrington2025improving}, we can overcome this issue by just setting one of
the $\ppsi$ values to the value for the observed direction $\dir(\widehat{\bD})$. Specifically, let $\ppsi^{(1)}=\dir(\widehat{\bD})$. Simulate $\ppsi^{(2)},
\cdots,\ppsi^{(M)}$ independently from the null distribution for $\ppsi$, and calculate the p-value as $\widetilde{P}_{\less}(\widehat{\bB};\wG_k,\wG_{k'})$ in \eqref{6}.
The following theorem establishes that the resulting post-selection p-value follows a uniform distribution on \([0,1]\) under the null hypothesis, regardless of the value of
$M$.
\begin{theorem}
For $\widetilde{P}_{\less}(\widehat{\bB};\wG_k,\wG_{k'})$ defined in \eqref{6}, given that there is one $m^*\in\{1,\cdots,M\}$ such that
$\ppsi^{(m^*)}$ corresponding to the observed direction $\dir(\widehat{\bD})$, while all other $\ppsi^{(m)}$ are  drawn independently from their null distribution.
Then under $H_0^{(\wG_k,\wG_{k'})}$, $\widetilde{P}_{\less}(\widehat{\bB};\wG_k,\wG_{k'})$ is valid  in the sense that
$$
\lim_{T\rightarrow\infty}\mbP_{H_0^{(\wG_k,\wG_{k'})}}\left(\widetilde{P}_{\less}({\bB};\wG_{k},\wG_{k'})\leq\alpha\Big\vert \wG_k,\wG_{k'}\in\text{Cluster}(\bB)\right)=\alpha,~~\forall~0\leq\alpha\leq1.
$$
\end{theorem}

\section{Extensions}\label{sec:4}
\subsection{Testing for differences due to a covariate}\label{sec:41}
In panel clustering, researchers  are often interested not only
in examining the overall differences across all covariates  collectively but also in identifying the specific explanatory variables that drive the differences between clusters \citep{chen2025testing}. On the other hand, in addition to the model \eqref{eq:1}, it is sensible to consider mixed linear panel models that contain
both homogeneous  and heterogeneous slope coefficients \citep{pesaran1999pooled},
\begin{equation}\label{eq:3}
    y_{it}=\bx_{1,it}\trans\bbeta_{1i}^0+\bx_{2,it}\trans\bbeta_{2}^0+\eta_i+u_{it},
\end{equation}
where $\bbeta_{1i}$ captures group-specific patterns, and $\bbeta_2$ is common across all individuals.
This special mixed-structure panel model has also been studied by  \cite{wang2021identifying}.
However, conducting valid post-selection inference for a specific parameter (e.g., the $j$th element) within $\bbeta_{1i}$  remains an under-explored problem.
These  two aspects collectively motivate the need for testing differences due to a covariates.

Specifically, for any pair of clusters $(\wG_k,\wG_{k'})$ and for a given feature $j$ ($j\in\{1,\cdots,p\}$), we aim to  test whether the group-specific coefficient of the $j$-th covariate differs significantly between two selected groups. Consider the hypothesis
\begin{equation}\label{eq:t2}
H_{0j}^{(\wG_k,\wG_{k'})}:{\alpha}_{\wG_k,j}={\alpha}_{\wG_{k'},j},~~~\text{versus}~~~
H_{1j}^{(\widehat{G}_k,\widehat{G}_{k'})}:{\alpha}_{\wG_k,j}\neq{\alpha}_{\wG_{k'},j},
\end{equation}
where ${\alpha}_{\wG_k,j}$ is the $j$th component of ${\aalpha}_{\wG_k}$.
Addressing this question helps to identify the important features that drive the separation between clusters, thereby enhancing the interpretability and practical utility of the clustering results. To control the selective type I error for hypothesis in \eqref{eq:t2},
a valid p-value similar to \eqref{eq:pbb} is given by
\begin{eqnarray}\label{eq:pb1}
P_j(\widehat{\bB};\wG_{k},\wG_{k'})=
\mathbb{P}\Big(
|\ttheta_j\trans\bB|\geq |\ttheta_j\trans\widehat{\bB}|~\Big|~
\widehat{G}_k,\wG_{k'}\in\text{Cluster}({\bB}),\mathcal{P}_{\ttheta_j}^{\perp}\bB=\mathcal{P}_{\ttheta_j}^{\perp}\widehat{\bB}
\Big),
\end{eqnarray}
where $\ttheta_j$ is the $j$th column of $\ttheta$ and $\cP_{\ttheta_j}=\XXi\ttheta_j(\ttheta_j\trans\XXi\ttheta_j)^{-1}\ttheta_j\trans$. This approach retains control of the selective Type I error rate (Definition \ref{def1}), as demonstrated by Proposition 3 in \cite{fithian2014optimal}. Analogously to \eqref{eq:bB}, we rewrite ${\bB}$ as
\begin{equation}\label{eq:bB1}
\bB(\kappa)=\XXi\ttheta_j(\ttheta_j\trans\XXi\ttheta_j)^{-1}\kappa+
\mathcal{P}_{\ttheta_j}^{\perp}\widehat{\bB},
\end{equation}
where the random scalar  $\kappa\in\mathbb{R}$ is chosen such that the clustering outcome remains unchanged when replacing the observed quantity $\ttheta_j\trans\widehat{\bB}$ with $\kappa$. That is
\begin{equation}\label{eq:cs1}
\cS_j=\{\kappa\in\mathbb{R}:~\text{Cluster}(\bB(\kappa))=\text{Cluster}(\widehat{\bB})\}.
\end{equation}
We now present the following theoretical results.
\begin{theorem}\label{thm2} 
Suppose $j\in\{1,\cdots,p\}$ and $\bB$ is the initial estimator generated from the panel model \eqref{eq:1}. The same fixed-\(N\) and \(T\to\infty\) framework as in Theorem \ref{thm1} is imposed.
Then,  for any $G_k,G_{k'}\in\text{Cluster}(\widehat{\bB})$, under the null hypothesis $H_{0j}^{(\wG_k,\wG_{k'})}$, the test statistic
$\ttheta_j\trans\bB$ converges to a truncated $\cN(0,\vartheta_j^2)$ distribution truncated to $\cS_{j}$, where $\vartheta_j^2=\bR_j\SSigma_{\widehat{\cG}}\bR_j\trans$, and  $\bR_j$ is the $j$th row vector of $\bR$.
In particular, the p-value defined in \eqref{eq:pb1} is
\begin{equation}\label{eq:sp1}
\lim_{T\rightarrow\infty}P_j(\widehat{\bB};\wG_{k},\wG_{k'})=1-\mathbb{F}\left(|\ttheta_j\trans\widehat{\bB}|;0,\vartheta_j^2,\cS_{j}\right)+
\mathbb{F}\left(-|\ttheta_j\trans\widehat{\bB}|;0,\vartheta_j^2,\cS_{j}\right)
\end{equation}
where $\mathbb{F}(t;\mu,\sigma^2,\cS_j)$ denotes the CDF of a $\cN(\mu,\sigma^2)$ distribution truncated to the set $\cS_{j}$.
Furthermore, the test that rejects $H_{0j}^{(\wG_k,\wG_{k'})}$ whenever $P_j(\widehat{\bB};\wG_{k},\wG_{k'})\leq\alpha$ controls the selective type I error rate at level $\alpha$ asymptotically, in the sense of Definition \ref{def1}.
\end{theorem}
Theorem \ref{thm2} implies that testing $H_{0j}^{(\wG_k,\wG_{k'})}$ using \eqref{eq:pb1} controls the selective Type I error rate. Similar to Theorem \ref{thm1}, if we use  $\vartheta_j^{-2}\left(\bR_j\aalpha_{\widehat{\cG}}\right)^2$ as the test statistic, then it follows a truncated $\chi_1^2$ distribution truncated to $\vartheta_j^{-2}\cS_{j}^2$. In addition, computing the selective p-value
in \eqref{eq:sp1} requires characterizing the truncation set in \eqref{eq:cs1}. Since the $k$-means clustering algorithm iteratively updates group assignments, we condition on all intermediate clusters:
\begin{equation}\label{eq:css}
\cS_j=\left\{
\kappa\in\mbR:\bigcap_{s=0}^S\bigcap_{i=1}^N
\left\{
g_i^{(s)}\left(\bB(\kappa)\right)=
g_i^{(s)}(\widehat{\bB})
\right\}
\right\},
\end{equation}
where $\bB(\kappa)$ is defined in \eqref{eq:bB1}.  Regarding computation, $P_j(\widehat{\bB};\wG_{k},\wG_{k'})$ can be evaluated using a  univariate truncated Gaussian distribution.  The approach outlined in Section \ref{sec:p} can be extended to efficiently compute the truncation set $\cS_j$. More details can be found in Section \ref{secb1} of the Appendix.

\subsection{The GMM framework}\label{sec:42}
 We now extend the post-selection inference framework to the generalized method of moments (GMM) estimation of the linear panel structure model in \eqref{eq:1}, where some regressors are lagged dependent variables or endogenous, such that $\mbE(\bx_{it} u_{it}) \neq {\bf0}$ for $t = 1, \dots, T$. To eliminate the individual effects $\eta_i$, we consider  the first-differenced system:
\begin{equation}\label{eq:fd}
\Delta y_{it}=\Delta\bx_{it}\trans\bbeta_i^0+\Delta u_{it},
\end{equation}
where $\Delta y_{it}=y_{it}-y_{i,t-1}$ for $t=1,\cdots,T$ and $i=1,\cdots,N$, and we assume initial observations  $y_{i0}$ and $\bx_{i0}$ are available. We further assume there exists a $q\times1$ vector of instruments $\bz_{it}$ with $q\geq p$ such that $\mbE(\bz_{it}\trans\Delta u_{it})=0$ for each $t=1,\cdots,T$.
Analogous to \eqref{eq:21}, the GMM objective function is
\begin{equation}\label{eq:31}
\min_{\underline{\aalpha},\bg_N}\sum_{i=1}^N
\left[\frac{1}{T}\sum_{t=1}\trans\bz_{it}(\Delta y_{it}-\Delta\bx_{it}\trans\aalpha_{g_i})\right]\trans\OOmega_{i,NT}\left[\frac{1}{T}\sum_{t=1}\trans\bz_{it}(\Delta y_{it}-\Delta\bx_{it}\trans\aalpha_{g_i})\right],
\end{equation}
where $\OOmega_{i,NT}$ is a $q\times q$ positive definite matrix.

The GMM objective function of the $i$th individual for a given $\bbeta_i\in\mathbb{R}^p$ is
\begin{eqnarray}\nonumber
&&\left[\sum_{t=1}\trans\bz_{it}(\Delta{y}_{it}-\Delta{\bx}_{it}\trans\bbeta_i)\right]\trans
\OOmega_{i,NT}\left[\sum_{t=1}\trans\bz_{it}(\Delta y_{it}-\Delta\bx_{it}\trans\bbeta_i)\right]\\\nonumber
&=&(\breve{\by}_i-\breve{\bx}_i\bbeta_i)\trans\bz_i\OOmega_{i,NT}\bz_i\trans\left(\breve{\by}_i-\breve{\bx}_i\bbeta_i\right)\\\nonumber
&=&(\breve{\by}_i-\breve{\bx}_i\widehat{\bbeta}_i^{\ini}+
\breve{\bx}_i\widehat{\bbeta}_i^{\ini}-\breve{\bx}_i{\bbeta}_i)\trans\bz_i\OOmega_{i,NT}\bz_i\trans
(\breve{\by}_i-\breve{\bx}_i\widehat{\bbeta}_i^{\ini}+
\breve{\bx}_i\widehat{\bbeta}_i^{\ini}-\breve{\bx}_i{\bbeta}_i)\\ \label{eq:32}
&=&(\widehat{\bbeta}_i^{\ini}-{\bbeta}_i)\trans\breve{\bx}_i\trans\bz_i\OOmega_{i,NT}\bz_i\trans\breve{\bx}_i
(\widehat{\bbeta}_i^{\ini}-{\bbeta}_i)+\text{constant},
\end{eqnarray}
where $\widehat{\bbeta}_i^{\ini}=(\breve{\bx}_i\trans\bz_i\OOmega_{i,NT}\bz_i\trans\breve{\bx}_i)^{-1}
\breve{\bx}_i\trans\bz_i\OOmega_{i,NT}\bz_i\trans\breve{\by}_i$ is the initial GMM estimate of the $i$th individual with  $\breve{\by}_i=(\Delta{y}_{i1},\cdots,\Delta{y}_{iT})\trans$,  $\breve{\bx}_i=(\Delta\bx_{i1},\cdots,\Delta\bx_{iT})\trans$ and $\bz_i=(\bz_{i1},\cdots,\bz_{iT})\trans$.
 The group-specific estimate for a given partition set $\cG=\{G_1,\cdots,G_K\}$ is
\begin{eqnarray}\nonumber
\widehat{\aalpha}_{G_k}&=&\left(\sum_{i\in G_k}\breve{\bx}_i\trans\bz_i\OOmega_{i,NT}\bz_i\trans\breve{\bx}_i\right)^{-1}\sum_{i\in G_k}\breve{\bx}_i\trans\bz_i\OOmega_{i,NT}\bz_i\trans\breve{\by}_i\\\nonumber
&=&\sum_{i\in G_k}\left\{
\left(\sum_{i'\in G_k}\breve{\bx}_{i'}\trans\bz_{i'}\OOmega_{i',NT}\bz_{i'}\trans\breve{\bx}_{i'}\right)^{-1}
\breve{\bx}_i\trans\bz_i\OOmega_{i,NT}\bz_i\trans\breve{\by}_i
\right\}\\\nonumber
&=&
\sum_{i\in G_k}\left\{
\left(\sum_{i'\in G_k}\breve{\bx}_{i'}\trans\bz_{i'}\OOmega_{i',NT}\bz_{i'}\trans\breve{\bx}_{i'}\right)^{-1}(\breve{\bx}_i\trans\bz_i\OOmega_{i,NT}\bz_i\trans\breve{\bx}_i)(\breve{\bx}_i\trans\bz_i\OOmega_{i,NT}\bz_i\trans\breve{\bx}_i)^{-1}
\breve{\bx}_i\trans\bz_i\OOmega_{i,NT}\bz_i\trans\breve{\by}_i
\right\}\\ \label{eq:33}
&=&\sum_{i:g_i=k}\bw_i(k)\widehat{\bbeta}_i^{\ini},
\end{eqnarray}
where $\bw_i(k)=\left(\sum_{i'\in G_k}\breve{\bx}_{i'}\trans\bz_{i'}\OOmega_{i',NT}\bz_{i'}\trans\breve{\bx}_{i'}\right)^{-1}(\breve{\bx}_i\trans\bz_i\OOmega_{i,NT}\bz_i\trans\breve{\bx}_i)$.
Thus, $\widehat{\aalpha}_{G_k}$ is a linear combination of initial  GMM estimates $\widehat{\bbeta}_i^{\ini}$ for $i\in G_k$.

The $k$-means procedure outlined in Algorithm \ref{alg:1} can be adapted to the GMM estimation with minor modifications.
Correspondingly,  the selective p-value for the GMM estimator takes a form similar to \eqref{eq:pbb}, with the initial estimate $\widehat{\bbeta}_1^{\ini}$ in $\widehat{\bB}$ replaced by the GMM initial estimates from \eqref{eq:32}, and the building block $\bw_i(k)$ in $\ttheta$ replaced by the weighting matrix defined in \eqref{eq:33}. Parallel to Theorem \ref{thm1},  the key properties of the selective p-value for the GMM estimator can be established.  Regarding computation, the approach in Section \ref{sec:p} can be extended to  efficiently compute the conditioning set by solving $O(NKS)$  quadratic inequalities. We leave these details to Section \ref{secb2} of the Appendix.


\section{Simulation Studies}\label{sec:5}
\subsection{Basic settings}
 In this section, we evaluate the finite-sample performance of the proposed conditional selective inference method through six Monte Carlo experiments. These experiments employ DGPs similar to those used in \cite{su2016identifying,liu2020identification} and \cite{mehrabani2023estimation}.
 The first three experiments focus on static panel data models and employ the LS estimation. The fourth experiment involves both the LS and GMM estimators in the context of dynamic panel data models containing a lagged dependent variable and several exogenous regressors.
 The last two experiments are designed to test the difference between the coefficients of a single regressor. For the first four experiments, we test the null
 hypothesis $H_0^{(\wG_k,\wG_{k'})}:\aalpha_{\wG_k}=\aalpha_{\wG_{k'}}$ versus  $H_1^{(\wG_k,\wG_{k'})}:{\aalpha}_{\wG_k}\neq{\aalpha}_{\wG_{k'}}$, where $\wG_k$ and $\wG_{k'}$ are a randomly-chosen pair of clusters $k\neq k'\in\{1,2,3\}$. The last two experiments are conducted for testing $H_{0j}^{(\wG_k,\wG_{k'})}:{\alpha}_{\wG_k,j}={\alpha}_{\wG_{k'},j}$ versus
 $H_{1j}^{(\wG_k,\wG_{k'})}:{\alpha}_{\wG_k,j}\neq{\alpha}_{\wG_{k'},j}$, where $j\in\{1,\cdots,p\}$ is selected randomly. In the computation process, the unknown quantities $\SSigma^0$ and $\XXi$ are obtained using the plug-in estimators.
 We consider the following DGPs:

\noindent{\bf DGP 1 (Static panel with two exogenous regressors):} The first DGP is based on the following static panel model with two exogenous regressors:
$$
y_{it}=x_{it,1}\beta_{i,1}+x_{it,2}\beta_{i,2}+\eta_i+u_{it},~~~i=1,\cdots,N,~~t=1,\cdots,T,
$$
where the regressors $\bx_{it}=(x_{it,1},x_{it,2})\trans$ are generated as  $x_{it,1}=0.2\eta_i+e_{it,1}$ and $x_{it,2}=0.2\eta_i+e_{it,2}$ where $e_{it,1}$ and $e_{it,2}$ are both i.i.d. $\cN(0,1)$ across $i$ and $t$. The individual fixed effects $\eta_i$ and the idiosyncratic errors $u_{it}$ follow a standard normal distribution, and are mutually independent. The true number of groups is $K^0=3$, with the true coefficients
$$
\left(
\aalpha_{G_1^0}^0,\aalpha_{G_2^0}^0,\aalpha_{G_3^0}^0
\right)=\left(
\left[
\begin{array}{c}
1-\delta\\
1
\end{array}
\right],
\left[
\begin{array}{c}
1\\
1+\sqrt{3}\delta
\end{array}
\right],
\left[
\begin{array}{c}
1+\delta\\
1
\end{array}
\right]
\right),
$$
where $\delta\in\{0, 0.2, 0.4,\cdots,2\}$.

\noindent{\bf DGP 2:} The same as DGP 1, but $u_{it}\sim\frac{t(3)}{\sqrt{3}}$.

\noindent{\bf DGP 3:} The same as DGP 1, but $u_{it}\sim\frac{\chi_3^2-3}{\sqrt{6}}$.

\noindent{\bf DGP 4 (Dynamic panel AR(1) with two exogenous regressors):} This DGP specifies a dynamic panel model with an autoregressive term and two exogenous regressors
$$
y_{it}=y_{i,t-1}\beta_{i,1}^0+x_{it,1}\beta_{i,2}^0+x_{it,2}\beta_{i,3}^0+\eta_i(1-\beta_{i,1}^0)+u_{it},
$$
where the exogenous regressors $x_{it,1}$ and $x_{it,2}$  are independently drawn from the standard normal distribution $\cN(0,1)$, and are mutually independent as well as independent of the error term. The initial observation is generated as $y_{i0}=\beta_{i,2}^0x_{it,1}+\beta_{i,3}^0x_{it,2}+\eta_i+u_{i0}$, which guarantees that the time series $(y_{i0},\cdots,y_{iT})$ is strictly stationary. Both the individual fixed effects $\eta_i$
  and the idiosyncratic errors  $u_{it}$
  follow the standard normal distribution $\cN(0,1)$, and are independent across individuals $i$ and time periods $t$. The true number of groups is  $K^0=3$, with the following group-specific coefficient vectors
  $$
\left(
\aalpha_{G_1^0}^0,\aalpha_{G_2^0}^0,\aalpha_{G_3^0}^0
\right)=\left(
\left[
\begin{array}{c}
0.6-\kappa\\
1-\delta\\
1
\end{array}
\right],
\left[
\begin{array}{c}
0.6\\
1\\
1+\sqrt{3}\delta
\end{array}
\right],
\left[
\begin{array}{c}
0.6+\kappa\\
1+\delta\\
1
\end{array}
\right]
\right).
$$

\noindent{\bf DGP 5 (Linear panel with  $p=4$):} The data are generated as
$$
y_{it}=\bx_{it}\trans\bbeta_i^0+\eta_i+u_{it},
$$
where $\bx_{it}$ is a $p\times1$ vector with the $j$th component given by $x_{it,j}=0.2\eta_i+e_{it,j}$, $j=1,\cdots,p$, and $e_{it,j}$, $u_{it}$, and the individual effects $\eta_i$ are all i.i.d. standard normal and mutually independent of each other.
The true coefficients $\bbeta_i^0$ can be categorized into three distinct groups with the group-specific parameter values given by
$$
\left(
\aalpha_{G_1^0}^0,\aalpha_{G_2^0}^0,\aalpha_{G_3^0}^0
\right)=\left(
\left[
\begin{array}{c}
(1-\delta)_{\frac{p}{2}\times1}\\
(1)_{\frac{p}{2}\times1}\\
\end{array}
\right],
\left[
\begin{array}{c}
(1)_{\frac{p}{2}\times1}\\
(1+\delta)_{\frac{p}{2}\times1}\\
\end{array}
\right],
\left[
\begin{array}{c}
(1+\delta)_{\frac{p}{2}\times1}\\
(1)_{\frac{p}{2}\times1}\\
\end{array}
\right]
\right),
$$
where $(a)_{n\times1}$ represents an $n$-dimensional vector with all elements being $a$. In this setup, we consider $p=4$.

\noindent{\bf DGP 6:} The same as DGP 5, but $p=6$.

For all DGPs,
we consider combinations of $(N,T)$ with $N=\{60,120\}$ and $T=\{15,25\}$.  The true group structures have equal number of individuals with $G_1^0=\{1,\cdots,\frac{N}{3}\}$,
 $G_2^0=\{\frac{N}{3}+1,\cdots,\frac{2N}{3}\}$ and $G_3^0=\{\frac{2N}{3}+1,\cdots,N\}$. Finally, the number of Monte Carlo simulations in all experiments
is set as $M=10000$.

\subsection{Selective type I error}


\begin{figure}[!h]
\centering
   \includegraphics[scale =0.4]{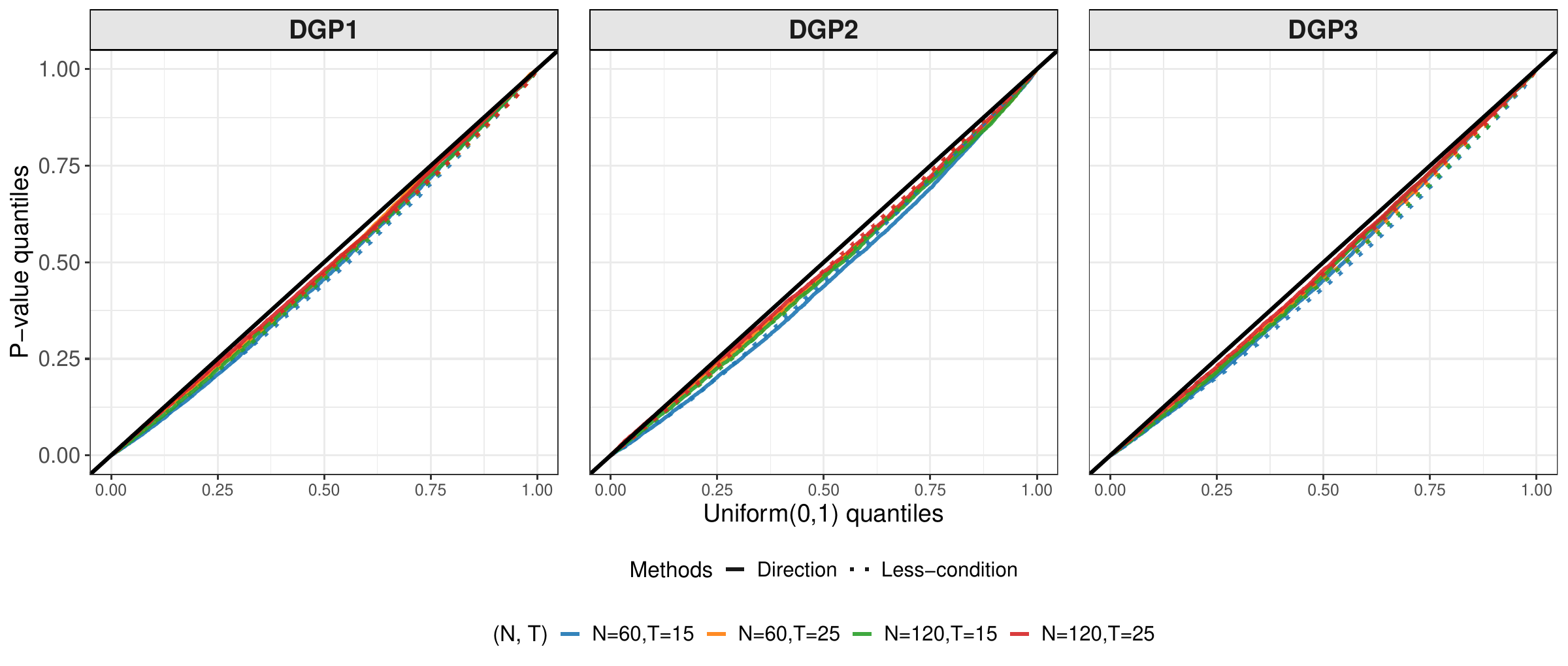}
	\caption{Q-Q plots of the conditional p-values  obtained from DGPs 1--3 under $H_0^{(\wG_k,\wG_{k'})}$ (i.e., $\delta=0$) using the test proposed in Section \ref{sec:31}}
	\label{fig:size123}
\end{figure}

 \begin{figure}[!h]
	\centering
   \includegraphics[scale =0.37]{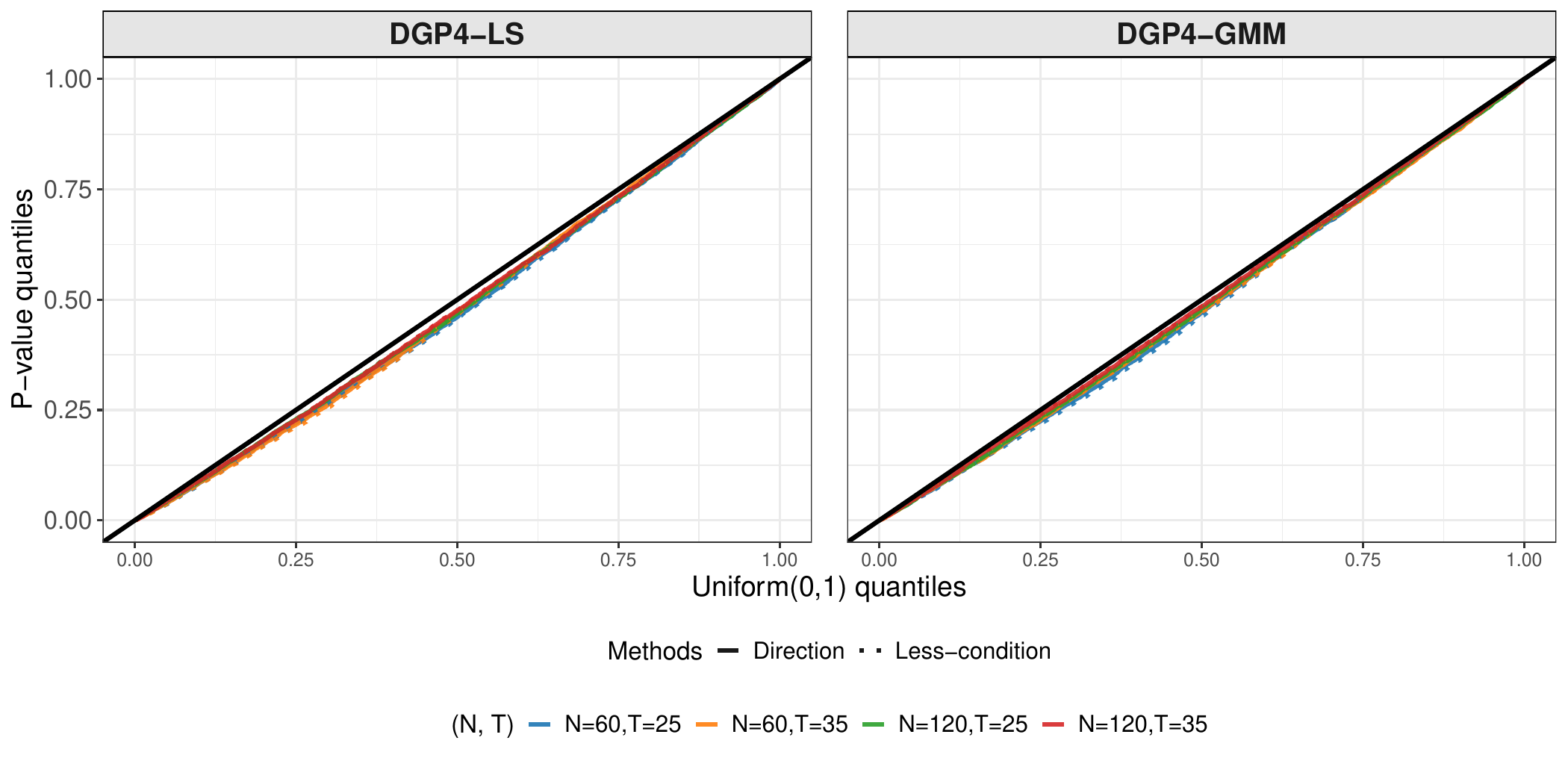}
	\caption{Q-Q plots of the conditional p-values  obtained from DGP 4 under $H_0^{(\wG_k,\wG_{k'})}$ (i.e., $\delta=0$ and $\kappa=0$) using the LS test in Section \ref{sec:31} and the GMM test in Section \ref{sec:42}.}
	\label{fig:size4}
\end{figure}

 \begin{figure}[!h]
	\centering
   \includegraphics[scale =0.35]{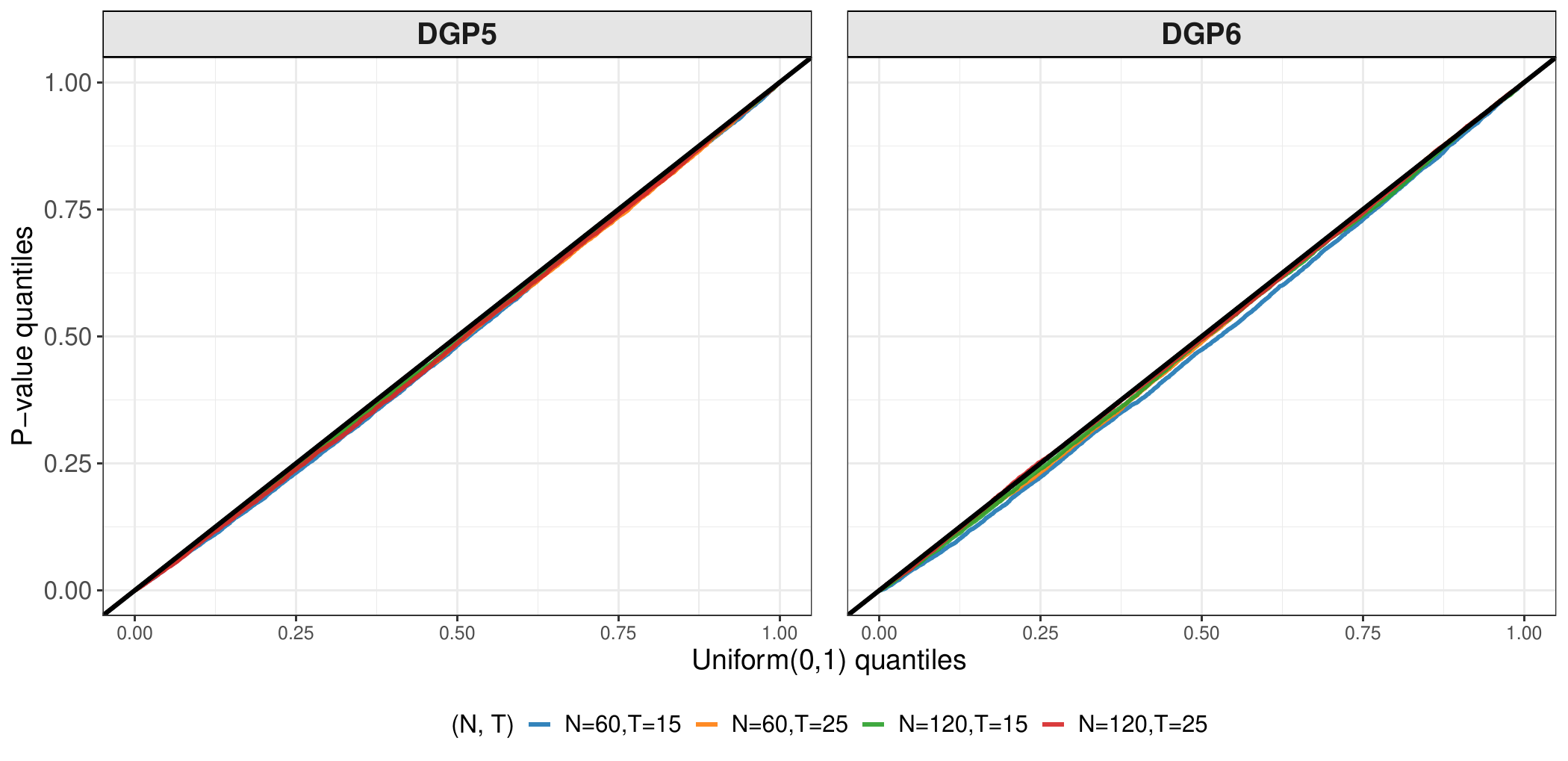}
	\caption{Q-Q plots of the conditional p-values  obtained from DGPs 5--6 under $H_0^{(\wG_k,\wG_{k'})}$ (i.e., $\delta=0$) using the test proposed in Section \ref{sec:42}.}
	\label{fig:size56}
\end{figure}

We compute the p-values for all DGPs under the null hypothesis ($\delta=0$ and $\kappa=0$ for DGP 4). Theoretically, a valid selective p-value should follow a Uniform$(0,1)$ distribution under the null hypothesis, as defined in Definition \ref{def1}. In addition to the direction-conditioned p-value in Section \ref{sec:3}, we also evaluate the less-conditioned p-value proposed in Section \ref{sec:direction-average-panel}. Since the latter integrates over the direction of the selected contrast rather than conditioning on its observed value, it is important to verify that the resulting power gain is not obtained at the cost of inflated selective type I error.

Figure \ref{fig:size123} displays the Q-Q plots of the empirical distribution of the p-values evaluated against the Uniform$(0,1)$ distribution under DGPs 1--3. The empirical quantiles of both the direction-conditioned and less-conditioned p-values are close to the $45^\circ$ line. This indicates that, for the linear panel model \eqref{eq:1}, the original conditional p-value defined in \eqref{eq:t1} and the less-conditioned p-value both effectively control the selective type I error. The result supports Theorem \ref{thm1} for the direction-conditioned procedure and the finite-sample Monte Carlo validity result in Section \ref{sec:direction-average-panel} for the less-conditioned procedure.
We next evaluate the type I error under DGP 4. Since DGP 4 is a dynamic panel, we report the p-value results based on the LS and GMM estimations in Figure \ref{fig:size4}. For all values of $N$ and $T$, the p-values based on the LS estimation are stochastically smaller than a Uniform$(0,1)$ random variable, suggesting that the LS-based test leads to an inflated type I error rate. This is expected because the LS estimator introduces bias in estimating the AR(1) parameter in dynamic panel models. In contrast, the GMM-based selective p-values, including their less-conditioned counterparts, align much more closely with the Uniform$(0,1)$ benchmark. Therefore, the GMM extension in Section \ref{sec:42} remains valid after replacing the direction-conditioned p-value with the less-conditioned p-value.
Finally, Figure \ref{fig:size56} displays the Q-Q plots of the observed p-values under DGPs 5--6, which are designed to test for differences in coefficients of a single regressor between a pair of selected groups. The empirical distributions remain close to Uniform$(0,1)$ for both conditioning schemes. This confirms that the covariate-specific selective inference procedure in Section \ref{sec:41} controls selective type I error even when the less-conditioned p-value is used. The conclusion continues to hold when the covariate dimension increases to six.

\subsection{Conditional power analysis}

This section demonstrates that our proposed test exhibits substantial power to reject the null hypothesis
when it is false. We generate panel data from DGPs 1--6 with varying $\delta\in\{0.25,0.5,\cdots,2.5\}$, representing effects ranging from weak
to strong. The $K^0=3$ groups are equidistant in the sense that the Euclidean distances between each pair of group-specific slope vectors are $2\delta$ in DGPs 1--3, $\sqrt{\kappa^2+4\delta^2}$ in DGP 4, and $\sqrt{p}\delta$ in DGPs 5--6. Since hypotheses are tested based on a pair of estimated clusters $\wG_k$ and $\wG_{k'}$,  which may not correspond to the true clusters, we evaluate the performance  using two metrics:  \emph{conditional power} and \emph{recovery probability} \citep{chen2023selective,gao2024selective}. The conditional  power is defined as the probability of rejecting the null hypotheses \eqref{eq:t} or \eqref{eq:t2} given that $\wG_k$ and $\wG_{k'}$ are true group memberships. Based on $M$ simulated datasets, we estimate it as:
\begin{equation}\label{eq:cp}
\text{Conditional power}=
\frac{
\sum_{m=1}^M{\bf1}\left\{
\left\{\wG_k^{(m)},\wG_{k'}^{(m)}\right\}\subseteq
\left\{G_1^0,\cdots,G_{K^0}^0\right\},P^{(m)}\leq\alpha
\right\}
}
{\sum_{m=1}^M{\bf1}\left\{
\left\{\wG_k^{(m)},\wG_{k'}^{(m)}\right\}\subseteq
\left\{G_1^0,\cdots,G_{K^0}^0\right\}
\right\}},
\end{equation}
where $\{G_1^0,\cdots,G_{K^0}^0\}$ are the true cluster, $\alpha$ is the significance level, and $P^{(m)}$ and $\wG_k^{(m)},\wG_{k'}^{(m)}$ correspond to the p-values and the chosen clusters in the $m$th simulation, respectively.  Since the quantity in \eqref{eq:cp} is conditional on the event that both $\wG_k$ and $\wG_{k'}$ correspond to the true groups, we also estimate the frequency of this event, referred to as the recovery probability:
\begin{equation}\label{eq:rp}
\text{Recovery probability}=
\frac{
\sum_{m=1}^M{\bf1}\left\{
\left\{\wG_k^{(m)},\wG_{k'}^{(m)}\right\}\subseteq
\left\{G_1^0,\cdots,G_{K^0}^0\right\}
\right\}
}
{M}.
\end{equation}

\begin{table}[!h]
  \centering
  \footnotesize
  \caption{Conditional power for DGPs 1--3.}
  \label{tab:p123}
  \begin{tabular}{c cccc cccc}
    \toprule
 \multirow{2}{*}{$\delta$}  &  \multicolumn{4}{c}{Direction-conditioned} & \multicolumn{4}{c}{Less-conditioned} \\
\cmidrule(lr){2-5} \cmidrule(lr){6-9}
 & $(60,15)$ & $(60,25)$ & $(120,15)$ & $(120,25)$ & $(60,15)$ & $(60,25)$ & $(120,15)$ & $(120,25)$ \\
    \midrule
 \multicolumn{9}{c}{DGP 1}\\
0.25	&	0.000 	&	0.000 	&	0.000 	&	0.000 	&	0.000 	&	0.000 	&	0.000 	&	0.000 	\\
0.5	&	0.694 	&	0.877 	&	0.000 	&	0.655 	&	0.839 	&	0.962 	&	0.000 	&	0.783 	\\
0.75	&	0.930 	&	0.991 	&	0.749 	&	0.856 	&	0.982 	&	0.999 	&	0.843 	&	0.942 	\\
1	&	0.990 	&	1.000 	&	0.867 	&	0.913 	&	0.999 	&	1.000 	&	0.984 	&	1.000 	\\
1.25	&	0.999 	&	1.000 	&	0.911 	&	0.940 	&	1.000 	&	1.000 	&	1.000 	&	1.000 	\\
1.5	&	1.000 	&	1.000 	&	0.935 	&	0.957 	&	1.000 	&	1.000 	&	1.000 	&	1.000 	\\
1.75	&	1.000 	&	1.000 	&	0.949 	&	0.963 	&	1.000 	&	1.000 	&	1.000 	&	1.000 	\\
2	&	1.000 	&	1.000 	&	0.957 	&	0.973 	&	1.000 	&	1.000 	&	1.000 	&	1.000 	\\
2.25	&	1.000 	&	1.000 	&	0.963 	&	0.975 	&	1.000 	&	1.000 	&	1.000 	&	1.000 	\\
2.5	&	1.000 	&	1.000 	&	0.970 	&	0.980 	&	1.000 	&	1.000 	&	1.000 	&	1.000 	\\
	 \multicolumn{9}{c}{DGP 2}\\																
0.25	&	0.000 	&	0.000 	&	0.000 	&	0.000 	&	0.000 	&	0.000 	&	0.000 	&	0.000 	\\
0.5	&	0.784 	&	0.882 	&	0.000 	&	0.699 	&	0.846 	&	0.952 	&	0.000 	&	0.805 	\\
0.75	&	0.921 	&	0.974 	&	0.776 	&	0.835 	&	0.976 	&	0.995 	&	0.858 	&	0.954 	\\
1	&	0.975 	&	0.990 	&	0.853 	&	0.899 	&	0.994 	&	0.999 	&	0.984 	&	0.999 	\\
1.25	&	0.987 	&	0.996 	&	0.900 	&	0.925 	&	0.998 	&	1.000 	&	0.999 	&	1.000 	\\
1.5	&	0.994 	&	0.998 	&	0.928 	&	0.945 	&	1.000 	&	1.000 	&	1.000 	&	1.000 	\\
1.75	&	0.997 	&	0.999 	&	0.942 	&	0.956 	&	1.000 	&	1.000 	&	1.000 	&	1.000 	\\
2	&	0.998 	&	0.999 	&	0.953 	&	0.965 	&	1.000 	&	1.000 	&	1.000 	&	1.000 	\\
2.25	&	0.999 	&	1.000 	&	0.966 	&	0.970 	&	1.000 	&	1.000 	&	1.000 	&	1.000 	\\
2.5	&	0.999 	&	1.000 	&	0.968 	&	0.974 	&	1.000 	&	1.000 	&	1.000 	&	1.000 	\\
		 \multicolumn{9}{c}{DGP 3}\\															
0.25	&	0.000 	&	0.000 	&	0.000 	&	0.000 	&	0.000 	&	0.000 	&	0.000 	&	0.000 	\\
0.5	&	0.704 	&	0.838 	&	0.000 	&	0.660 	&	0.789 	&	0.934 	&	0.000 	&	0.767 	\\
0.75	&	0.911 	&	0.982 	&	0.729 	&	0.833 	&	0.971 	&	0.998 	&	0.828 	&	0.932 	\\
1	&	0.978 	&	0.998 	&	0.850 	&	0.903 	&	0.997 	&	1.000 	&	0.980 	&	1.000 	\\
1.25	&	0.994 	&	1.000 	&	0.900 	&	0.931 	&	0.999 	&	1.000 	&	1.000 	&	1.000 	\\
1.5	&	0.998 	&	1.000 	&	0.924 	&	0.949 	&	1.000 	&	1.000 	&	1.000 	&	1.000 	\\
1.75	&	1.000 	&	1.000 	&	0.942 	&	0.957 	&	1.000 	&	1.000 	&	1.000 	&	1.000 	\\
2	&	1.000 	&	1.000 	&	0.953 	&	0.964 	&	1.000 	&	1.000 	&	1.000 	&	1.000 	\\
2.25	&	1.000 	&	1.000 	&	0.961 	&	0.969 	&	1.000 	&	1.000 	&	1.000 	&	1.000 	\\
2.5	&	1.000 	&	1.000 	&	0.964 	&	0.973 	&	1.000 	&	1.000 	&	1.000 	&	1.000 	\\

   \bottomrule
  \end{tabular}
\end{table}

\begin{table}[!h]
  \centering
  \footnotesize
  \caption{Conditional power for DGP 4.}
  \label{tab:p4}
  \begin{tabular}{c cccc cccc}
    \toprule
 \multirow{2}{*}{$\delta$}  &  \multicolumn{4}{c}{Direction-conditioned} & \multicolumn{4}{c}{Less-conditioned} \\
\cmidrule(lr){2-5} \cmidrule(lr){6-9}
 & $(60,25)$ & $(60,35)$ & $(120,25)$ & $(120,35)$ & $(60,25)$ & $(60,35)$ & $(120,25)$ & $(120,35)$ \\
    \midrule
 \multicolumn{9}{c}{LS}\\
0.25	&	0.000 	&	0.000 	&	0.000 	&	0.000 	&	0.000 	&	0.000 	&	0.000 	&	0.000 	\\
0.5	&	0.503 	&	0.837 	&	0.000 	&	0.687 	&	0.588 	&	0.996 	&	0.000 	&	0.695 	\\
0.75	&	0.876 	&	0.978 	&	0.700 	&	0.850 	&	0.936 	&	1.000 	&	0.701 	&	0.850 	\\
1	&	0.961 	&	0.997 	&	0.819 	&	0.897 	&	0.986 	&	1.000 	&	0.813 	&	0.940 	\\
1.25	&	0.982 	&	0.999 	&	0.855 	&	0.921 	&	0.994 	&	1.000 	&	0.860 	&	0.992 	\\
1.5	&	0.991 	&	1.000 	&	0.884 	&	0.930 	&	0.998 	&	1.000 	&	0.927 	&	1.000 	\\
1.75	&	0.995 	&	1.000 	&	0.896 	&	0.937 	&	0.998 	&	1.000 	&	0.968 	&	1.000 	\\
2	&	0.996 	&	1.000 	&	0.908 	&	0.942 	&	0.998 	&	1.000 	&	0.986 	&	1.000 	\\
2.25	&	0.998 	&	1.000 	&	0.912 	&	0.947 	&	0.999 	&	1.000 	&	0.995 	&	1.000 	\\
2.5	&	0.998 	&	1.000 	&	0.919 	&	0.947 	&	0.999 	&	1.000 	&	0.998 	&	1.000 	\\
			 \multicolumn{9}{c}{GMM}\\														
0.25	&	0.500 	&	0.754 	&	0.234 	&	0.435 	&	0.583 	&	0.870 	&	0.238 	&	0.443 	\\
0.5	&	0.918 	&	0.987 	&	0.692 	&	0.799 	&	0.977 	&	0.999 	&	0.695 	&	0.833 	\\
0.75	&	0.993 	&	1.000 	&	0.852 	&	0.906 	&	1.000 	&	1.000 	&	0.966 	&	1.000 	\\
1	&	1.000 	&	1.000 	&	0.917 	&	0.940 	&	1.000 	&	1.000 	&	1.000 	&	1.000 	\\
1.25	&	1.000 	&	1.000 	&	0.942 	&	0.956 	&	1.000 	&	1.000 	&	1.000 	&	1.000 	\\
1.5	&	1.000 	&	1.000 	&	0.960 	&	0.970 	&	1.000 	&	1.000 	&	1.000 	&	1.000 	\\
1.75	&	1.000 	&	1.000 	&	0.965 	&	0.975 	&	1.000 	&	1.000 	&	1.000 	&	1.000 	\\
2	&	1.000 	&	1.000 	&	0.974 	&	0.978 	&	1.000 	&	1.000 	&	1.000 	&	1.000 	\\
2.25	&	1.000 	&	1.000 	&	0.976 	&	0.986 	&	1.000 	&	1.000 	&	1.000 	&	1.000 	\\
2.5	&	1.000 	&	1.000 	&	0.981 	&	0.986 	&	1.000 	&	1.000 	&	1.000 	&	1.000 	\\
   \bottomrule
  \end{tabular}
\end{table}

\begin{table}[!h]
  \centering
  \footnotesize
  \caption{Conditional power for DGPs 5--6.}
  \label{tab:p56}
  \begin{tabular}{c cccc cccc}
    \toprule
\multirow{2}{*}{$\delta$}&  \multicolumn{4}{c}{DGP 5} & \multicolumn{4}{c}{DGP 6} \\
\cmidrule(lr){2-5} \cmidrule(lr){6-9}
  & $(60,15)$ & $(60,25)$ & $(120,15)$ & $(120,25)$ & $(60,15)$ & $(60,25)$ & $(120,15)$ & $(120,25)$ \\
    \midrule
 0.25	&	0.000 	&	0.000 	&	0.000 	&	0.000 	&	0.000 	&	0.308 	&	0.000 	&	0.000 	\\
0.5	&	0.321 	&	0.531 	&	0.160 	&	0.534 	&	0.374 	&	0.607 	&	0.380 	&	0.604 	\\
0.75	&	0.583 	&	0.695 	&	0.568 	&	0.686 	&	0.622 	&	0.722 	&	0.627 	&	0.721 	\\
1	&	0.694 	&	0.740 	&	0.690 	&	0.747 	&	0.703 	&	0.761 	&	0.706 	&	0.760 	\\
1.25	&	0.734 	&	0.783 	&	0.734 	&	0.780 	&	0.744 	&	0.786 	&	0.759 	&	0.783 	\\
1.5	&	0.761 	&	0.786 	&	0.758 	&	0.793 	&	0.775 	&	0.803 	&	0.771 	&	0.795 	\\
1.75	&	0.782 	&	0.799 	&	0.783 	&	0.798 	&	0.788 	&	0.805 	&	0.779 	&	0.810 	\\
2	&	0.791 	&	0.808 	&	0.786 	&	0.807 	&	0.797 	&	0.806 	&	0.792 	&	0.810 	\\
2.25	&	0.791 	&	0.815 	&	0.793 	&	0.810 	&	0.801 	&	0.813 	&	0.802 	&	0.812 	\\
2.5	&	0.808 	&	0.816 	&	0.805 	&	0.811 	&	0.802 	&	0.822 	&	0.809 	&	0.821 	\\
   \bottomrule
  \end{tabular}
\end{table}

The recovery probability results are reported in Appendix E. They complement the conditional power results below by showing how often the selected clusters coincide with the true latent groups.
Table \ref{tab:p123} reports the conditional power for the linear panel models in DGPs 1--3. The table compares the original direction-conditioned p-value with the less-conditioned p-value proposed in Section \ref{sec:direction-average-panel}. Several patterns are evident. First, for all three DGPs, conditional power increases monotonically as the separation parameter $\delta$ becomes larger. When the signal is weak, for example $\delta=0.25$, both procedures have little power, whereas the rejection probabilities quickly approach one as $\delta$ increases. Second, increasing the time dimension $T$ generally improves conditional power, because each individual slope estimator is more accurately estimated from a longer panel. Third, and most importantly, the less-conditioned procedure uniformly improves or matches the direction-conditioned procedure in almost all designs. The gain is especially visible at moderate signal strengths. For instance, under DGP 1 with $(N,T)=(60,15)$ and $\delta=0.5$, the conditional power increases from 0.694 for the direction-conditioned p-value to 0.839 for the less-conditioned p-value. Similar improvements appear under DGPs 2 and 3. When $\delta$ is large, both procedures have power close to one, and the numerical difference naturally becomes small.
These results provide finite-sample support for the theoretical motivation in Section \ref{sec:direction-average-panel}. Conditioning on the observed direction is useful for reducing the selective inference problem to a one-dimensional calculation, but it also removes information that may help distinguish the null from the alternative. By integrating over the direction, the less-conditioned p-value preserves selective validity while using more of the information contained in the selected contrast. The robustness of this improvement across DGPs 1--3 also suggests that the power gain is not driven by a particular error distribution.

Table \ref{tab:p4} presents the conditional power results for the dynamic panel model in DGP 4 under both LS and GMM estimation. For DGP 4, the GMM estimator uses $(y_{i,t-2},y_{i,t-3},\Delta x_{it,1},\Delta x_{it,2})$ as instruments in the first-differenced estimation. The same qualitative pattern continues to hold: conditional power increases with $\delta$, and the less-conditioned p-value generally dominates the direction-conditioned p-value. The improvement is particularly clear for the GMM estimator when the signal is moderate.
The LS results in Table \ref{tab:p4} should be interpreted together with the type I error analysis in the previous subsection. Although LS sometimes produces high conditional power, its null p-values are stochastically smaller than Uniform$(0,1)$ in the dynamic panel setting, leading to inflated selective type I error. Therefore, the apparent power advantage of LS is not a valid improvement. In contrast, the GMM-based procedure controls selective type I error and still exhibits strong conditional power, especially after replacing the direction-conditioned p-value with the less-conditioned p-value.

Table \ref{tab:p56} reports the conditional power for DGPs 5--6, which focus on testing differences in the coefficient of a single regressor between two selected groups. The conditional power again increases with $\delta$ and tends to be higher when $T=25$ than when $T=15$. Compared with DGP 5, DGP 6 often delivers slightly higher power, especially at smaller and moderate values of $\delta$, reflecting the fact that the richer covariate structure provides additional information for separating the latent groups. Overall, the results in Table \ref{tab:p56} confirm that the proposed covariate-specific selective inference procedure remains powerful in finite samples.

In summary, Tables \ref{tab:p123}--\ref{tab:p56} show that the proposed selective inference procedures have satisfactory conditional power once the selected groups correspond to true latent groups. More importantly, the less-conditioned p-value improves power relative to the direction-conditioned p-value while maintaining selective type I error control. This confirms the main empirical implication of the conditioning-on-less approach: avoiding unnecessary conditioning on the direction of the selected contrast can lead to more powerful post-selection inference.

\section{Empirical Application}\label{sec:6}
\subsection{The model and data}
Understanding the relationship between the economic growth and the emission of
 chemicals (mainly the carbon dioxide (CO$_2$)) is a  long-standing research interest in environmental economics.
 Theoretical advances and empirical studies on this topic have been developed over many years; see \cite{grossman1995economic,azomahou2006economic,du2012economic,liao2013does,kaika2013environmental,baloch2019effect}, among many others. Empirical analyses in this field usually employ standard panel data techniques to account for heterogeneity, or simply
  partition countries into distinct groups based on prior information such as geographic locations.
  For instance, \cite{sharma2010relationship}  examined the  relationship between energy consumption and economic growth using a panel of 66 countries, employing a regional classification that revealed significant heterogeneity in growth patterns across different geographical groups. In this section, we re-analyze this empirical problem by applying the methodology developed in the paper.

To capture the potential heterogeneous  relationship between energy consumption and economic growth, we consider the following econometric model
$$
C_{it}=G_{it}\beta_{1i}+I_{it}\beta_{2i}+\eta_i+u_{it},
$$
where $C_{it}$ is the logarithm of per capita CO$_2$ emissions, $G_{it}$ is the logarithm of per capita gross domestic product (GDP), $I_{it}$ is the  proportion of imports of goods and services to GDP, $\eta_i$ is a fixed effect, and $u_{it}$ is an  idiosyncratic error term. The effect of the imports of goods and services on CO$_2$ emissions is well documented \citep{copeland2004trade}, therefore added to the model specification.  Data are collected from the World Development Indicators, a  comprehensive database published by the World Bank.  Following \cite{li2016panel}, we exclude countries with populations below six million and those with missing observations during the sample period. Using the periods 1996--2010, we construct a balanced panel of 70 countries. A list of these countries is provided in Table \ref{tab:em2}.

\subsection{Classification and estimation results.}
\begin{figure}[h]
	\centering
   \includegraphics[scale =0.48]{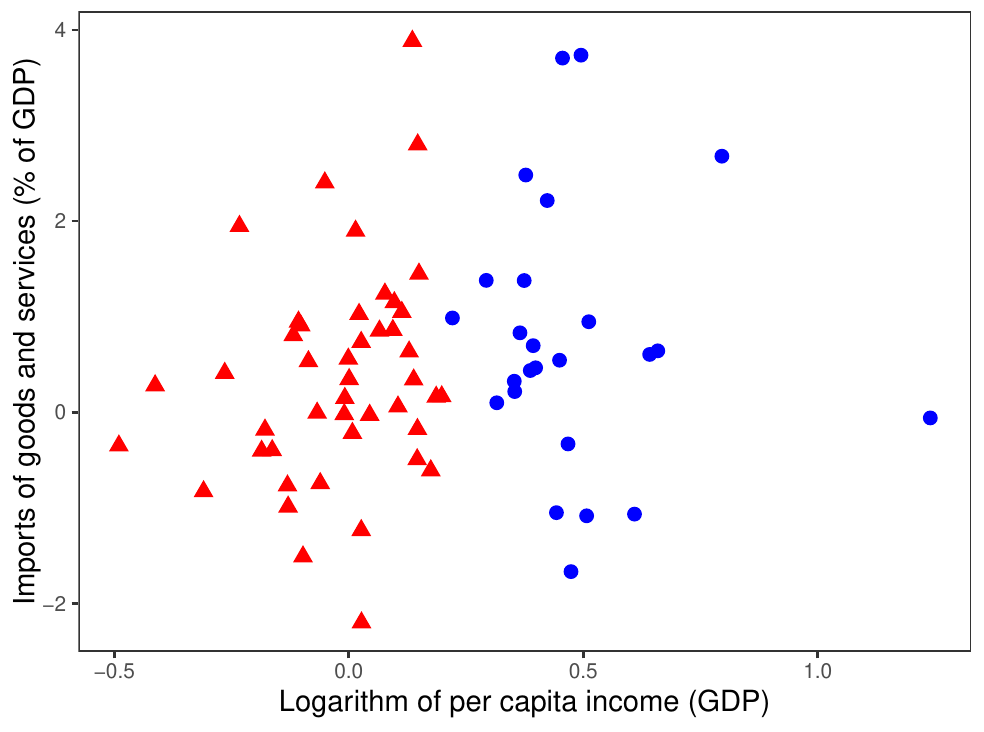}
	\caption{Scatter plot for the preliminary estimates of $\bbeta_i$'s in the application. }
	\label{fig:emp}
\end{figure}

We first examine the presence of heterogeneity in the dataset by using the test statistic in \cite{pesaran2008testing}. The resulting p-value is very close to 0, indicating substantial heterogeneity among countries. To determine the latent group structures, we apply the information criterion in \cite{liu2020identification} to select $K$. This procedure yields two estimated clusters identified by the $k$-means clustering algorithm \ref{alg:1}.
 Table \ref{tab:em2} reports the estimation and classification results for the  identified groups.
Notably, the classification reveals a clear dichotomy between developing and developed economies. For example, we observe a strong collection of $25$ developing countries in Group 1, including India, Indonesia, Thailand, and Egypt. Group 2 mainly comprises some  developed countries such as United States, United Kingdom, France, Japan, and Australia, along with some smaller economies. Figure \ref{fig:emp} displays the scatter diagrams for the preliminary estimates of all $\bbeta_i$'s, from which we observe a distinct separation between the two groups.

From Table \ref{tab:em2}, we observe substantial
differences between the group-specific slope estimates after classification.  Compared to Group 2, all coefficient estimates for Group 1 are statistically significant at the 1\% level. Moreover, the magnitudes of the coefficients  for Group 1 are substantially larger than the corresponding coefficients for Group 2. This suggests that, for developing countries, which have smaller economic scales, both GDP growth and the import ratio  have a strong positive relationship with CO$_2$ emissions. In contrast, for developed countries, the import ratio  does not significantly affect CO$_2$ emissions, and GDP growth
 has only a weak effect.
In summary, the heterogeneity between Group 1 and Group 2 is evident in both the magnitude and statistical significance of the parameter estimates.

\begin{table}[!h]
  \centering
  \footnotesize
  \caption{Estimation and classification results of the CO$_2$ emission dataset.}
  \label{tab:em2}
  \begin{tabular}{c cc cc}
    \toprule
   &  \multicolumn{2}{c}{Group 1} & \multicolumn{2}{c}{Group 2} \\
\midrule
No. of Countries & \multicolumn{2}{c}{25} & \multicolumn{2}{c}{45}\\
$\beta_{1i}$ &      \multicolumn{2}{c}{0.5018***} & \multicolumn{2}{c}{0.0349*}\\
    S.E.      & \multicolumn{2}{c}{(0.0180)} & \multicolumn{2}{c}{(0.0122)}\\
$\beta_{2i}$ & \multicolumn{2}{c}{0.4686***} & \multicolumn{2}{c}{0.0736}\\
    S.E.      & \multicolumn{2}{c}{(0.0922)} & \multicolumn{2}{c}{(0.0572)}\\
    \hline
\multicolumn{5}{c}{Group 1}\\
Benin	&	Burkina Faso	&	Bangladesh	&	Bolivia	&	Chile	\\
Dominican Republic	&	Egypt	&	Guatemala	&	Honduras	&	Indonesia	\\
India	&	South Korea	&	Sri Lanka	&	Morocco	&	Mozambique	\\
Malaysia	&	Pakistan	&	Peru	&	Sudan	&	El Salvador	\\
Chad	&	Thailand	&	Tunisia		&	Uganda	& Zimbabwe \\
\hline
\multicolumn{5}{c}{Group 2}\\
Argentina	&	Australia	&	Austria	&	Belgium	&	Bulgaria	\\
Brazil	&	Canada	&	Switzerland	&	Ivory Coast	&	Cameroon	\\
Colombia	&	Cuba	&	Algeria	&	Ecuador	&	Spain	\\
France	&	United Kingdom	&	Ghana	&	Greece	& Hong Kong (China)\\
Hungary	&	Israel	&	Italy	&	Jordan	& Japan\\
Kenya	&	Madagascar	&	Mexico	&	Mali & Malawi	\\
Niger	&	Nigeria	&	Netherlands	&	Philippines	& Portugal\\
Rwanda	&	Saudi Arabia	&	Senegal	&	Sweden	& Togo\\
Turkey	&	United States	&	Venezuela	&	South Africa	&	Zambia	\\
   \bottomrule
  \end{tabular}
  \begin{tablenotes}
\item Note: *, **, and *** refer to significance at 10\%, 5\% and 1\% level, respectively.
\end{tablenotes}
\end{table}

\subsection{Inference for the estimated groups}

However, it remains to check
whether the slopes of the two identified groups are statistically different. We test $H_0^{(\wG_1,\wG_2)}$ for the estimated clusters $\wG_1$ and $\wG_2$ using the proposed method in Section \ref{sec:3}, and compare it with the naive Wald test described in \eqref{eq:n}. We emphasize that the latter does not account for the fact the clusters are estimated from the data, and thus fails to control the selective type I error rate. In addition to the overall effect, we are also interested in identifying which specific covariates drive the differences by testing $H_{0j}^{(\wG_1,\wG_2)}$ for $j=1,2$ using the test proposed in Section \ref{sec:41}. Results are in Table \ref{tab:em1}.
In this dataset, the naive Wald p-values are extremely small, while our proposed selective p-values are relatively large. For the joint test of both covariates (log per capita GDP $G_{it}$  and import ratio $I_{it}$),  the selective p-value is $1.033\times10^{-16}$, leading to strong rejection of the null hypothesis.
The less-conditioned p-value is $1.002\times10^{-16}$, which leads to the same conclusion and is slightly smaller than the direction-conditioned p-value, consistent with the power improvement discussed in Section \ref{sec:direction-average-panel}.
Examining the selective p-values of each covariate, we find that the p-value for $G_{it}$ is $8.427\times10^{-5}$ (significant at the 1\% level), while the p-value for $I_{it}$ is 0.221 (indicating no significance). Thus, we conclude that log per capita GDP is the primary driver of the differences between groups.

\begin{table}[!h]
  \centering
  \footnotesize
  \caption{Results from applying the test of $H_0^{(\wG_1,\wG_2)}$ and $H_{0j}^{(\wG_1,\wG_2)}$, $j=1,2$.}
  \label{tab:em1}
  \begin{tabular}{c ccc ccc}
    \toprule
  & \multicolumn{3}{c}{Full sample (70 individuals)} & \multicolumn{3}{c}{Group 2 (45 individuals)} \\
   \cmidrule(lr){2-4} \cmidrule(lr){5-7}
Covariates  &  $(G_{it},I_{it})$  & $G_{it}$ & $I_{it}$ &  $(G_{it},I_{it})$  & $G_{it}$ & $I_{it}$\\
\midrule
Test statistic & 595.296 & 314.788 & 11.650 & 130.011 & 99.271 & 5.994\\
Our p-value (Direction) & $1.033\times10^{-16}$ & $8.427\times10^{-5}$ & 0.221 & 0.113 & 0.249 & 0.950 \\
Our p-value (Less) & $1.002\times10^{-16}$ & $-$ & $-$ & 0.104 & $-$ & $-$ \\
Wald p-value & $<10^{-307}$ & $<10^{-307}$ &  $6.420\times10^{-4}$ & $<10^{-307}$ & $<10^{-307}$ & $0.014$\\
   \bottomrule
  \end{tabular}
\end{table}

 To further examine the robustness and reliability of our empirical results, we perform post-selection inference specifically on Group 2, which has the largest number of countries. We apply $k$-means algorithm to divide it into two groups,  and compute the corresponding post-selection  p-values. As shown in the right column of Table \ref{tab:em1}, the selective p-values proposed in this paper do not reject  the null hypothesis in any case. The less-conditioned p-value for the joint test is 0.104, close to the direction-conditioned value 0.113, and therefore supports the same non-rejection conclusion.
 In contrast, the naive Wald test  incorrectly reject the null hypothesis for two of the covariates and their combination, even though visual inspection suggests no significant difference within Group 2.  This confirms that our proposed approach is a valid tool to correctly reject the selected null hypothesis  when it does not hold.

\section{Concluding Remarks}\label{sec:7}

In this paper, we propose a selective inference framework for testing the null hypothesis of no difference in slopes between two estimated groups in panel model \eqref{eq:11}. We extend this framework in three important directions: (i) a less-conditioned selective p-value that improves power by avoiding unnecessary conditioning on the direction of the selected contrast, (ii) inference on the coefficients of a specific explanatory variable between two identified groups, and (iii) generalization to the GMM estimation framework, accommodating  dynamic effects and endogeneity. Unlike existing methods such as \cite{chen2023selective} and \cite{gao2024selective}, which rely critically on normality assumption of observations, our approach only requires the asymptotic normality property of the initial individual estimators, a condition easily satisfied in panel models. This makes our framework less sensitive to data distributional assumptions, as confirmed by its robustness in diverse simulation settings. The simulation and empirical results show that the less-conditioned p-value maintains selective type I error control and can deliver higher power than the direction-conditioned benchmark.

Our work opens several promising research avenues, which we hope to take on as
 future investigations. First, while we compute selective p-value in Section \ref{sec:p} using the $k$-means clustering algorithm,  other detection methods such as the C-LASSO or SBSA could be explored. However, it may be challenging to compute the feasible set $\cS$ for $\phi$. A potential solution to avoid computing $\cS$ is to apply the Monte Carlo approximation with importance sampling, as detailed in Section 4.1 of \cite{chen2023selective}. Second, although we focus on linear panel models \eqref{eq:1} and \eqref{eq:fd}, extending our approach to nonlinear models (e.g., probit, tobit) is highly relevant. This extension faces technical hurdles, such as the inability to eliminate individual fixed effects $\eta_i$ via concentration or differencing. We plan to address this by leveraging algorithm stability theory \citep{zrnic2023post} to enable valid post-selection inference in broader panel model contexts.

\section*{Acknowledgments}
Xu thanks the support from the National Natural Science Foundation of China (NNSFC) (72473118, 72333001, 71988101). Wan’s research was supported by
National Natural Science Foundation of China (NNSFC) (12301344, 12571284, 72373074) and China Post
doctoral Science Foundation (2024M761153).


\section*{Declaration of generative AI and AI-assisted technologies in the writing process}

During the preparation of this work, the authors used ChatGPT, DeepSeek and Gemini in order to enhance the writing quality. After using these tools, the authors have reviewed and edited the content as needed and take full responsibility for the content of the publication.

\bibliographystyle{chicago}
\bibliography{conselective}

\clearpage
\setcounter{section}{0}
\renewcommand\thesection{\Alph{section}}
\setcounter{page}{1}
\setcounter{footnote}{0}
{\Large \bf
	\begin{center}
		Appendix to ``Conditional Selective Inference for the Selected Groups in Panel Data''
	\end{center}
}

This appendix includes an explanation of how to interpret the notation $\bB(\phi)$, additional illustrations for  Sections~\ref{sec:direction-average-panel}--\ref{sec:4}, and the proofs of the main theorems in the paper.

\renewcommand{\theequation}{A.\arabic{equation}}
\renewcommand{\thefigure}{A.\arabic{figure}}
\renewcommand{\thetable}{A.\arabic{table}}
\renewcommand{\thealgorithm}{A.\arabic{algorithm}}
\renewcommand{\thetheorem}{A.\arabic{theorem}}
\renewcommand{\thesection}{A}
\setcounter{figure}{0}

\noindent

\section{Interpreting $\bB(\phi)$ in Section \ref{sec:3}}\label{ap0}

We now illustrate how to interpret $\bB(\phi)$. By simple algebra, the $i$th individual estimate of $\bB(\phi)$ in \eqref{eq:bB} is
\begin{eqnarray}\label{eq:ip}
\bbeta_i^{\ini}(\phi)=\left\{
\begin{array}{ll}
\widehat{\bbeta}_i^{\ini}+\frac{\XXi_i\bw_i(k)^{\mbox{\tiny{T}}}(\ttheta\trans\XXi\ttheta)^{-1}(\widehat{\aalpha}_{\wG_k}-\widehat{\aalpha}_{\wG_{k'}})}{\|
\cP_{\ttheta}\widehat{\bB}\|_2}\cdot(\phi-\|
\cP_{\ttheta}\widehat{\bB}\|_2), & \text{if}~i\in\wG_k,\\
\widehat{\bbeta}_i^{\ini}-\frac{\XXi_i\bw_i({k'})^{\mbox{\tiny{T}}}(\ttheta\trans\XXi\ttheta)^{-1}(\widehat{\aalpha}_{\wG_k}-\widehat{\aalpha}_{\wG_{k'}})}{\|
\cP_{\ttheta}\widehat{\bB}\|_2}\cdot(\phi-\|
\cP_{\ttheta}\widehat{\bB}\|_2), & \text{if}~i\in\wG_{k'},\\
\widehat{\bbeta}_i^{\ini}, & \text{if}~i\notin\wG_k\cup\wG_{k'},
\end{array}
\right.
\end{eqnarray}
We can interpret $\bB(\phi)$  as a perturbed version of $\widehat{\bB}$, where the randomization originates from the random response $\tbY$. Specifically, for each $\bbeta_i^{\ini}(\phi)$, we can always find a proper $\tby_i(\phi)$ such that $\bbeta_i^{\ini}(\phi)=(\tbx_i\trans\tbx_i)^{-1}\tbx_i\trans\tby_i(\phi)$. From \eqref{eq:ip}, the individuals in clusters
$\wG_k$ and $\wG_{k'}$ have been ``pulled apart" if $\phi>\|\cP_{\ttheta}\widehat{\bB}\|_2$ or
``pushed together" if $0\leq \phi\leq \|\cP_{\ttheta}\widehat{\bB}\|_2$ in the direction of $\widehat{\aalpha}_{\wG_k}-\widehat{\aalpha}_{\wG_{k'}}$.
Furthermore, $\cS$  represents the set of non-negative values of $\phi$ for which applying the $k$-means clustering to $\bB(\phi)$ yields the same clusters $\wG_k$ and $\wG_{k'}$. To  illustrate this interpretation graphically, we generate
a new dataset from \eqref{eq:1} with $N=T=30$ and $p=2$ regressors. We set the number of groups to be three (i.e., $K^0=3$), and number of individuals in each group is $N/3=10$. The true group-specific parameters are $\aalpha_1^0=(-0.65,0)\trans$, $\aalpha_2^0=(0.65,0)\trans$
and $\aalpha_3^0=(0,1.5)\trans$, respectively.

 \begin{figure}[h]
	\centering
   \includegraphics[width =\textwidth]{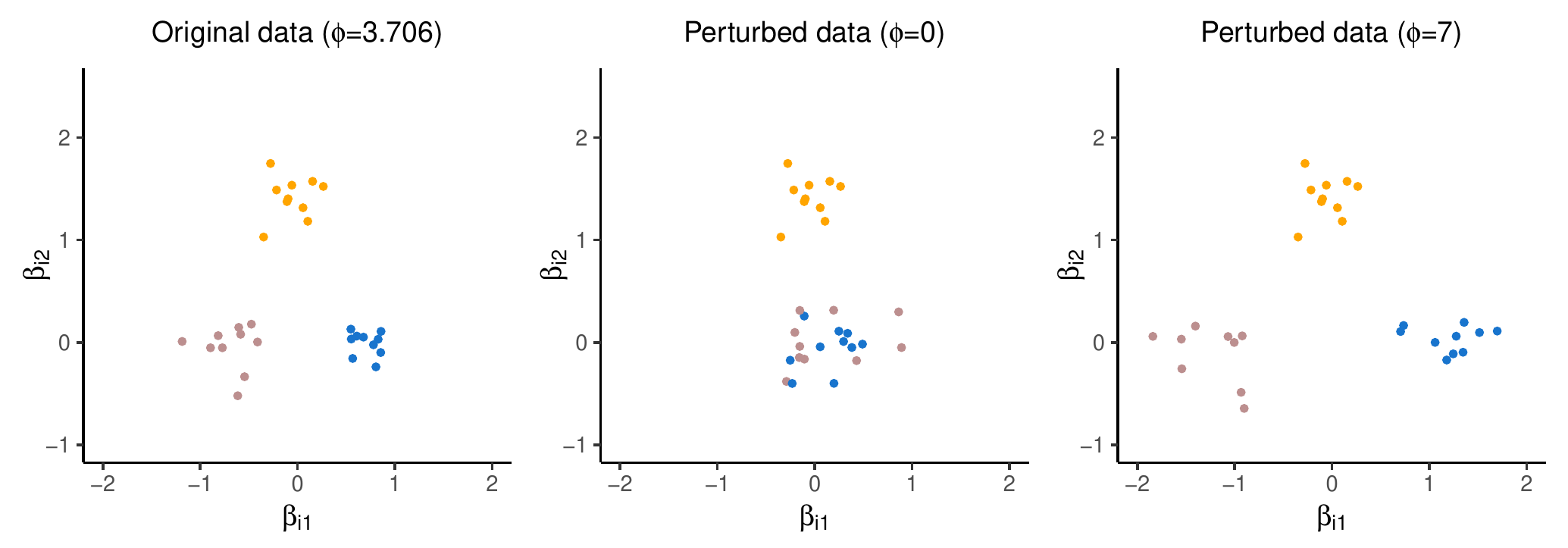}
	\caption{One simulated panel dataset generated from model \eqref{eq:1} with $N=30$ individuals and $\bbeta_i^0={\bf1}\{1\leq i\leq 10\}(-0.65,0)\trans+
{\bf1}\{11\leq i\leq 20\}(0.65,0)\trans+{\bf1}\{21\leq i\leq 30\}(0,1.5)\trans$. \emph{(a)}: the original initial
estimate $\widehat{\bB}=\bB(\phi)$ with $\phi=\|\cP_{\ttheta}\widehat{\bB}\|_2=3.706$; \emph{(b)}: a perturbed initial estimate $\bB(\phi)$ with $\phi=0$; \emph{(c)}: a perturbed initial estimate $\bB(\phi)$ with $\phi=7$.}
	\label{fig:phi}
\end{figure}

Figure \ref{fig:phi} displays $\widehat{\bB}=\bB(\phi)$ with $\phi=\|
\cP_{\ttheta}\widehat{\bB}\|_2=3.706$, along with $\bB(\phi)$ for $\phi=0$ and $\phi=7$, respectively.
 The selected two clusters $\widehat{G}_1,\wG_2\in\Cluster(\widehat{\bB})$ are displayed in blue and brown.
As  observed in panel (a) of Figure \ref{fig:phi}, the $k$-means algorithm successfully partitions the original panel data into three clusters. In panel (b), the blue and brown individuals have been ``pushed together", eliminating any difference in slopes between $\wG_1$ and $\wG_2$. Consequently, the $k$-means algorithm cannot distinguish these clusters. By contrast,
 panel (c) shows that the blue and brown clusters have been ``pulled apart", and the $k$-means algorithm  can
identify them more easily. Putting panels (a)--(c) together, we find that the orange cluster remains unchanged, since this cluster is not involved in the test.
 In this example, $\widehat{\cS}=[2.036,\infty)$.

\renewcommand{\theequation}{B.\arabic{equation}}
\renewcommand{\thefigure}{B.\arabic{figure}}
\renewcommand{\thetable}{B.\arabic{table}}
\renewcommand{\thealgorithm}{B.\arabic{algorithm}}
\renewcommand{\thetheorem}{B.\arabic{theorem}}
\renewcommand{\thesection}{B}
\setcounter{figure}{0}
\setcounter{theorem}{0}

\section{Additional  Analysis in Section \ref{sec:direction-average-panel}}\label{sec:ap4}
\subsection{Power improvement by conditioning on less}
This section provides theoretical support for the power improvement achieved
by the proposed less-conditioned selective inference procedure. The main idea
is that conditioning on the observed direction of the selected contrast, as in
the conventional direction-conditioned approach, removes part of the random
variation that may be informative under the alternative. By contrast, our
procedure conditions only on the selection event and the nuisance component
orthogonal to the contrast of interest, while averaging over the null
distribution of the direction. This leads to a valid selective p-value and
retains additional information for detecting departures from the null.

Recall that the conventional
direction-conditioned procedure conditions on
\[
\mathcal C_{\mathrm{dir}}
=\left\{\cP_{\ttheta}^{\perp}\bB = \cP_{\ttheta}^\perp\widehat{\bB},~ \ppsi=\dir(\widehat{\bD}),~\Cluster(\bB)=\Cluster(\widehat{\bB})\right\},
\]
whereas the proposed less-conditioned procedure conditions on
\[
\mathcal C_{\mathrm{less}}
=
\left\{
\cP_{\ttheta}^{\perp}\bB = \cP_{\ttheta}^\perp\widehat{\bB},~\Cluster(\bB)=\Cluster(\widehat{\bB})
\right\}.
\]
The following theorem compares the best attainable selective power under the two
conditioning schemes. It does not require the realized less-conditioned
p-value to be smaller than the direction-conditioned
p-value for every sample. Rather, it shows that, after averaging over the
additional direction, a direction-conditioned valid test induces a valid test
under the less-conditioned experiment. Hence removing the unnecessary
conditioning on \(\psi\) cannot reduce optimal selective power.

\begin{theorem}
\label{thm:power-less-conditioning}
Let \(\mathcal T_{\mathrm{dir}}(\alpha)\) be the class of selective tests with
conditional level \(\alpha\) under \(\mathcal C_{\mathrm{dir}}\), and let
\(\mathcal T_{\mathrm{less}}(\alpha)\) be the class of selective tests with
conditional level \(\alpha\) under \(\mathcal C_{\mathrm{less}}\).
Then, for
any fixed alternative \(H_1^{(\wG_k,\wG_{k'})}:\ttheta\trans\bB\neq 0\),
\[
\sup_{\varphi\in\mathcal T_{\mathrm{less}}(\alpha)}
\mathbb E_{H_1^{(\wG_k,\wG_{k'})}}[\varphi]
\ge
\sup_{\varphi\in\mathcal T_{\mathrm{dir}}(\alpha)}
\mathbb E_{H_1^{(\wG_k,\wG_{k'})}}[\varphi].
\]
\end{theorem}

\begin{proof}
Since $\mathcal C_{\mathrm{dir}}$ contains all information in $\mathcal C_{\mathrm{less}}$
plus the additional direction \(\psi\), the direction-conditioned experiment is
a further conditioning of the less-conditioned experiment.

Take any test
\(\varphi_{\mathrm{dir}}\in\mathcal T_{\mathrm{dir}}(\alpha)\). By definition,
\[
\mathbb E_{H_0^{(\wG_k,\wG_{k'})}}
\left[
\varphi_{\mathrm{dir}}
\Big|
\cP_{\ttheta}^{\perp} \bB,\ppsi,\Cluster({\bB})
\right]
\le \alpha
\quad\text{a.s.}
\]
Since $\sigma\{\cP_{\ttheta}^{\perp} \bB,\operatorname{Cluster}( \bB)\}
\subseteq
\sigma\{\cP_{\ttheta}^{\perp} \bB,\ppsi,\operatorname{Cluster}( \bB)\}$,
the law of iterated expectations gives
\begin{eqnarray*}
&&\mathbb E_{H_0^{(\wG_k,\wG_{k'})}}
\left[
\varphi_{\mathrm{dir}}
\Big|
\cP_{\ttheta}^{\perp} \bB,\Cluster({\bB})
\right]\\
&=&
\mathbb E_{H_0^{(\wG_k,\wG_{k'})}}
\left[
\mathbb E_{H_0^{(\wG_k,\wG_{k'})}}
\left[
\varphi_{\mathrm{dir}}
\Big|
\cP_{\ttheta}^{\perp} \bB,\ppsi,\Cluster({\bB})
\right]
\Big|
\cP_{\ttheta}^{\perp} \bB,\Cluster({\bB})
\right]
\le \alpha .
\end{eqnarray*}
Hence every direction-conditioned level-\(\alpha\) test is also a valid
level-\(\alpha\) test after averaging over the direction. Therefore, after
the natural marginalization over \(\psi\), any test in
\(\mathcal T_{\mathrm{dir}}(\alpha)\) induces an element of
\(\mathcal T_{\mathrm{less}}(\alpha)\).

It follows immediately that the maximal attainable power under less
conditioning is no smaller:
\[
\sup_{\varphi\in\mathcal T_{\mathrm{less}}(\alpha)}
\mathbb E_{H_1^{(\wG_k,\wG_{k'})}}[\varphi]
\ge
\sup_{\varphi\in\mathcal T_{\mathrm{dir}}(\alpha)}
\mathbb E_{H_1^{(\wG_k,\wG_{k'})}}[\varphi].
\]

Finally, suppose that the conditional distribution of \(\psi\) under \(H_1^{(\wG_k,\wG_{k'})}\)
differs from its conditional distribution under \(H_0^{(\wG_k,\wG_{k'})}\) given
\((\cP_{\ttheta}^{\perp} \bB,\Cluster({\bB}))\). Then \(\psi\) contains
non-degenerate information about the alternative. Conditioning on \(\psi\)
removes this information from the testing problem, whereas the
less-conditioned procedure preserves it through integration over the null
direction distribution. Therefore the less-conditioned experiment admits a test
that exploits this additional information, yielding strictly larger power for
such alternatives.
\end{proof}

\subsection{Proof of Theorem 3}

\begin{proof}
By construction, one of these directions, say  \(\ppsi^{(I)}\), corresponds to
the observed direction, whereas the remaining directions are independently
drawn from the null distribution of \(\ppsi\). Since the proposed p-value
\(\widetilde{P}_{\less}\) is invariant to permutations of the labels of
\(\ppsi^{(1)},\ldots,\ppsi^{(M)}\), it suffices to condition on the unordered
collection
\[
\ppsi^{(1:M)}=\{\ppsi^{(1)},\ldots,\ppsi^{(M)}\}
\]
and then average over the random label \(I\).

For a fixed direction \(\ppsi^{(m)}\), let
\[
\cR_{\ppsi^{(m)}}
=
\left\{
\xi:
\operatorname{Cluster}
\left(\bB(\ppsi^{(m)},\xi)
\right)
=
\operatorname{Cluster}(\widehat \bB)
\right\}
\]
be the selection region for the radial statistic along direction
\(\ppsi^{(m)}\). Let \(f\) denote the null density of \(\xi\), and let \(g\)
denote the null density of each component of \(\ppsi\). Conditional on
\(\ppsi^{(1:M)}\), the joint density of \((\xi,\ppsi^{(1:M)},I)\) after
selection is proportional to
\[
f(\xi)
\left\{
\prod_{m=1}^{M} g(\ppsi^{(m)})
\right\}
\mathbf 1\{\xi\in\mathcal R_{\ppsi^{(I)}}\}.
\]
If we condition on $\ppsi^{(1:M)}$, then we get
\[
f(\xi,I\mid \ppsi^{(1:M)},\mathcal R)
\propto
f(\xi)\mathbf 1\{\xi\in\mathcal R_{\ppsi^{(I)}}\}.
\]

To normalize this conditional density, define
\[
w_m
=
\mathbb P\{\xi\in\mathcal R_{\ppsi^{(m)}}\}
=
\int f(\xi)\mathbf 1\{\xi\in\mathcal R_{\ppsi^{(m)}}\}\,d\xi.
\]
It follows that
\[
f(\xi,I\mid \ppsi^{(1:M)},\mathcal R)
=
\frac{1}{\sum_{m=1}^{M} w_m}
f(\xi)\mathbf 1\{\xi\in\mathcal R_{\ppsi^{(I)}}\}.
\]
Marginalizing over the unknown label \(I\), we obtain
\[
\begin{aligned}
f(\xi\mid~\ppsi^{(1:M)},\mathcal R)
&=
\sum_{I=1}^{M}
f(\xi,I\mid \ppsi^{(1:M)},\mathcal R)  \\
&=
\frac{1}{\sum_{m=1}^{M} w_m}
\sum_{I=1}^{M}
f(\xi)\mathbf 1\{\xi\in\mathcal R_{\ppsi^{(I)}}\}  \\
&=
\frac{1}{\sum_{m=1}^{M} w_m}
\sum_{I=1}^{M}
w_I
\frac{
f(\xi)\mathbf 1\{\xi\in\mathcal R_{\ppsi^{(I)}}\}
}{
w_I
}.
\end{aligned}
\]
Thus the conditional distribution of \(\xi\), given
\(\ppsi^{(1:M)}\) and the selection event, is a mixture of the truncated null
distributions corresponding to the directions
\(\ppsi^{(1)},\ldots,\ppsi^{(M)}\), with mixture weights \(w_I/\sum_{m=1}^{M} w_m\).

Consequently, for the observed value \(\widehat{W}_{k,k'}\),
\[
\begin{aligned}
\mathbb P_{H_0^{(\wG_k,\wG_{k'})}}
\left(
W\ge \widehat{W}_{k,k'}
\Big |
\ppsi^{(1:M)},\mathcal R
\right)
&=
\frac{1}{\sum_{m=1}^{M} w_m}
\sum_{I=1}^{M}
w_I
\mathbb P_{H_0^{(\wG_k,\wG_{k'})}}
\left(
W\ge \widehat{W}_{k,k'}
\Big |
\xi\in\mathcal R_{\ppsi^{(I)}}
\right)  \\
&=
\frac{
\sum_{I=1}^{M}
\mathbb P_{H_0^{(\wG_k,\wG_{k'})}}
\left(
W\ge \widehat{W}_{k,k'},
\xi\in\mathcal R_{\ppsi^{(I)}}
\right)
}{
\sum_{I=1}^{M}
\mathbb P_{H_0^{(\wG_k,\wG_{k'})}}
\left(
\xi\in\mathcal R_{\ppsi^{(I)}}
\right)
}.
\end{aligned}
\]
The right-hand side is exactly the proposed
p-value ${P}_{\mathrm{less}}$.
Here, for each fixed direction \(\ppsi^{(I)}\), the Wald statistic is a
monotone function of the radius \(\xi\). Equivalently, the event
\(\{W\ge \widehat W_{k,k'}\}\) can be represented by the corresponding radial
tail event along that direction. Thus the above mixture tail probability is
the same truncated radial tail probability used in the definition of
\(\widetilde P_{\less}\).

Therefore,
\[
\widetilde{P}_{\less}({\bB};\wG_{k},\wG_{k'})
=
\mathbb P_{H_0^{(\wG_k,\wG_{k'})}}
\left(
W\ge \widehat{W}_{k,k'}
\Big |
\ppsi^{(1:M)},\mathcal R
\right).
\]
By the probability integral transform, this conditional survival probability
is uniformly distributed on \([0,1]\) under the selected null hypothesis,
provided that the conditional null distribution is continuous. Hence, for
every \(0\le \alpha\le 1\),
\[
\mathbb P_{H_0^{(\wG_k,\wG_{k'})}}
\left(
\widetilde{P}_{\less}({\bB};\wG_{k},\wG_{k'})\le \alpha
\mid
\ppsi^{(1:M)},\mathcal R
\right)
=
\alpha.
\]
Averaging over \(\ppsi^{(1:M)}\) gives
\[
\mathbb P_{H_0^{(\wG_k,\wG_{k'})}}
\left(
\widetilde{P}_{\less}({\bB};\wG_{k},\wG_{k'})\le \alpha
\mid
\mathcal R
\right)
=
\alpha.
\]
This establishes the selective validity of the proposed less-conditioned
p-value.
\end{proof}


\renewcommand{\theequation}{C.\arabic{equation}}
\renewcommand{\thefigure}{C.\arabic{figure}}
\renewcommand{\thetable}{C.\arabic{table}}
\renewcommand{\thealgorithm}{C.\arabic{algorithm}}
\renewcommand{\thetheorem}{C.\arabic{theorem}}
\renewcommand{\thesection}{C}
\setcounter{figure}{0}
\setcounter{algorithm}{0}

\section{Additional illustrations  in Section \ref{sec:4}}
\subsection{Testing for a difference in coefficients due to a covariate}\label{secb1}

Similar to Equation \eqref{eq:ip}, we also provide illustration on how to interpret $\bB(\kappa)$.
For the $j$th covariate,  the $\bB(\kappa)$ can be rewritten as
\begin{eqnarray}\label{eq:ipj}
\bbeta_i^{\ini}(\kappa)=\left\{
\begin{array}{ll}
\widehat{\bbeta}_i^{\ini}+\XXi_i\bw_i(k)^{\mbox{\tiny{T}}}\be_j(\ttheta_j\trans\XXi\ttheta_j)^{-1}(\kappa-\ttheta_j\trans\widehat{\bB}), & \text{if}~i\in\wG_k,\\
\widehat{\bbeta}_i^{\ini}-\XXi_i\bw_i(k')^{\mbox{\tiny{T}}}\be_j(\ttheta_j\trans\XXi\ttheta_j)^{-1}(\kappa-\ttheta_j\trans\widehat{\bB}), & \text{if}~i\in\wG_{k'},\\
\widehat{\bbeta}_i^{\ini}, & \text{if}~i\notin\wG_k\cup\wG_{k'},
\end{array}
\right.
\end{eqnarray}
where $\be_j$ is a $p\times1$ vector with the $j$th element being 1 and all other elements being 0.

\begin{figure}[h]
	\centering
   \includegraphics[width =\textwidth]{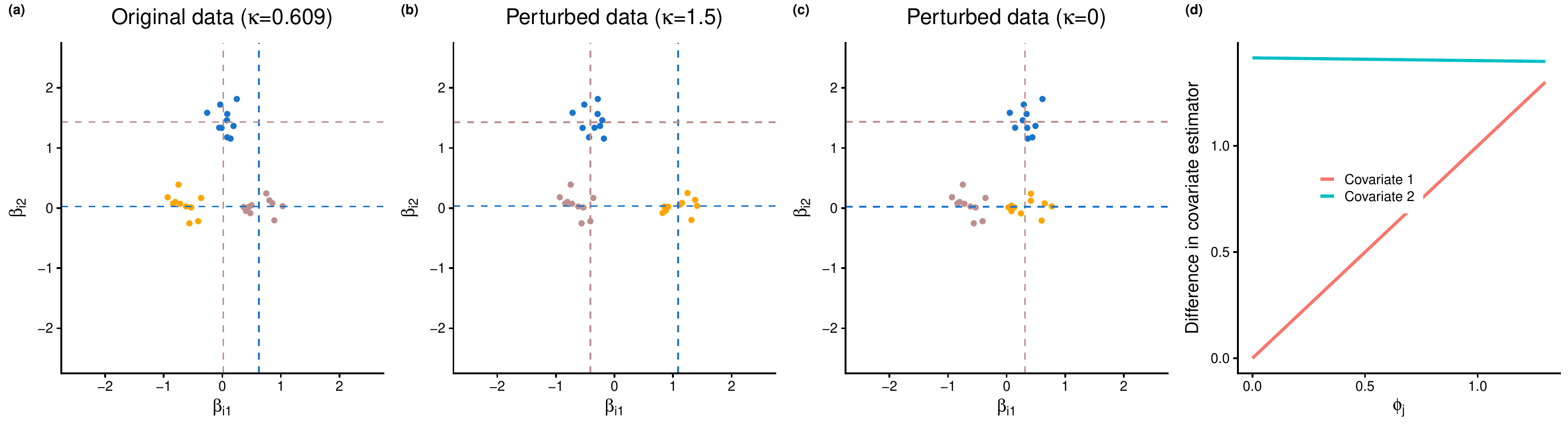}
	\caption{One simulated panel data generated from model \eqref{eq:1} with $N=30$ individuals and $\bbeta_i^0={\bf1}\{1\leq i\leq 10\}(-0.65,0)\trans+
{\bf1}\{11\leq i\leq 20\}(0.65,0)\trans+{\bf1}\{21\leq i\leq 30\}(0,1.5)\trans$. \emph{(a)}: the original initial
estimate $\widehat{\bB}=\bB(\kappa)$ with $\kappa=\ttheta_j\trans\widehat{\bB}=0.609$; \emph{(b)}: a perturbed initial estimate $\bB(\kappa)$ with $\kappa=0$; \emph{(c)}: a perturbed initial estimate $\bB(\kappa)$ with $\kappa=1.5$; \emph{(d)}: the empirical difference in slope for two regressors $x_{it,1}$ (Covariate 1)
and $x_{it,2}$ (Covariate 2).}
	\label{fig:phi2}
\end{figure}

Figure \ref{fig:phi2} illustrates the $k$-means clustering results for the panel model \eqref{eq:1} with $N=T=30$, $p=2$ regressors, and $K^0=3$ latent groups of equal size ($N/3=10$). The true group-specific parameter vectors are $\aalpha_1^0=(-0.65,0)\trans$, $\aalpha_2^0=(0.65,0)\trans$, and $\aalpha_3^0=(0,1.5)\trans$. Suppose we focus on the first regressor and test for a difference in slopes between the selected groups $\wG_k$ (blue) and $\wG_{k'}$ (brown).   Panel (a) displays the observed initial estimates
for the $N$ individual, corresponding to $\bB(\kappa)$ with $\kappa=\ttheta_j\trans\widehat{\bB}=0.609$. Panels (b) and (c)  show $\bB(\kappa)$ with $\kappa=0$ and $\kappa=1.5$, respectively. In panel (b) ($\kappa=0$),  the blue and rosy brown clusters are ``pushed together" along the first covariate, thus leading to $\ttheta_j\trans\bB(\kappa)=0$, i.e., no statistical difference in the first covariate (x-axis) between the  two clusters. By contrast, in panel (c) ($\kappa=1.5$), the blue and brown groups are ``pulled apart", resulting in an increasing gap between the blue and the brown clusters. Panel (d) illustrates how the slope differences for both covariates  vary with  $\kappa$.  It can be seen that the difference in the coefficient of Covariate 1 (red line) increases linearly with $\kappa$, whereas the difference for Covariate 2 (teal line) remains nearly constant.

Recall that the feasible set
$$
\cS_j=\{\kappa\in\mbR:\Cluster(\widehat{\bB})=\text{Cluster}(\bB(\kappa))\}.
$$
Similar to Theorem 2, we have
\begin{align}\label{eq:sj1}
\cS_j=&\left(
\bigcap_{i=1}^N\bigcap_{k=1}^K
\left\{
\phi_j:\left\|{\bbeta}_i^{\ini}(\kappa)-\widehat{\bmm}_{\hat{g}_i^{(0)}(\widehat{\bB})}^{(0)}\left(\bB(\kappa)\right)\right\|_{\tbx_i}^2
\leq
\left\|{\bbeta}_i^{\ini}(\kappa)-\widehat{\bmm}_{k}^{(0)}\left(\bB(\kappa)\right)\right\|_{\tbx_i}^2
\right\}
\right)\\\nonumber
&\cap\left(
\bigcap_{s=1}^S\bigcap_{i=1}^N\bigcap_{k=1}^K
\left\{
\phi_j:\left\|{\bbeta}_i^{\ini}(\kappa)-\sum_{i'=1}^N\bw_{i'}^{(s-1)}\left(
\hat{g}_i^{(s)}(\widehat{\bB})
\right)\bbeta_i^{\ini}(\kappa)\right\|_{\tbx_i}^2\right.\right.\\\label{eq:sj2}
&~~~~~~~~~~~~~~~~~~~~~~~~~~~~~~~~~~~~~~~~\left.\left.\leq
\left\|{\bbeta}_i^{\ini}(\kappa)-\sum_{i'=1}^N\bw_{i'}^{(s-1)}\left(
k\right)\bbeta_i^{\ini}(\kappa)\right\|_{\tbx_i}^2
\right\}
\right).
\end{align}

It turns out that computing the p-value $P_j(\widehat{\bB},\widehat{G}_k,\widehat{G}_{k'})$  reduces to characterizing the
 set $\cS_j$. Inspired by the results in Theorem 2, which tests for a difference in the slopes for all covariates, the computation of this p-value admits a computationally-efficient recipe. Consequently, it suffices to characterize the inequalities in \eqref{eq:sj1} and \eqref{eq:sj2}. We now present two important lemmas as follows.

\begin{lemma}\label{lem11}
For $\bB(\kappa)$ defined in \eqref{eq:ipj}, we have that $\|{\bbeta}_i^{\ini}(\kappa)-\bbeta_{i'}^{\ini}(\kappa)\|_{\tbx_i}^2=a_j\kappa^2+b_j\kappa+\gamma_j$, where
\begin{eqnarray*}
a_j&=&
\left\|
\{v_i\XXi_i\bw_i(\hat{g}_i)^{\mbox{\tiny{T}}}-
v_{i'}\XXi_{i'}\bw_{i'}(\hat{g}_{i'})^{\mbox{\tiny{T}}}
\}\be_j(\ttheta_j\trans\XXi\ttheta_j)^{-1}\right\|_{\tbx_i}^2,\\
b_j&=&2\left[
\{v_i\XXi_i\bw_i(\hat{g}_i)^{\mbox{\tiny{T}}}-
v_{i'}\XXi_{i'}\bw_{i'}(\hat{g}_{i'})^{\mbox{\tiny{T}}}
\}\be_j(\ttheta_j\trans\XXi\ttheta_j)^{-1}
\right]\trans(\tbx_i\trans\tbx_i)\times\\
&&
\left[
\widehat{\bbeta}_i^{\ini}-\widehat{\bbeta}_{i'}^{\ini}-
\{v_i\XXi_i\bw_i(\hat{g}_i)^{\mbox{\tiny{T}}}-
v_{i'}\XXi_{i'}\bw_{i'}(\hat{g}_{i'})^{\mbox{\tiny{T}}}
\}\be_j(\ttheta_j\trans\XXi\ttheta_j)^{-1}\bR_j\widehat{\aalpha}_{\widehat{\cG}}
\right],\\
\gamma_j&=&\left\|\widehat{\bbeta}_i^{\ini}-\widehat{\bbeta}_{i'}^{\ini}
-\left\{
v_i\XXi_i\bw_i(\hat{g}_i)^{\mbox{\tiny{T}}}-
v_{i'}\XXi_{i'}\bw_{i'}(\hat{g}_{i'})^{\mbox{\tiny{T}}}
\right\}
\be_j(\ttheta_j\trans\XXi\ttheta_j)^{-1}\bR_j\widehat{\aalpha}_{\widehat{\cG}}\right\|_{\tbx_i}^2.
\end{eqnarray*}
\end{lemma}

\begin{lemma}\label{lem22}
For $\bB(\kappa)$ defined in \eqref{eq:ipj}, we have that $\|\bbeta_i^{\ini}(\kappa)-
\sum_{i'=1}^N\bw_{i'}^{(s-1)}(k)\bbeta_{i'}^{\ini}(\kappa)\|_{\tbx_i}^2=\tilde{a}_j\kappa^2+\tilde{b}_j\kappa+\tilde{\gamma}_j$, where
\begin{eqnarray*}
\tilde{a}_j&=&\left\|
\left\{
v_i\XXi_i\bw_i(\hat{g}_i)^{\mbox{\tiny{T}}}-
\sum_{i'=1}^Nv_{i'}\bw_{i'}^{(s-1)}(k)\XXi_{i'}\bw_{i'}(\hat{g}_{i'})^{\mbox{\tiny{T}}}
\right\}
\be_j(\ttheta_j\trans\XXi\ttheta_j)^{-1}
\right\|_{\tbx_i}^2,\\
\tilde{b}_j&=&2\left[
\left\{
v_i\XXi_i\bw_i(\hat{g}_i)^{\mbox{\tiny{T}}}-
\sum_{i'=1}^Nv_{i'}\bw_{i'}^{(s-1)}(k)\XXi_{i'}\bw_{i'}(\hat{g}_{i'})^{\mbox{\tiny{T}}}
\right\}
\be_j(\ttheta_j\trans\XXi\ttheta_j)^{-1}
\right]\trans(\tbx_i\trans\tbx_i)\times\\
&& \left[
\widehat{\bbeta}_i^{\ini}-\sum_{i'=1}^N\bw_{i'}^{(s-1)}(k)\widehat{\bbeta}_{i'}^{\ini}
-\left\{
v_i\XXi_i\bw_i(\hat{g}_i)^{\mbox{\tiny{T}}}-
\sum_{i'=1}^Nv_{i'}\bw_{i'}^{(s-1)}(k)\XXi_{i'}\bw_{i'}(\hat{g}_{i'})^{\mbox{\tiny{T}}}
\right\}
\be_j(\ttheta_j\trans\XXi\ttheta_j)^{-1}\bR_j\widehat{\aalpha}_{\widehat{\cG}}
\right],\\
\tilde{\gamma}_j&=&\left\|
\widehat{\bbeta}_i^{\ini}-\sum_{i'=1}^N\bw_{i'}^{(s-1)}(k)\widehat{\bbeta}_{i'}^{\ini}
-\left\{
v_i\XXi_i\bw_i(\hat{g}_i)^{\mbox{\tiny{T}}}-
\sum_{i'=1}^Nv_{i'}\bw_{i'}^{(s-1)}(k)\XXi_{i'}\bw_{i'}(\hat{g}_{i'})^{\mbox{\tiny{T}}}
\right\}
\be_j(\ttheta_j\trans\XXi\ttheta_j)^{-1}\bR_j\widehat{\aalpha}_{\widehat{\cG}}
\right\|_{\tbx_i}^2.
\end{eqnarray*}
\end{lemma}

From Lemmas \ref{lem11} and \ref{lem22}, all of the inequalities in \eqref{eq:sj1} and \eqref{eq:sj2} are quadratic in $\kappa$, with coefficients that can be analytically computed. We omit the proofs because  they closely mirror the arguments presented in the proofs of Lemmas 1 and 2.

\subsection{Selective p-value for the GMM estimator}\label{secb2}
We briefly review the $k$-means clustering algorithm based on the GMM estimator. Recall that the individual GMM estimators $\widehat{\bbeta}_i^{\ini}=(\breve{\bx}_i\trans\bz_i\OOmega_{i,NT}\bz_i\trans\breve{\bx}_i)^{-1}
\breve{\bx}_i\trans\bz_i\OOmega_{i,NT}\bz_i\trans\breve{\by}_i$, $i=1,\cdots,N$, and
the  group-specific estimator is determined by a function of $\widehat{\bbeta}_i^{\ini}$, i.e., $\widehat{\aalpha}_{G_k}=
\sum_{i:g_i=k}\bw_i(k)\widehat{\bbeta}_i^{\ini}$, where
$\bw_i(k)=\left(\sum_{i'\in G_k}\breve{\bx}_{i'}\trans\bz_{i'}\OOmega_{i',NT}\bz_{i'}\trans\breve{\bx}_{i'}\right)^{-1}(\breve{\bx}_i\trans\bz_i
\OOmega_{i,NT}\bz_i\trans\breve{\bx}_i)$.
 For a positive integer $K$, $k$-means clustering partitions the $N$ individuals into disjoint subsets $\widehat{G}_1,\cdots,\widehat{G}_K$ by solving the following optimization problem
\begin{eqnarray}\nonumber
&&\underset{G_1,\cdots,G_K}{\text{minimize}}
\left\{
\sum_{k=1}^K\sum_{i\in G_k}\left(\widehat{\bbeta}_i^{\ini}-\sum_{i\in G_k}\bw_i(k)\widehat{\bbeta}_i^{\ini}\right)\trans\breve{\bx}_i\trans
\bz_i\OOmega_{i,NT}\bz_i\trans\breve{\bx}_i\left(
\widehat{\bbeta}_i^{\ini}-\sum_{i\in G_k}\bw_i(k)\widehat{\bbeta}_i^{\ini}\right)
\right\},\\ \label{eq:333}
&&~~~~~~~~~~\text{subject to}~~\cup_{k=1}^KG_k=[N],~~G_k\cap G_{k'}=\emptyset,~~\forall\text{$k\neq k'$.}
\end{eqnarray}
We summarize the main procedures in Algorithm \ref{alg:2}.

\begin{algorithm*}[!h]
\caption{$k$-means clustering algorithm based on the GMM objective function.}
\label{alg:2}
\begin{algorithmic}[1]
\STATE Initialize the centroids $(\widehat{\bmm}_1^{(0)},\cdots,\widehat{\bmm}_K^{(0)})$ by sampling $K$ individuals from
$\widehat{\bbeta}_1^{\ini},\cdots,\widehat{\bbeta}_N^{\ini}$ without replacement using a random seed.
\STATE Compute assignments $\hat{g}_i^{(0)}\leftarrow\arg\min_{k\in[K]}
(\widehat{\bbeta}_i^{\ini}-\widehat{\bmm}_k^{(0)})\trans\breve{\bx}_i\trans\bz_i\OOmega_{i,NT}\bz_i\breve{\bx}_i(\widehat{\bbeta}_i^{\ini}-\widehat{\bmm}_k^{(0)})$;
\STATE Initialize $s=0$:
\WHILE{$s\leq S$}
\STATE Update centroids: $\widehat{\bmm}_{k}^{(s+1)}=\sum_{i=\hat{g}_i^{(s)}}\bw_i^{\hat{g}_i^{(s)}}\widehat{\bbeta}_i^{\ini}$; 
\STATE Update assignment: $\hat{g}_i^{(s+1)}\leftarrow\arg\min_{k\in[K]}
(\widehat{\bbeta}_i^{\ini}-\widehat{\bmm}_{k}^{(s+1)})\trans\breve{\bx}_i\trans\bz_i\OOmega_{i,NT}\bz_i\breve{\bx}_i(\widehat{\bbeta}_i^{
\ini}-\widehat{\bmm}_{k}^{(s+1)})$;
\IF{$\hat{g}_i^{(s+1)}=\hat{g}_i^{(s)}$ for all $1\leq i\leq N$}
\STATE break
\ELSE
\STATE $s\leftarrow s+1$
\ENDIF
\ENDWHILE
\RETURN $\{\widehat{\aalpha}_{\wG_k}=\widehat{\bmm}_k^{(S)}\}_{k=1}^K$ and $\wG_k=\{i:\hat{g}_i=k\}$ with $\hat{g}_i=\hat{g}_i^{(S)}$.
\end{algorithmic}
\end{algorithm*}

Let $\widehat{\bB}=(\widehat{\bbeta}_1^{\ini\tiny{\mbox{T}}},\cdots,\widehat{\bbeta}_N^{\ini\tiny{\mbox{T}}})\trans$ be a $Np\times1$ vector where $\widehat{\bbeta}_i^{\ini}$ is replaced by the initial GMM estimators, and $\ttheta=\left(v_1\bw_1({\hat{g}_1}),\cdots,v_N\bw_N({\hat{g}_N})\right)\trans$ be a $Np\times p$ matrix where $\hat{g}_i$ is given in Algorithm \ref{alg:2}. It suffices to characterize the one-dimensional set
\begin{equation}\label{eq:sg}
\cS=\left\{
\phi\geq0:\text{Cluster}(\widehat{\bB})=
\text{Cluster}({\bB}(\phi))
\right\},
\end{equation}
where $\bB(\phi)=\dir(\cP_{\ttheta}\widehat{\bB})\cdot\phi+\cP_{\ttheta}^{\perp}\widehat{\bB}$.
Proposition \ref{thm:g1} below is a direct extension of Theorem 1 to the case of GMM estimator.

\begin{proposition}\label{thm:g1}
Let $\bB$ be an initial GMM estimator, and impose the fixed-\(N\) and \(T\to\infty\) framework. Then, conditional on $\text{Cluster}(\widehat{\bB})$, $\dir(\cP_{\ttheta}\widehat{\bB})$, and $\cP_{\ttheta}^{\perp}\widehat{\bB}$, and under the null hypothesis $H_0^{(G_k,G_{k'})}$ for any $G_k,G_{k'}\in\text{Cluster}(\widehat{\bB})$, the test statistic $W_{k,k'}=(\ttheta\trans\bB)\trans(\bR\SSigma_{\widehat{\cG}}
\bR\trans)^{-1}(\ttheta\trans\bB)$ converges to a truncated $\chi_p^2$ distribution over the set  $\cR$ where
$$
\cR=\left\{\omega:
\omega=\phi^2\left(\ttheta\trans\dir(\cP_{\ttheta}\widehat{\bB})\right)\trans
\left(\bR\widehat{\SSigma}_{\widehat{\cG}}\bR\right)^{-1}\left(\ttheta\trans\dir(\cP_{\ttheta}\widehat{\bB})\right),\phi\in\cS~
\text{defined in \eqref{eq:sg}}\right\}.
$$
 In particular, the selective p-value is given by
\begin{equation}\label{eq:gp1}
\lim_{T\rightarrow\infty}P(\widehat{\bB};G_k,G_{k'})=1-\mathbb{F}(\widehat{W}_{k,k'};\cR).
\end{equation}
Moreover, the test that reject $H_0^{(\wG_k,\wG_{k'})}:\widehat{\aalpha}_{\wG_k}=\widehat{\aalpha}_{\wG_{k'}}$ when $P(\widehat{\bB};\wG_k,\wG_{k'})\leq\alpha$ controls the selective Type
 I error at level $\alpha$.
\end{proposition}

From Proposition \ref{thm:g1}, we know that to compute $P(\widehat{\bB};G_k,G_{k'})$ in \eqref{eq:gp1}, it suffices to characterize the set
\begin{equation}\label{eq:sg1}
\cS=\left\{
\phi\geq0:\bigcap_{s=0}^S\bigcap_{i=1}^N
\left\{
g_i^{(s)}\left(\bB(\phi)\right)=
g_i^{(s)}(\widehat{\bB})
\right\}
\right\},
\end{equation}
where $g_i^{(s)}(\cdot)$ is given in Algorithm \ref{alg:2}. Parallel to Theorem 3.2, we provide the following proposition to illustrate how to characterize set $\cS$ in \eqref{eq:sg1}.

\begin{proposition}\label{prop2}
Suppose that we apply the $k$-means clustering (Algorithm \ref{alg:2}) to the GMM coefficient vector $\widehat{\bB}$, yielding $K$ clusters in at most $S$ iterations. Define
\begin{equation}\label{eq:wsm}
\bw_i^{(s)}(k)=\left(\sum_{\{i':\hat{g}_{i'}^{(s)}(\widehat{\bB})=k\}}\breve{\bx}_{i'}\trans\bz_{i'}\OOmega_{i',NT}
\bz_{i'}\trans\breve{\bx}_{i'}\right)^{-1}(\breve{\bx}_i\trans\bz_i
\OOmega_{i,NT}\bz_i\trans\breve{\bx}_i)\cdot1\left
\{\hat{g}_i^{(s)}(\widehat{\bB})=k\right\}.
\end{equation}
Then, for $\cS$ defined in \eqref{eq:sg1}, we have
\begin{align}\label{eq:smm1}
\cS=&\left(
\bigcap_{i=1}^N\bigcap_{k=1}^K
\left\{
\phi:\left\|{\bbeta}_i^{\ini}(\phi)-\widehat{\bmm}_{\hat{g}_i^{(0)}(\widehat{\bB})}^{(0)}\left(\bB(\phi)\right)\right\|_{\OOmega_{i,NT}^{1/2}\bz_i\trans\breve{\bx}_i}^2
\leq
\left\|{\bbeta}_i^{\ini}(\phi)-\widehat{\bmm}_{k}^{(0)}\left(\bB(\phi)\right)\right\|_{\OOmega_{i,NT}^{1/2}\bz_i\trans\breve{\bx}_i}^2
\right\}
\right)\\\nonumber
&\cap\left(
\bigcap_{s=1}^S\bigcap_{i=1}^N\bigcap_{k=1}^K
\left\{
\phi:\left\|{\bbeta}_i^{\ini}(\phi)-\sum_{i'=1}^N\bw_{i'}^{(s-1)}\left(
\hat{g}_i^{(s)}(\widehat{\bB})
\right)\bbeta_i^{\ini}(\phi)\right\|_{\OOmega_{i,NT}^{1/2}\bz_i\trans\breve{\bx}_i}^2\right.\right.\\\label{eq:smm2}
&~~~~~~~~~~~~~~~~~~~~~~~~~~~~~~~~~~~\left.\left.\leq
\left\|{\bbeta}_i^{\ini}(\phi)-\sum_{i'=1}^N\bw_{i'}^{(s-1)}\left(
k\right)\bbeta_i^{\ini}(\phi)\right\|_{\OOmega_{i,NT}^{1/2}\bz_i\trans\breve{\bx}_i}^2
\right\}
\right).
\end{align}
\end{proposition}
Proposition \ref{prop2} says that $\cS$ can be yielded by characterizing the sets in \eqref{eq:smm1} and \eqref{eq:smm2}. Similar to Lemmas 1 and 2 in the main text, we present the following two lemmas for the GMM estimators.

\begin{lemma}\label{lemm1}
For $\bB(\phi)$ defined in this section, we have that $\|{\bbeta}_i^{\ini}(\phi)-\bbeta_{i'}^{\ini}(\phi)\|_{\OOmega_{i,NT}^{1/2}\bz_i\trans\breve{\bx}_i}^2=a\phi^2+b\phi+\gamma$, where
\begin{eqnarray*}
a&=&\left\|\frac{\{v_i\XXi_i\bw_i(\hat{g}_i)^{\mbox{\tiny{T}}}
-v_{i'}\XXi_{i'}\bw_{i'}(\hat{g}_{i'})^{\mbox{\tiny{T}}}
\}(\ttheta\trans\XXi\ttheta)^{-1}\bR\widehat{\aalpha}_{\widehat{\cG}}
}
{\|\cP_{\ttheta}\widehat{\bB}\|_2}\right\|_{\OOmega_{i,NT}^{1/2}\bz_i\trans\breve{\bx}_i}^2,\\
b&=&2\frac{
\left[
\{v_i\XXi_i\bw_i(\hat{g}_i)^{\mbox{\tiny{T}}}
-v_{i'}\XXi_{i'}\bw_{i'}(\hat{g}_{i'})^{\mbox{\tiny{T}}}
\}(\ttheta\trans\XXi\ttheta)^{-1}\bR\widehat{\aalpha}_{\widehat{\cG}}
\right]\trans}
{\|\cP_{\ttheta}\widehat{\bB}\|_2}
(\breve{\bx}_i\trans\bz_i
\OOmega_{i,NT}\bz_i\trans\breve{\bx}_i)\\
&&\times \left[
\widehat{\bbeta}_i^{\ini}-\widehat{\bbeta}_{i'}^{\ini}
-\left\{
v_i\XXi_i\bw_i(\hat{g}_i)^{\mbox{\tiny{T}}}-
v_{i'}\XXi_{i'}\bw_{i'}(\hat{g}_{i'})^{\mbox{\tiny{T}}}
\right\}(\ttheta\trans\XXi\ttheta)^{-1}\bR\widehat{\aalpha}_{\widehat{\cG}}
\right],\\
\gamma&=&\left\|\widehat{\bbeta}_i^{\ini}-\widehat{\bbeta}_{i'}^{\ini}
-\left\{
v_i\XXi_i\bw_i(\hat{g}_i)^{\mbox{\tiny{T}}}-
v_{i'}\XXi_{i'}\bw_{i'}(\hat{g}_{i'})^{\mbox{\tiny{T}}}
\right\}(\ttheta\trans\XXi\ttheta)^{-1}\bR\widehat{\aalpha}_{\widehat{\cG}}\right\|_{\OOmega_{i,NT}^{1/2}\bz_i\trans\breve{\bx}_i}^2.
\end{eqnarray*}
\end{lemma}

\begin{lemma}\label{lemm2}
For $\bB(\phi)$ defined in this section and $\bw_i^{(s)}(k)$ in \eqref{eq:wsm}, we have $\|\bbeta_i^{\ini}(\phi)-
\sum_{i'=1}^N\bw_{i'}^{(s-1)}(k)\bbeta_{i'}^{\ini}(\phi)$ $\|_{\OOmega_{i,NT}^{1/2}\bz_i\trans\breve{\bx}_i}^2=\tilde{a}\phi^2+\tilde{b}\phi+\tilde{\gamma}$, where
\begin{eqnarray*}
\tilde{a}&=&\left\|\frac{
\left\{
v_i\XXi_i\bw_i(\hat{g}_i)^{\mbox{\tiny{T}}}-
\sum_{i'=1}^N
v_{i'}\bw_{i'}^{(s-1)}(k)\XXi_{i'}\bw_{i'}(\hat{g}_{i'})^{\mbox{\tiny{T}}}
\right\}
(\ttheta\trans\XXi\ttheta)^{-1}\bR\widehat{\aalpha}_{\widehat{\cG}}
}
{\|\cP_{\ttheta}\widehat{\bB}\|_2}\right\|_{\OOmega_{i,NT}^{1/2}\bz_i\trans\breve{\bx}_i}^2,\\
\tilde{b}&=&2\frac{
\left[
\left\{
v_i\XXi_i\bw_i(\hat{g}_i)^{\mbox{\tiny{T}}}-
\sum_{i'=1}^N
v_{i'}\bw_{i'}^{(s-1)}(k)\XXi_{i'}\bw_{i'}(\hat{g}_{i'})^{\mbox{\tiny{T}}}
\right\}
(\ttheta\trans\XXi\ttheta)^{-1}\bR\widehat{\aalpha}_{\widehat{\cG}}
\right]\trans
}
{\|\cP_{\ttheta}\widehat{\bB}\|_2}(\breve{\bx}_i\trans\bz_i
\OOmega_{i,NT}\bz_i\trans\breve{\bx}_i)\\
&&\times \left[
\widehat{\bbeta}_i^{\ini}-\sum_{i'=1}^N\bw_{i'}^{(s-1)}(k)\widehat{\bbeta}_{i'}^{\ini}-
\left\{
v_i\XXi_i\bw_i(\hat{g}_i)^{\mbox{\tiny{T}}}
-\sum_{i'=1}^Nv_{i'}\bw_{i'}^{(s-1)}(k)\XXi_{i'}\bw_{i'}(\hat{g}_{i'})^{\mbox{\tiny{T}}}
\right\}
(\ttheta\trans\XXi\ttheta)^{-1}\bR\widehat{\aalpha}_{\widehat{\cG}}
\right],\\
\tilde{\gamma}&=&\left\|
\widehat{\bbeta}_i^{\ini}-\sum_{i'=1}^N\bw_{i'}^{(s-1)}(k)\widehat{\bbeta}_{i'}^{\ini}-
\left\{
v_i\XXi_i\bw_i(\hat{g}_i)^{\mbox{\tiny{T}}}
-\sum_{i'=1}^Nv_{i'}\bw_{i'}^{(s-1)}(k)\XXi_{i'}\bw_{i'}(\hat{g}_{i'})^{\mbox{\tiny{T}}}
\right\}
(\ttheta\trans\XXi\ttheta)^{-1}\bR\widehat{\aalpha}_{\widehat{\cG}}
\right\|_{\OOmega_{i,NT}^{1/2}\bz_i\trans\breve{\bx}_i}^2.
\end{eqnarray*}
\end{lemma}

We do not provide proofs for the corollaries and lemmas in this section, as they are similar to those of the theoretical results established in Section \ref{sec:3} of the main text and proven in Appendix \ref{A}.

\renewcommand{\theequation}{D.\arabic{equation}}
\renewcommand{\thefigure}{D.\arabic{figure}}
\renewcommand{\thetable}{D.\arabic{table}}
\renewcommand{\thealgorithm}{D.\arabic{algorithm}}
\renewcommand{\thetheorem}{D.\arabic{theorem}}
\renewcommand{\thesection}{D}
\setcounter{figure}{0}


\section{Proofs}\label{A}


\subsection{Proof of Theorem 1}

In this proof, recall that  $\bB$ and $\widehat{\bB}$ represent the $Np\times1$-vector of initial random coefficients and its observed realization, respectively. We work under the fixed-\(N\), fixed-\(p\), fixed-\(K\), and fixed-\(S\) framework stated in Theorem \ref{thm1}, with \(T\to\infty\). Hence \(\bB\) is fixed-dimensional, and the selection event generated by the recorded \(k\)-means path contains only finitely many inequalities.

We consider the Mahalanobis-type statistic
\begin{align}
W = NT (\ttheta\trans\bB)\trans (\ttheta\trans\XXi\ttheta)^{-1} (\ttheta\trans\bB). \label{eq:quad_form_W}
\end{align}

We decompose $\bB$ into the following two components \citep{yun2023selective}:
\[\bB=\mathcal{P}_{\ttheta}\bB+\mathcal{P}_{\ttheta}^{\perp}\bB=\frac{\XXi\ttheta(\ttheta\trans\XXi\ttheta)^{-1}\ttheta\trans\bB}{
\|\XXi\ttheta(\ttheta\trans\XXi\ttheta)^{-1}\ttheta\trans\bB\|_2}\cdot\|\XXi\ttheta(
\ttheta\trans\XXi\ttheta)^{-1}\ttheta\trans\bB\|_2+\mathcal{P}_{\ttheta}^{\perp}\bB.\]

Substituting the above equality into the definition of $P(\widehat{\bB};\wG_{k},\wG_{k'})$ yields
\begin{eqnarray*}\nonumber
P(\widehat{\bB};\wG_{k},\wG_{k'})&=&
\mathbb{P}\Big(
W\geq NT (\ttheta\trans\widehat{\bB})\trans (\ttheta\trans\XXi\ttheta)^{-1} (\ttheta\trans\widehat{\bB})~\Big|~\mathcal{P}_{\ttheta}^{\perp}\bB=\mathcal{P}_{\ttheta}^{\perp}\widehat{\bB},
\dir(\cP_{\ttheta}\bB)=
\dir(\cP_{\ttheta}\widehat{\bB}),
\\\label{eq:pb}
&&~~~~~~~\ \text{Cluster}(\widehat{\bB})= \text{Cluster}(\dir(\cP_{\ttheta}\bB)\cdot\|\cP_{\ttheta}\bB\|_2+\mathcal{P}_{\ttheta}^{\perp}\bB)
\Big)
\end{eqnarray*}
To simplify the above equation, we now show that
\begin{equation}
\label{eq: S1}
    W \perp\!\!\!\perp \mathcal{P}_{\ttheta}^{\perp}\bB, \quad \mathrm{as~}T\to \infty,
\end{equation}
where the notation $\perp\!\!\!\perp$ denotes independence in the limiting fixed-dimensional Gaussian experiment. It is enough to show that the Gaussian limits of $\ttheta\trans \bB$ and $\mathcal{P}_{\ttheta}^{\perp}\bB$ have zero cross-covariance. Under the fixed-dimensional CLT in Theorem \ref{thm1}, the joint limiting distribution of
\[
\sqrt{NT}\,\ttheta\trans(\bB-\bB^0)
\quad\text{and}\quad
\sqrt{NT}\,\mathcal{P}_{\ttheta}^{\perp}(\bB-\bB^0)
\]
is Gaussian. The cross-covariance of these two limiting components is
\begin{equation*}
\begin{aligned}
    \text{Cov}\left(\ttheta\trans \bB, \mathcal{P}_{\ttheta}^{\perp}\bB\right)
    &= \ttheta\trans \XXi (\mathbf{I} - \mathcal{P}_{\ttheta})\trans \\
    &= \ttheta\trans \XXi - \ttheta\trans \XXi
    \left(\XXi\ttheta(\ttheta\trans\XXi\ttheta)^{-1}\ttheta\trans\right)\trans\\
    &= \ttheta\trans \XXi - \ttheta\trans \XXi \ttheta
    (\ttheta\trans\XXi\ttheta)^{-1}\ttheta\trans\XXi\\
    &= \mathbf{0}.
\end{aligned}
\end{equation*}
Because the limiting vector is Gaussian and fixed-dimensional, zero cross-covariance implies independence in the limit:
\begin{equation*}
    \ttheta\trans \bB \perp\!\!\perp \mathcal{P}_{\ttheta}^{\perp}\bB, \quad \text{as } T \to \infty.
\end{equation*}
This justifies Equation \eqref{eq: S1}.

Under $H_0^{(\wG_k,\wG_{k'})}$, the fixed-dimensional CLT implies $\sqrt{NT}\ttheta\trans\bB \xrightarrow{d} \mathcal{N}(\mathbf{0}, \ttheta\trans\XXi\ttheta)$. Therefore, the statistic $W$ asymptotically follows a chi-square distribution
\begin{equation*}
    W \xrightarrow{d} \chi^2_p.
\end{equation*}
By the properties of the multivariate standard normal distribution, the magnitude $W$ and the direction of the projected vector are asymptotically independent
\begin{equation}
    \label{eq: S2}
    W \perp\!\!\!\perp \dir(\cP_{\ttheta}\bB), \quad \mathrm{as~}T\to \infty.
\end{equation}

It immediately implies that
\begin{eqnarray*}\nonumber
&&\lim_{T\to \infty} P(\widehat{\bB};\wG_{k},\wG_{k'})\\
 &=&\lim_{T\to \infty}
\mathbb{P}_{H_0^{(\wG_k,\wG_{k'})}}\Big(
W\geq NT (\ttheta\trans\widehat{\bB})\trans (\ttheta\trans\XXi\ttheta)^{-1} (\ttheta\trans\widehat{\bB})~\Big|\text{Cluster}(\widehat{\bB})=\text{Cluster}(\dir(\cP_{\ttheta}\widehat{\bB})\cdot\|\cP_{\ttheta}\bB\|_2+\mathcal{P}_{\ttheta}^{\perp}\widehat{\bB})
\Big)\\
 &=& \lim_{T\to \infty}\mathbb{P}_{H_0^{(\wG_k,\wG_{k'})}}\Big(
W\geq NT (\ttheta\trans\widehat{\bB})\trans (\ttheta\trans\XXi\ttheta)^{-1} (\ttheta\trans\widehat{\bB})~\Big|\|\cP_{\ttheta}\bB\|_2\in \mathcal{S}(\widehat{\bB})
\Big),
\end{eqnarray*}
where the first equality holds due to the above two limiting independences, and $\mathcal{S}(\widehat{\bB})=\{\phi\geq0:\text{Cluster}(\widehat{\bB})=\text{Cluster}({\bB}(\phi))\}$. Because \(N,K\), and \(S\) are fixed, this conditioning event is a fixed measurable truncation region in the finite-dimensional Gaussian limit.

Note that $\cP_{\ttheta}\bB = \|\cP_{\ttheta}\bB\|_2 \cdot \dir(\cP_{\ttheta}\bB)$, and by the property of the projection, $\ttheta\trans \bB = \ttheta\trans (\cP_{\ttheta} \bB) = \|\cP_{\ttheta}\bB\|_2 \cdot (\ttheta\trans \dir(\cP_{\ttheta}\bB))$. Substituting this into \eqref{eq:quad_form_W} gives
\begin{equation*}
W = \|\cP_{\ttheta}\bB\|_2^2 \cdot \left[ NT (\ttheta\trans \dir(\cP_{\ttheta}\bB))\trans (\ttheta\trans \XXi \ttheta)^{-1} (\ttheta\trans \dir(\cP_{\ttheta}\bB)) \right] := \|\cP_{\ttheta}\bB\|_2^2 \cdot C(\dir(\cP_{\ttheta}\bB)),
\end{equation*}
where $C(\dir(\cP_{\ttheta}\bB))$ is a constant given the fixed direction $\dir(\cP_{\ttheta}\bB)$. We can thus define an equivalent truncation set for $W$ as $\mathcal{R}(\widehat{\bB}) = \{ \omega \geq 0 : \sqrt{\omega / C(\dir(\cP_{\ttheta}\widehat{\bB}))} \in \mathcal{S}(\widehat{\bB}) \}$.

It follows that the conditional event $\|\cP_{\ttheta}\bB\|_2 \in \mathcal{S}(\widehat{\bB})$ is equivalent to $W \in \mathcal{S}_W(\widehat{\bB})$. Substituting this into the previous equation, we have
\begin{eqnarray*}\nonumber
&&\lim_{T\to \infty} P(\widehat{\bB};\wG_{k},\wG_{k'})\\
 &=& \lim_{T\to \infty}\mathbb{P}_{H_0^{(\wG_k,\wG_{k'})}}\Big(
W\geq NT (\ttheta\trans\widehat{\bB})\trans (\ttheta\trans\XXi\ttheta)^{-1} (\ttheta\trans\widehat{\bB}) ~\Big| W \in \mathcal{R}(\widehat{\bB})
\Big).
\end{eqnarray*}

Since $W \xrightarrow{d} \chi^2_p$, the selective p-value is computed as
\begin{equation*}
\lim_{T\to \infty} P(\widehat{\bB};\wG_k, \wG_{k'})= 1-\mathbb{F}_{\chi^2_p}\left(NT (\ttheta\trans\widehat{\bB})\trans (\ttheta\trans\XXi\ttheta)^{-1} (\ttheta\trans\widehat{\bB});\mathcal{R}(\widehat{\bB}) \right),
\end{equation*}
where $\mathbb{F}_{\chi^2_p}$ is the CDF of the $\chi^2_p$ distribution and $\mathcal{R}$ is the truncation set.

Finally, we will prove the second conclusion in Theorem 1. By the definition of $P(\bB;G_{k},G_{k'})$, we have
\begin{equation}
\label{eq: S4}
\lim_{T\to \infty} \mathbb{P}_{H_0^{(G_k,G_{k'})}}\Big(
P(\bB;G_k,G_{k'})\leq \alpha~\Big|~\mathcal{P}_{\ttheta}^{\perp}\bB,
\dir(\cP_{\ttheta}\bB),\text{Cluster}(\bB)
\Big) = \alpha.
\end{equation}

Therefore, we have
\begin{equation*}
    \begin{aligned}
        &\lim_{T\to \infty}\mathbb{P}_{H_0^{(G_k,G_{k'})}}\Big(
P(\bB;G_k,G_{k'})\leq \alpha~\Big|\text{Cluster}(\bB)
\Big)\\
&=\lim_{T\to \infty}\mathbb{E}_{H_0^{(G_k,G_{k'})}}\Big({\bf 1}\{
P(\bB;G_k,G_{k'})\leq \alpha\}~\Big|\text{Cluster}(\bB)
\Big)\\
&=\lim_{T\to \infty}\mathbb{E}_{H_0^{(G_k,G_{k'})}}\Big[\mathbb{E}_{H_0^{(G_k,G_{k'})}}\Big({\bf 1}\{
P(\bB;G_k,G_{k'})\leq \alpha\}~\Big|~\mathcal{P}_{\ttheta}^{\perp}\bB,
\dir(\cP_{\ttheta}\bB),\text{Cluster}(\bB)
\Big)\\
&~~~~~~~~~~~~~~~~~~~~~~~~~~~~~~~~\Big|\text{Cluster}(\bB)\Big]\\
&=\lim_{T\to \infty}\mathbb{E}_{H_0^{(G_k,G_{k'})}}\Big( \alpha~\Big|\text{Cluster}(\bB)
\Big)\\
&=\alpha,
    \end{aligned}
\end{equation*}
where the second equality holds due to the law of iterated expectation, and the third equality follows from Equation \eqref{eq: S4}.

\subsection{Proof of Theorem 2}

We begin with two lemmas, which serve as the ``base case'' of the induction argument of the proof.

\begin{lemma}
\label{lem: S1}
    Recall that $\hat{g}_i^{(s)}$ denotes the group index of individual $i$ at the $s$th iteration obtained by Algorithm 1 of the paper, and that $\widehat{\bmm}_{k}^{(0)}$ denotes the $k$th centroid. Define
    \begin{equation}
        \label{eq: lemS1_1}
        \mathcal{S}_0=\bigcap_{i=1}^N \left\{\phi\in \mathbb{R}:\hat{g}_i^{(0)}\left(\bB(\phi)\right)= \hat{g}_i^{(0)}(\widehat{\bB})\right\}.
    \end{equation}
    We have
    \begin{equation}
        \label{eq: lemS1_2}
        \mathcal{S}_0=\bigcap_{i=1}^N \bigcap_{k=1}^K \left\{\phi:\left\|{\bbeta}_i^{\ini}(\phi)-\widehat{\bmm}_{\hat{g}_i^{(0)}(\widehat{\bB})}^{(0)}\left(\bB(\phi)\right)\right\|_{\tbx_i}^2 \leq \left\|{\bbeta}_i^{\ini}(\phi)-\widehat{\bmm}_{k}^{(0)}\left(\bB(\phi)\right)\right\|_{\tbx_i}^2\right\}.
    \end{equation}
\end{lemma}
\begin{proof}
    We first prove that the set in Equation \eqref{eq: lemS1_1} is a subset of the right hand side in Equation \eqref{eq: lemS1_2}, denoted $\cS_0'$. For any $\phi_0\in \cS_0$ and any $i=1,\dots,N$, we have
    \begin{equation}
        \begin{aligned}
            \label{eq: lem_S1_a}\hat{g}_i^{(0)}\left(\bB(\phi_0)\right)&= \arg\min_{k\in[K]} (\widehat{\bbeta}_i^{\ini}(\phi_0)-\widehat{\bmm}_k^{(0)}(\phi_0))\trans\tilde{\bx}_i\trans\tilde{\bx}_i(\widehat{\bbeta}_i^{\ini}(\phi_0)-\widehat{\bmm}_k^{(0)}(\phi_0))\\
            &:=\arg\min_{k\in[K]} \Big\|\widehat{\bbeta}_i^{\ini}(\phi_0)-\widehat{\bmm}_k^{(0)}(\phi_0)\Big\|_{\tilde{\mathbf{x}}_i}^2,
        \end{aligned}
    \end{equation}
    By the definition of the argmin function, it implies that
    \[ \Big\|\widehat{\bbeta}_i^{\ini}(\phi_0)-\widehat{\bmm}_{\hat{g}_i^{(0)}\left(\bB(\phi_0)\right)}^{(0)}(\phi_0)\Big\|_{\tilde{\mathbf{x}}_i}^2\leq \Big\|\widehat{\bbeta}_i^{\ini}(\phi_0)-\widehat{\bmm}_k^{(0)}(\phi_0)\Big\|_{\tilde{\mathbf{x}}_i}^2, \forall k\in [K]. \]
    Since $\hat{g}_i^{(0)}\left(\bB(\phi)\right)= \hat{g}_i^{(0)}(\widehat{\bB})$ by the definition of $\cS_0$, we have
    \[ \Big\|\widehat{\bbeta}_i^{\ini}(\phi_0)-\widehat{\bmm}_{\hat{g}_i^{(0)}(\widehat{\bB})}^{(0)}(\phi_0)\Big\|_{\tilde{\mathbf{x}}_i}^2\leq \Big\|\widehat{\bbeta}_i^{\ini}(\phi_0)-\widehat{\bmm}_k^{(0)}(\phi_0)\Big\|_{\tilde{\mathbf{x}}_i}^2, \forall k\in [K]. \]

    Next, we prove that $\cS_0'\subseteq \cS_0$. For any $\phi_0\in \cS_0'$ and any $i=1,\dots,N$, we have
    \[ \left\|{\bbeta}_i^{\ini}(\phi_0)-\widehat{\bmm}_{\hat{g}_i^{(0)}(\widehat{\bB})}^{(0)}\left(\bB(\phi_0)\right)\right\|_{\tbx_i}^2 \leq \left\|{\bbeta}_i^{\ini}(\phi_0)-\widehat{\bmm}_{k}^{(0)}\left(\bB(\phi_0)\right)\right\|_{\tbx_i}^2, \forall k\in [K]. \]
    It immediately implies that
    \[ \hat{g}_i^{(0)}(\widehat{\bB})=\arg\min_{k\in[K]} \Big\|\widehat{\bbeta}_i^{\ini}(\phi_0)-\widehat{\bmm}_k^{(0)}(\phi_0)\Big\|_{\tilde{\mathbf{x}}_i}^2. \]
    By Equation \eqref{eq: lem_S1_a}, we have $\hat{g}_i^{(0)}(\widehat{\bB})=\hat{g}_i^{(0)}\left(\bB(\phi_0)\right)$.

    The claim holds by combining these two directions.
\end{proof}

\begin{lemma}
    \label{lem: S2}
    Recall that $\hat{g}_i^{(s)}$ denotes the group index of individual $i$ at the $s$th iteration obtained by Algorithm  1 of the paper, and that $\widehat{\bmm}_{k}^{(0)}$ denotes the $k$th centroid. For $\mathcal{S}_1$ defined as
    \begin{equation}
        \label{eq: lemS2_1}
        \mathcal{S}_1=\bigcap_{s=0}^1\bigcap_{i=1}^N \left\{\phi\in \mathbb{R}:\hat{g}_i^{(s)}\left(\bB(\phi)\right)= \hat{g}_i^{(s)}(\widehat{\bB})\right\},
    \end{equation}
    and $\bw_i^{(s)}(k)=\left(\sum_{\{i':\hat{g}_{i'}^{(s)}(\widehat{\bB})=k\}}\tbx_{i'}\trans\tbx_{i'}\right)^{-1}(\tbx_i\trans\tbx_i)\cdot1\left
\{\hat{g}_i^{(s)}(\widehat{\bB})=k\right\}$, we have
    \begin{equation}
        \begin{aligned}
            \label{eq: lemS2_2}
            \mathcal{S}_1&=\bigcap_{i=1}^N \bigcap_{k=1}^K \left\{\phi:\left\|{\bbeta}_i^{\ini}(\phi)-\widehat{\bmm}_{\hat{g}_i^{(0)}(\widehat{\bB})}^{(0)}\left(\bB(\phi)\right)\right\|_{\tbx_i}^2 \leq \left\|{\bbeta}_i^{\ini}(\phi)-\widehat{\bmm}_{k}^{(0)}\left(\bB(\phi)\right)\right\|_{\tbx_i}^2\right\}\cap\\
            &\left(\bigcap_{i=1}^N \bigcap_{k=1}^K \left\{\phi:\left\|{\bbeta}_i^{\ini}(\phi)-\sum_{i'=1}^N\bw_{i'}^{(0)}\left(\hat{g}_i^{(1)}(\widehat{\bB})\right)\bbeta_i^{\ini}(\phi)\right\|_{\tbx_i}^2\leq\left\|{\bbeta}_i^{\ini}(\phi)-\sum_{i'=1}^N\bw_{i'}^{(0)}\left(k\right)\bbeta_i^{\ini}(\phi)\right\|_{\tbx_i}^2\right\}\right).
        \end{aligned}
    \end{equation}
\end{lemma}

\begin{proof}
    We first prove that the set in Equation \eqref{eq: lemS2_1} is a subset of the right hand side in Equation \eqref{eq: lemS2_2}, denoted $\cS_1'$. For any $\phi_0\in \cS_1$ and any $i=1,\dots,N$, we have
    \begin{equation*}
        \begin{aligned}
        &\hat{g}_i^{(1)}\left(\bB(\phi_0)\right)=\arg\min_{k\in[K]} \Big\|\widehat{\bbeta}_i^{\ini}(\phi_0)-\widehat{\bmm}_k^{(1)}(\phi_0)\Big\|_{\tilde{\mathbf{x}}_i}^2\\
        \overset{(a)}{\implies}& \hat{g}_i^{(1)}(\widehat{\bB})=\arg\min_{k\in[K]} \Big\|\widehat{\bbeta}_i^{\ini}(\phi_0)-\widehat{\bmm}_k^{(1)}(\phi_0)\Big\|_{\tilde{\mathbf{x}}_i}^2\\
        \overset{(b)}{\implies}& \hat{g}_i^{(1)}(\widehat{\bB})=\arg\min_{k\in[K]} \left\|\widehat{\bbeta}_i^{\ini}(\phi_0)-\sum_{{i'}=1}^N\left(\sum_{\{j:\hat{g}_j^{(0)}(\bB(\phi_0))=k\}}
        \tbx_j\trans\tbx_j\right)^{-1}(\tbx_{i'}\trans\tbx_{i'})\times\right.\\
        &~~~~~~~~~~~~~~~~~~~~~~~~~~~~~\left.1\left\{\hat{g}_{i'}^{(0)}(\bB(\phi_0))=
        k\right\}\widehat{\bbeta}_i^{\ini}(\phi_0)\right\|_{\tilde{\mathbf{x}}_i}^2\\
        \overset{(c)}{\implies}& \left\|\widehat{\bbeta}_i^{\ini}(\phi_0)-\sum_{{i'}=1}^N\left(\sum_{\{j:\hat{g}_j^{(0)}(\bB(\phi_0))=
        \hat{g}_i^{(1)}(\widehat{\bB})\}}\tbx_j\trans\tbx_j\right)^{-1}(\tbx_{i'}\trans\tbx_{i'})\times\right.\\
        &~~~~\left.1\left
        \{\hat{g}_{i'}^{(0)}(\bB(\phi_0))=\hat{g}_i^{(1)}(\widehat{\bB})\right\}\widehat{\bbeta}_i^{\ini}(\phi_0)
        \right\|_{\tilde{\mathbf{x}}_i}^2\\
        &\leq \left\|\widehat{\bbeta}_i^{\ini}(\phi_0)-\sum_{{i'}=1}^N\left(\sum_{\{j:\hat{g}_j^{(0)}(\bB(\phi_0))=k\}}\tbx_j\trans\tbx_j\right)^{-1}(\tbx_{i'}\trans\tbx_{i'})\cdot1\left\{\hat{g}_{i'}^{(0)}(\bB(\phi_0))=k\right\}\widehat{\bbeta}_i^{\ini}(\phi_0)\right\|_{\tilde{\mathbf{x}}_i}^2,\\
        &~~~~~~~\forall k\in [K]\\
        \overset{(d)}{\implies}& \left\|\widehat{\bbeta}_i^{\ini}(\phi_0)-\sum_{{i'}=1}^N\left(\sum_{\{j:\hat{g}_j^{(0)}(\widehat{\bB})=\hat{g}_i^{(1)}(\widehat{\bB})\}}\tbx_j\trans\tbx_j\right)^{-1}(\tbx_{i'}\trans\tbx_{i'})\cdot1\left\{\hat{g}_{i'}^{(0)}(\widehat{\bB})=\hat{g}_i^{(1)}(\widehat{\bB})\right\}\widehat{\bbeta}_i^{\ini}(\phi_0)\right\|_{\tilde{\mathbf{x}}_i}^2\\
        &\leq \left\|\widehat{\bbeta}_i^{\ini}(\phi_0)-\sum_{{i'}=1}^N\left(\sum_{\{j:\hat{g}_j^{(0)}(\widehat{\bB})=k\}}\tbx_j\trans\tbx_j\right)^{-1}(\tbx_{i'}\trans\tbx_{i'})\cdot1\left\{\hat{g}_{i'}^{(0)}(\widehat{\bB})=k\right\}\widehat{\bbeta}_i^{\ini}(\phi_0)\right\|_{\tilde{\mathbf{x}}_i}^2, \forall k\in [K]\\
        \overset{(e)}{\implies}& \left\|\widehat{\bbeta}_i^{\ini}(\phi_0)-\sum_{{i'}=1}^N\bw_{i'}^{(0)}(\hat{g}_i^{(1)}(\widehat{\bB}))\widehat{\bbeta}_i^{\ini}(\phi_0)\right\|_{\tilde{\mathbf{x}}_i}^2\leq \left\|\widehat{\bbeta}_i^{\ini}(\phi_0)-\sum_{{i'}=1}^N\bw_{i'}^{(0)}(k)\widehat{\bbeta}_i^{\ini}(\phi_0)\right\|_{\tilde{\mathbf{x}}_i}^2, \forall k\in [K],\\
        \end{aligned}
    \end{equation*}
    where
    \begin{itemize}
        \item Step (a) follows from the definition of Equation \eqref{eq: lemS2_1} ($\hat{g}_i^{(1)}\left(\bB(\phi_0)\right)=\hat{g}_i^{(1)}(\widehat{\bB})$);
        \item Step (b) follows from the definition of $\widehat{\bmm}_k^{(1)}(\phi_0)$;
        \item Step (c) follows from the definition of argmin function;
        \item Step (d) follows from the definition of Equation \eqref{eq: lemS2_1} ($\hat{g}_i^{(1)}\left(\bB(\phi_0)\right)=\hat{g}_i^{(1)}(\widehat{\bB})$);
        \item Step (e) follows from the definition of $\bw_i^{(s)}(k)$ defined in Equation (\ref{eq:ws}) in the paper.
    \end{itemize}
    Moreover, $\phi_0 \in \cS_1$ implies that $\phi_0 \in \cS_0$, which, according to Lemma \ref{lem: S1}, further implies that $\phi_0$ is an element of the first set in the intersection in Equation \eqref{eq: lemS2_2}. To summarize, we have proven that $\cS_1 \subseteq \cS_1'$.

    Next, we proceed to prove the other direction. For any $\phi_0\in \cS_1'$ and any $i=1,\dots,N$, we have
    \begin{equation*}
        \begin{aligned}
        & \left\|\widehat{\bbeta}_i^{\ini}(\phi_0)-\sum_{{i'}=1}^N\left(\sum_{\{j:\hat{g}_j^{(0)}(\widehat{\bB})=\hat{g}_i^{(1)}(\widehat{\bB})\}}\tbx_j\trans\tbx_j\right)^{-1}(\tbx_{i'}\trans\tbx_{i'})\cdot1\left\{\hat{g}_{i'}^{(0)}(\widehat{\bB})=\hat{g}_i^{(1)}(\widehat{\bB})\right\}\widehat{\bbeta}_i^{\ini}(\phi_0)\right\|_{\tilde{\mathbf{x}}_i}^2\\
        &\leq \left\|\widehat{\bbeta}_i^{\ini}(\phi_0)-\sum_{{i'}=1}^N\left(\sum_{\{j:\hat{g}_j^{(0)}(\widehat{\bB})=k\}}\tbx_j\trans\tbx_j\right)^{-1}(\tbx_{i'}\trans\tbx_{i'})\cdot1\left\{\hat{g}_{i'}^{(0)}(\widehat{\bB})=k\right\}\widehat{\bbeta}_i^{\ini}(\phi_0)\right\|_{\tilde{\mathbf{x}}_i}^2, \forall k\in [K]\\
        \overset{(a)}{\implies}& \left\|\widehat{\bbeta}_i^{\ini}(\phi_0)-\sum_{{i'}=1}^N\left(\sum_{\{j:\hat{g}_j^{(0)}(\bB(\phi_0))=\hat{g}_i^{(1)}(\widehat{\bB})\}}\tbx_j\trans\tbx_j\right)^{-1}(\tbx_{i'}\trans\tbx_{i'})\cdot1\left\{\hat{g}_{i'}^{(0)}(\bB(\phi_0))=\hat{g}_i^{(1)}(\widehat{\bB})\right\}\widehat{\bbeta}_i^{\ini}(\phi_0)\right\|_{\tilde{\mathbf{x}}_i}^2\\
        &\leq \left\|\widehat{\bbeta}_i^{\ini}(\phi_0)-\sum_{{i'}=1}^N\left(\sum_{\{j:\hat{g}_j^{(0)}(\bB(\phi_0))=k\}}\tbx_j\trans\tbx_j\right)^{-1}(\tbx_{i'}\trans\tbx_{i'})\cdot1\left\{\hat{g}_{i'}^{(0)}(\bB(\phi_0))=k\right\}\widehat{\bbeta}_i^{\ini}(\phi_0)\right\|_{\tilde{\mathbf{x}}_i}^2,\\ &~~~~~~~~\forall k\in [K]\\
        \overset{(b)}{\implies}& \hat{g}_i^{(1)}(\widehat{\bB})=\arg\min_{k\in[K]} \left\|\widehat{\bbeta}_i^{\ini}(\phi_0)-\sum_{{i'}=1}^N\left(\sum_{\{j:\hat{g}_j^{(0)}(\bB(\phi_0))=k\}}\tbx_j\trans\tbx_j\right)^{-1}
        (\tbx_{i'}\trans\tbx_{i'})\times\right. \\ &\left.~~~~~1\left\{\hat{g}_{i'}^{(0)}(\bB(\phi_0))=k\right\}\widehat{\bbeta}_i^{\ini}(\phi_0)\right\|_{\tilde{\mathbf{x}}_i}^2\\
        \overset{(c)}{\implies}& \hat{g}_i^{(1)}(\widehat{\bB})=\arg\min_{k\in[K]} \Big\|\widehat{\bbeta}_i^{\ini}(\phi_0)-\widehat{\bmm}_k^{(1)}(\phi_0)\Big\|_{\tilde{\mathbf{x}}_i}^2\equiv\hat{g}_i^{(1)}\left(\bB(\phi_0)\right),
        \end{aligned}
    \end{equation*}
    where
    \begin{itemize}
        \item In step (a), we first apply Lemma \ref{lem: S1}, which implies that $\cS_1'\subseteq \cS_0'$. Therefore, $\phi_0\in \cS_1'$ implies $\hat{g}_i^{(0)}\left(\bB(\phi_0)\right)=\hat{g}_i^{(0)}(\widehat{\bB})$ by the definition of $\cS_0'$;
        \item Step (b) follows from the definition of argmin function;
        \item Step (c) follows from the definition of $\widehat{\bmm}_k^{(s)}(\phi_0)$ and $\hat{g}_i^{(1)}\left(\bB(\phi_0)\right)$.
    \end{itemize}
    We conclude the proof by combining these results above.
\end{proof}

Next, we prove the inductive step in the proof of Theorem 2, which relies on the following claim.

\begin{lemma}
    \label{lem: S3}
    Recall that $\hat{g}_i^{(s)}$ denotes the group index of individual $i$ at the $s$th iteration obtained by Algorithm 1, and that $\widehat{\bmm}_{k}^{(0)}$ denotes the $k$th centroid in Algorithm 1 of the paper. For any $1\leq \tilde{S}\leq S-1$, define
    \begin{equation}
        \label{eq: lemS3_1}
        \mathcal{S}_{\tilde{S}}=\left\{\phi\in \mathbb{R}:\bigcap_{s=0}^{\tilde{S}}\bigcap_{i=1}^N \left\{\hat{g}_i^{(s)}\left(\bB(\phi)\right)= \hat{g}_i^{(s)}(\widehat{\bB})\right\} \right\},
    \end{equation}
and
    \begin{equation}
        \begin{aligned}
            \label{eq: lemS3_2}
            \mathcal{S}_{\tilde{S}}'&=\bigcap_{i=1}^N \bigcap_{k=1}^K \left\{\phi:\left\|{\bbeta}_i^{\ini}(\phi)-\widehat{\bmm}_{\hat{g}_i^{(0)}(\widehat{\bB})}^{(0)}\left(\bB(\phi)\right)
            \right\|_{\tbx_i}^2 \leq \left\|{\bbeta}_i^{\ini}(\phi)-\widehat{\bmm}_{k}^{(0)}\left(\bB(\phi)\right)\right\|_{\tbx_i}^2\right\}\cap\\
            &\left(\bigcap_{s=1}^{\tilde{S}}\bigcap_{i=1}^N \bigcap_{k=1}^K \left\{\phi:\left\|{\bbeta}_i^{\ini}(\phi)-\sum_{i'=1}^N\bw_{i'}^{(s-1)}\left(\hat{g}_i^{(s)}(\widehat{\bB})\right)
            \bbeta_i^{\ini}(\phi)\right\|_{\tbx_i}^2\leq\right.\right.\\
            &~~~~~~~\left.\left. \left\|{\bbeta}_i^{\ini}(\phi)-\sum_{i'=1}^N\bw_{i'}^{(s-1)}
            \left(k\right)\bbeta_i^{\ini}(\phi)\right\|_{\tbx_i}^2\right\}\right),
        \end{aligned}
    \end{equation}
    where $\bw_{i}^{(s)}\left(\cdot\right)$ is defined in Equation (\ref{eq:ws}) in the paper. Then, if $\mathcal{S}_{\tilde{S}}=\mathcal{S}_{\tilde{S}}'$, then $\mathcal{S}_{\tilde{S}+1}=\mathcal{S}_{\tilde{S}+1}'$.
\end{lemma}

\begin{proof}
    By the definition of $\mathcal{S}_{\tilde{S}}$ and $\mathcal{S}_{\tilde{S}+1}$, we have
    \begin{equation}
        \label{eq: lemS3_5}
        \mathcal{S}_{\tilde{S}+1}=\mathcal{S}_{\tilde{S}}\cap \left\{\bigcap_{i=1}^N \left\{\hat{g}_i^{(\tilde{S}+1)}\left(\bB(\phi)\right)= \hat{g}_i^{(\tilde{S}+1)}(\widehat{\bB})\right\} \right\}.
    \end{equation}
    Therefore, it suffices to prove that the Equation \eqref{eq: lemS3_5} is equal to $\mathcal{S}_{\tilde{S}+1}'$.
    We now prove that Equation \eqref{eq: lemS3_5} is a subset of $\mathcal{S}_{\tilde{S}+1}'$. For any $\phi_0\in \mathcal{S}_{\tilde{S}+1}$ and any $i=1,\dots,N$, we have
    \begin{equation*}
        \begin{aligned}
            &\hat{g}_i^{(\tilde{S}+1)}\left(\bB(\phi)\right)=\hat{g}_i^{(\tilde{S}+1)}(\widehat{\bB})\overset{(a)}{\implies}\hat{g}_i^{
            (\tilde{S}+1)}(\widehat{\bB})=\arg\min_{k\in[K]} \Big\|\widehat{\bbeta}_i^{\ini}(\phi_0)-\widehat{\bmm}_k^{(\tilde{S}+1)}(\phi_0)\Big\|_{\tilde{\mathbf{x}}_i}^2\\
            \overset{(b)}{\implies}&\Bigg\|\widehat{\bbeta}_i^{\ini}(\phi_0)-\sum_{{i'}=1}^N\left(\sum_{\{j:\hat{g}_j^{(\tilde{S})}
            (\bB(\phi_0))=\hat{g}_i^{(\tilde{S}+1)}(\widehat{\bB})\}}\tbx_j\trans\tbx_j\right)^{-1}(\tbx_{i'}\trans\tbx_{i'})\times\\
            &~~~~~~~~~~~~~~~~~~~~~~~~~~~~~~~~ 1\left\{\hat{g}_{i'}^{(\tilde{S})}(\bB(\phi_0))=\hat{g}_i^{(\tilde{S}+1)}(\widehat{\bB})\right\}
            \widehat{\bbeta}_i^{\ini}(\phi_0)\Bigg\|_{\tilde{\mathbf{x}}_i}^2\\
        &\leq \left\|\widehat{\bbeta}_i^{\ini}(\phi_0)-\sum_{{i'}=1}^N\left(\sum_{\{j:\hat{g}_j^{(\tilde{S})}(\bB(\phi_0))=k\}}\tbx_j\trans\tbx_j
        \right)^{-1}(\tbx_{i'}\trans\tbx_{i'})\cdot1\left\{\hat{g}_{i'}^{(\tilde{S})}(\bB(\phi_0))=k\right\}\widehat{\bbeta}_i^{\ini}(
        \phi_0)\right\|_{\tilde{\mathbf{x}}_i}^2,\\
        &~~~~~~~~~~~~~~~~ \forall k\in [K]\\
        \overset{(c)}{\implies}& \left\|\widehat{\bbeta}_i^{\ini}(\phi_0)-\sum_{{i'}=1}^N\left(\sum_{\{j:\hat{g}_j^{(\tilde{S})}(\widehat{\bB})=\hat{g}_i^{(\tilde{S}+1)}(\widehat{\bB})\}}\tbx_j\trans\tbx_j\right)^{-1}(\tbx_{i'}\trans\tbx_{i'})\cdot1\left\{\hat{g}_{i'}^{(\tilde{S})}(\widehat{\bB})=\hat{g}_i^{(\tilde{S}+1)}(\widehat{\bB})\right\}\widehat{\bbeta}_i^{\ini}(\phi_0)\right\|_{\tilde{\mathbf{x}}_i}^2\\
        &\leq \left\|\widehat{\bbeta}_i^{\ini}(\phi_0)-\sum_{{i'}=1}^N\left(\sum_{\{j:\hat{g}_j^{(\tilde{S})}(\widehat{\bB})=k\}}\tbx_j\trans\tbx_j\right)^{-1}(\tbx_{i'}\trans\tbx_{i'})\cdot1\left\{\hat{g}_{i'}^{(\tilde{S})}(\widehat{\bB})=k\right\}\widehat{\bbeta}_i^{\ini}(\phi_0)\right\|_{\tilde{\mathbf{x}}_i}^2, \forall k\in [K]\\
        \overset{(d)}{\implies}& \phi_0\in\mathcal{S}_{\tilde{S}+1}',
        \end{aligned}
    \end{equation*}
    where
    \begin{itemize}
        \item Step (a) follows from the definition of $\hat{g}_i^{(\tilde{S}+1)}\left(\bB(\phi_0)\right)$;
        \item Step (b) follows from the definition of $\widehat{\bmm}_k^{(\tilde{S}+1)}(\phi_0)$;
        \item Step (c) follows from $\phi_0\in$ Equation \eqref{eq: lemS3_5}, which implies $\hat{g}_i^{(\tilde{S}+1)}\left(\bB(\phi)\right)=\hat{g}_i^{(\tilde{S}+1)}(\widehat{\bB})$;
        \item Step (d) follows from the definition of $\bw_{i}^{(s)}\left(\cdot\right)$.
    \end{itemize}
    Next, we prove the reverse direction. By the definition of $\mathcal{S}_{\tilde{S}+1}$, together with the definition of $\mathcal{S}_{\tilde{S}+1}'$, we have
    \begin{equation}
    \begin{aligned}
        \label{eq: lemS3_6}
        &\mathcal{S}_{\tilde{S}+1}'=\mathcal{S}_{\tilde{S}}\cap\left(\bigcap_{i=1}^N\bigcap_{k=1}^K\left\{\left\|\widehat{\bbeta}_i^{\ini}(\phi_0)-\sum_{{i'}=1}^N
        \left(\sum_{\{j:\hat{g}_j^{(\tilde{S})}(\widehat{\bB})=\hat{g}_i^{(\tilde{S}+1)}(\widehat{\bB})\}}
        \tbx_j\trans\tbx_j\right)^{-1}(\tbx_{i'}\trans\tbx_{i'})\times \right.\right.\right. \\
        &~~~~~~~~~~~~~~~~~~~~~\left.\left.\left. 1\left\{\hat{g}_{i'}^{(\tilde{S})}
        (\widehat{\bB})=\hat{g}_i^{(\tilde{S}+1)}(\widehat{\bB})\right\}\widehat{\bbeta}_i^{\ini}(\phi_0)
        \right\|_{\tilde{\mathbf{x}}_i}^2\right.\right.\\
        &\leq \left.\left.\left\|\widehat{\bbeta}_i^{\ini}(\phi_0)-\sum_{{i'}=1}^N\left(\sum_{\{j:\hat{g}_j^{(\tilde{S})}(\widehat{\bB})=k\}}\tbx_j\trans\tbx_j\right)^{-1}(\tbx_{i'}\trans\tbx_{i'})\cdot1\left\{\hat{g}_{i'}^{(\tilde{S})}(\widehat{\bB})=k\right\}\widehat{\bbeta}_i^{\ini}(\phi_0)\right\|_{\tilde{\mathbf{x}}_i}^2\right\}\right).
    \end{aligned}
    \end{equation}
    For any $\phi_0\in\mathcal{S}_{\tilde{S}+1}'$  and $i=1,\dots,N$, we have
    \begin{equation*}
        \begin{aligned}
            &\Bigg\|\widehat{\bbeta}_i^{\ini}(\phi_0)-\sum_{{i'}=1}^N\left(\sum_{\{j:\hat{g}_j^{(\tilde{S})}(\widehat{\bB})=
            \hat{g}_i^{(\tilde{S}+1)}(\widehat{\bB})\}}\tbx_j\trans\tbx_j\right)^{-1}
            (\tbx_{i'}\trans\tbx_{i'})1\left\{\hat{g}_{i'}^{(\tilde{S})}(\widehat{\bB})=
            \hat{g}_i^{(\tilde{S}+1)}(\widehat{\bB})\right\}\widehat{\bbeta}_i^{\ini}(\phi_0)\Bigg\|_{\tilde{\mathbf{x}}_i}^2\\
        &\leq \left\|\widehat{\bbeta}_i^{\ini}(\phi_0)-\sum_{{i'}=1}^N\left(\sum_{\{j:\hat{g}_j^{(\tilde{S})}(\widehat{\bB})=k\}}\tbx_j\trans\tbx_j\right)^{-1}(\tbx_{i'}\trans\tbx_{i'})\cdot1\left\{\hat{g}_{i'}^{(\tilde{S})}(\widehat{\bB})=k\right\}\widehat{\bbeta}_i^{\ini}(\phi_0)\right\|_{\tilde{\mathbf{x}}_i}^2, \forall k\in [K]\\
        \overset{(a)}{\implies} &\Bigg\|\widehat{\bbeta}_i^{\ini}(\phi_0)-\sum_{{i'}=1}^N\left(\sum_{\{j:\hat{g}_j^{(\tilde{S})}(\bB(\phi_0))=
        \hat{g}_i^{(\tilde{S}+1)}(\widehat{\bB})\}}\tbx_j\trans\tbx_j\right)^{-1}
        (\tbx_{i'}\trans\tbx_{i'})\times \\
        &~~~~~~~~~ 1\left\{\hat{g}_{i'}^{(\tilde{S})}(\bB(\phi_0))
        =\hat{g}_i^{(\tilde{S}+1)}(\widehat{\bB})\right\}\widehat{\bbeta}_i^{\ini}(\phi_0)\Bigg\|_{\tilde{\mathbf{x}}_i}^2\\
        &\leq \Bigg\|\widehat{\bbeta}_i^{\ini}(\phi_0)-\sum_{{i'}=1}^N\left(\sum_{\{j:\hat{g}_j^{(\tilde{S})}(\bB(\phi_0))=k\}}
        \tbx_j\trans\tbx_j\right)^{-1}(\tbx_{i'}\trans\tbx_{i'})\cdot1\left\{\hat{g}_{i'}^{(\tilde{S})}(\bB(\phi_0))=k\right\}
        \widehat{\bbeta}_i^{\ini}(\phi_0)\Bigg\|_{\tilde{\mathbf{x}}_i}^2,\\
        &~~~~~~~~~ \forall k\in [K]\\
        \overset{(b)}{\implies} &\hat{g}_i^{(\tilde{S}+1)}(\widehat{\bB})=\arg\min_{k\in[K]} \Big\|\widehat{\bbeta}_i^{\ini}(\phi_0)-\widehat{\bmm}_k^{(\tilde{S}+1)}(\phi_0)\Big\|_{\tilde{\mathbf{x}}_i}^2\\
        \overset{(c)}{\implies}& \hat{g}_i^{(\tilde{S}+1)}(\widehat{\bB})=\hat{g}_i^{(\tilde{S}+1)}\left(\bB(\phi_0)\right),
        \end{aligned},
    \end{equation*}
    where
    \begin{itemize}
        \item Step (a) follows from Equation \eqref{eq: lemS3_6}, any element $\phi_0$ in $\mathcal{S}_{\tilde{S}+1}'$ is also an element of $\mathcal{S}_{\tilde{S}}$. Therefore, using the definition of $\mathcal{S}_{\tilde{S}}$ in Equation \eqref{eq: lemS3_1}, we have $\bigcap_{s=1}^{\tilde{S}}\{\hat{g}_i^{(s)}\left(\bB(\phi)\right)=\hat{g}_i^{(s)}(\widehat{\bB})\}$;
        \item Step (b) follows from the definition of $\widehat{\bmm}_k^{(\tilde{S}+1)}(\phi_0)$ and the definition of argmin function;
        \item Step (c) follows from the definition of argmin function.
    \end{itemize}

    Combining the above results, we complete the proof.
\end{proof}
The inductive proof of Theorem 2 follows from combining Lemmas \ref{lem: S1}, \ref{lem: S2} and \ref{lem: S3}.

\subsection{Proof of Lemma 1}
\begin{proof}
    In this proof, we will deal with many block matrix, e.g., $\cP_{\ttheta}\widehat{\bB}$, and $(\cP_{\ttheta}\widehat{\bB})_i$ denotes its $i$-th block.
    Note that we have
    \begin{equation*}
        \begin{aligned}
            \|{\bbeta}_i^{\ini}(\phi)-\bbeta_j^{\ini}(\phi)\|_{\tbx_i}^2&=\left({\bbeta}_i^{\ini}(\phi)-\bbeta_j^{\ini}(\phi)\right)\trans \tbx_i\trans \tbx_i \left({\bbeta}_i^{\ini}(\phi)-\bbeta_j^{\ini}(\phi)\right).
        \end{aligned}
    \end{equation*}
    Now we express ${\bbeta}_i^{\ini}(\phi)-{\bbeta}_j^{\ini}(\phi)$ as a function of $\phi$. Since ${\bB}(\phi)=\dir(\cP_{\ttheta}\widehat{\bB})\cdot\phi+\mathcal{P}_{\ttheta}^{\perp}\widehat{\bB}$, we have ${\bbeta}_i^{\ini}(\phi)=\dir(\cP_{\ttheta}\widehat{\bB})_i\cdot\phi+(\mathcal{P}_{\ttheta}^{\perp}\widehat{\bB})_i=
    \frac{v_i\XXi_i\bw_i(\hat{g}_i)^{{\tiny{\mbox{T}}}}(\ttheta\trans\XXi\ttheta)^{-1}\ttheta\trans\widehat{\bB}}{\Vert \cP_{\ttheta}\widehat{\bB}\Vert_2 }\cdot\phi + \widehat{\bbeta}_i-v_i\XXi_i\bw_i(\hat{g}_i)^{{\tiny{\mbox{T}}}}(\ttheta\trans\XXi \ttheta)^{-1}\ttheta\trans\widehat{\bB}$.
    Then we have
    \begin{equation*}
        \begin{aligned}
            {\bbeta}_i^{\ini}(\phi)-{\bbeta}_j^{\ini}(\phi)&= \left(\frac{v_i\XXi_i\bw_i(\hat{g}_i)^{{\tiny{\mbox{T}}}}(\ttheta\trans\XXi \ttheta)^{-1}\ttheta\trans\widehat{\bB}}{\Vert \cP_{\ttheta}\widehat{\bB}\Vert_2 }-\frac{v_j\XXi_j\bw_j(\hat{g}_j)^{{\tiny{\mbox{T}}}}(\ttheta\trans\XXi \ttheta)^{-1}\ttheta\trans\widehat{\bB}}{\Vert \cP_{\ttheta}\widehat{\bB}\Vert_2 }\right)\cdot\phi \\&+ \widehat{\bbeta}_i-\widehat{\bbeta}_j+v_i\XXi_i\bw_i(\hat{g}_i)^{{\tiny{\mbox{T}}}}(\ttheta\trans \XXi \ttheta)^{-1}\ttheta\trans\widehat{\bB}-v_j\XXi_j\bw_j(\hat{g}_j)^{{\tiny{\mbox{T}}}}(\ttheta\trans \XXi \ttheta)^{-1}\ttheta\trans\widehat{\bB}.
        \end{aligned}
    \end{equation*}

    Since $\ttheta\trans\bB=\bR\aalpha_{\widehat{\cG}}$ and by the definition of $a$, $b$ and $\gamma$, we have
    \begin{equation*}
        \begin{aligned}
            a=\left\|\frac{v_i\XXi_i\bw_i(\hat{g}_i)^{{\tiny{\mbox{T}}}}(\ttheta\trans\XXi \ttheta)^{-1}\bR\aalpha_{\widehat{\cG}}-v_j\XXi_j\bw_j(\hat{g}_j)^{{\tiny{\mbox{T}}}}(\ttheta\trans\XXi \ttheta)^{-1}\bR\aalpha_{\widehat{\cG}}}{\Vert \cP_{\ttheta}\widehat{\bB}\Vert_2}\right\|_{\tbx_i}^2,
        \end{aligned}
    \end{equation*}

    \begin{equation*}
        \begin{aligned}
            b&=2\frac{\left(v_i\XXi_i\bw_i(\hat{g}_i)^{\tiny{\mbox{T}}}(\ttheta\trans\XXi\ttheta)^{-1}\bR\widehat{\aalpha}_{
            \widehat{\cG}}-v_j\XXi_j\bw_j(\hat{g}_j)^{\tiny{\mbox{T}}}(\ttheta\trans\XXi\ttheta)^{-1}\bR\widehat{\aalpha}_{\widehat{\cG}}\right)\trans}{\|\cP_{\ttheta}\widehat{\bB}\|_2}(\tbx_i\trans\tbx_i)\\
            &~~~~\times \left(\widehat{\bbeta}_i-\widehat{\bbeta}_j+v_j\XXi_j\bw_j(\hat{g}_j)^{\tiny{\mbox{T}}}(\ttheta\trans\XXi\ttheta)^{-1}
            \bR\widehat{\aalpha}_{\widehat{\cG}}-v_i\XXi_i\bw_i(\hat{g}_i)^{\tiny{\mbox{T}}}(\ttheta\trans\XXi\ttheta)^{-1}\bR\widehat{\aalpha}_{\widehat{\cG}}\right),
        \end{aligned}
    \end{equation*}

    \begin{equation*}
        \begin{aligned}
            \gamma=\left\|\widehat{\bbeta}_i-\widehat{\bbeta}_j+v_j\XXi_j
            \bw_j(\hat{g}_j)^{\tiny{\mbox{T}}}(\ttheta\trans\ttheta)^{-1}\bR\widehat{\aalpha}_{\widehat{\cG}}-
            v_i\XXi_i\bw_i(\hat{g}_i)^{\tiny{\mbox{T}}}(\ttheta\trans\XXi\ttheta)^{-1}\bR\widehat{\aalpha}_{\widehat{\cG}}\right\|_{\tbx_i}^2.
        \end{aligned}
    \end{equation*}
    This completes the proof.
\end{proof}

\subsection{Proof of Lemma 2}
\begin{proof}
    This result follows immediately from the same arguments used in the proof of Lemma 1: simply substitute $\bbeta_j^{\ini}(\phi)$ with $\sum_{i'=1}^N\bw_{i'}^{(s-1)}(k)\bbeta_{i'}^{\ini}(\phi)$ throughout. The calculations are identical, so we omit the details.
\end{proof}

\subsection{Proof of Theorem 4}
\begin{proof}
We decompose $\bB$ into the following two components:
\begin{equation*}
    \bB=\cP_{\ttheta_j}\bB+\mathcal{P}_{\ttheta_j}^{\perp}\bB=\XXi\ttheta_j(\ttheta_j\trans\XXi\ttheta_j)^{-1}\ttheta_j\trans\bB+\mathcal{P}_{\ttheta_j}^{\perp}\bB
\end{equation*}
Substituting the above equality into the definition of $P_j(\widehat{\bB};\wG_{k},\wG_{k'})$ yields
\begin{eqnarray}\nonumber
    P_j(\widehat{\bB};\wG_{k},\wG_{k'})&=&
    \mathbb{P}\Big(
    |\ttheta_j\trans\bB|\geq |\ttheta_j\trans\widehat{\bB}|~\Big|~
    \text{Cluster}(\XXi\ttheta_j(\ttheta_j\trans\XXi\ttheta_j)^{-1}\ttheta_j\trans\bB+\mathcal{P}_{\ttheta_j}^{\perp}\bB)=\text{Cluster}(\widehat\bB),\\
    &&~~~~~~~\mathcal{P}_{\ttheta_j}^{\perp}\bB=\mathcal{P}_{\ttheta_j}^{\perp}\widehat{\bB}
    \Big),
\end{eqnarray}

To simplify the above equation, we now show that
\begin{equation}
    \label{eq: thm4.1_1}
    |\ttheta_j\trans\bB|\perp\!\!\!\perp \mathcal{P}_{\ttheta_j}^{\perp}\bB, \quad \mathrm{as~}T \to \infty,
\end{equation}
where the notation $\perp\!\!\!\perp$ denotes independence in the limiting fixed-dimensional Gaussian experiment. Recall that $\mathcal{P}_{\ttheta_j}^{\perp}$ is the covariance-adjusted projection onto the complement of the contrast direction. Under the fixed-\(N\), \(T\to\infty\) framework, the joint limit of \(\ttheta_j\trans\bB\) and \(\mathcal{P}_{\ttheta_j}^{\perp}\bB\) is Gaussian with zero cross-covariance by construction of \(\mathcal{P}_{\ttheta_j}^{\perp}\). Hence the two components are independent in the limiting Gaussian experiment.

It immediately implies that
\begin{eqnarray*}\nonumber
&&\lim_{T\to \infty}P_j(\widehat{\bB};\wG_{k},\wG_{k'})\\
&=&
\lim_{T\to \infty}\mathbb{P}_{H_{0j}^{(\wG_k,\wG_{k'})}}\Big(
|\ttheta_j\trans\bB|\geq |\ttheta_j\trans\widehat{\bB}|~\Big|\text{Cluster}(\widehat{\bB})=\text{Cluster}(
\XXi\ttheta_j(\ttheta_j\trans\XXi\ttheta_j)^{-1}\ttheta_j\trans\bB+\mathcal{P}_{\ttheta_j}^{\perp}\widehat{\bB})
\Big)\\
&=& \lim_{T\to \infty}\mathbb{P}_{H_{0j}^{(\wG_k,\wG_{k'})}}\Big(
|\ttheta_j\trans\bB|\geq |\ttheta_j\trans\widehat{\bB}|~\Big|\|\cP_{\ttheta_j}\bB\|_2\in \mathcal{S}_j(\widehat{\bB})
\Big),
\end{eqnarray*}
where $\mathcal{S}_j(\widehat{\bB})=\left\{\kappa\in\mathbb{R}:\text{Cluster}(\widehat{\bB})=\text{Cluster}({\bB}(\kappa))\right\}$.

Under the stated asymptotics, we have
\[
\sqrt{NT}(\bR_j\aalpha_{\widehat{\cG}}-\bR_j \aalpha^0_{\cG^0})
\;\overset{d}{\longrightarrow}\; \cN(0,\; \bR_j\SSigma^0\bR_j\trans).
\]
So the test statistic
$\ttheta_j\trans\bB$ follows a $N(0,\vartheta_j^2)$ distribution, where $\vartheta_j^2=\bR_j\SSigma^0\bR_j\trans$ and $\bR_j$ is the $j$th row of $\bR$. Therefore,
\begin{equation*}
\lim_{T\to \infty}P_j(\widehat{\bB};\wG_{k},\wG_{k'})=1-\mathbb{F}\left(|\ttheta_j\trans\widehat{\bB}|;0,\vartheta_j^2,\cS_{j}\right)+
\mathbb{F}\left(-|\ttheta_j\trans\widehat{\bB}|;0,\vartheta_j^2,\cS_{j}\right),
\end{equation*}
where $\mathbb{F}(t;\mu,\sigma^2,\cS_j)$ denotes the CDF of a $\cN(\mu,\sigma^2)$ distribution truncated to the set $\cS_j$.

Finally, we will show the test that rejects $H_{0j}^{(\wG_k,\wG_{k'})}$ whenever $P_j(\widehat{\bB};\wG_{k},\wG_{k'})\leq\alpha$ controls the selective type I error rate at level $\alpha$, in the sense of Definition 1.

By the definition of $P_j(\bB;G_{k},G_{k'})$, we have
\begin{equation}
\label{eq: thm4.1_2}
\lim_{T\to\infty}\mathbb{P}_{H_{0j}^{(\wG_k,\wG_{k'})}}\Big(
P_j(\bB;G_k,G_{k'})\leq \alpha~\Big|~\mathcal{P}_{\ttheta_j}^{\perp}\bB,
\text{Cluster}(\bB)
\Big)=\alpha
\end{equation}

Therefore, we have
\begin{equation*}
    \begin{aligned}
        &\lim_{T\to\infty}\mathbb{P}_{H_{0j}^{(G_k,G_{k'})}}\Big(
P_j(\bB;G_k,G_{k'})\leq \alpha~\Big|\text{Cluster}(\bB)
\Big)\\
&=\lim_{T\to\infty}\mathbb{E}_{H_{0j}^{(G_k,G_{k'})}}\Big({\bf 1}\{
P_j(\bB;G_k,G_{k'})\leq \alpha\}~\Big|\text{Cluster}(\bB)
\Big)\\
&=\lim_{T\to\infty}\mathbb{E}_{H_{0j}^{(G_k,G_{k'})}}\Big[\mathbb{E}_{H_{0j}^{(G_k,G_{k'})}}\Big({\bf 1}\{
P_j(\bB;G_k,G_{k'})\leq \alpha\}~\Big|~\mathcal{P}_{\ttheta_j}^{\perp}\bB,\text{Cluster}(\bB)
\Big)\\
&~~~~~~~~~~~~~~~~~~~~~~~~~~~~~~~~\Big|~\text{Cluster}(\bB)\Big]\\
&=\lim_{T\to\infty}\mathbb{E}_{H_{0j}^{(G_k,G_{k'})}}\Big( \alpha~\Big|\text{Cluster}(\bB)
\Big)\\
&=\alpha,
    \end{aligned}
\end{equation*}
where the second equality holds due to the law of total expectation; and the third equality follows from Equation \eqref{eq: thm4.1_2}.
\end{proof}

\renewcommand{\theequation}{E.\arabic{equation}}
\renewcommand{\thefigure}{E.\arabic{figure}}
\renewcommand{\thetable}{E.\arabic{table}}
\renewcommand{\thealgorithm}{E.\arabic{algorithm}}
\renewcommand{\thetheorem}{E.\arabic{theorem}}
\renewcommand{\thesection}{E}
\setcounter{figure}{0}

\section{Additional Simulation Results}


 \begin{figure}[!h]
	\centering
   \includegraphics[scale =0.4]{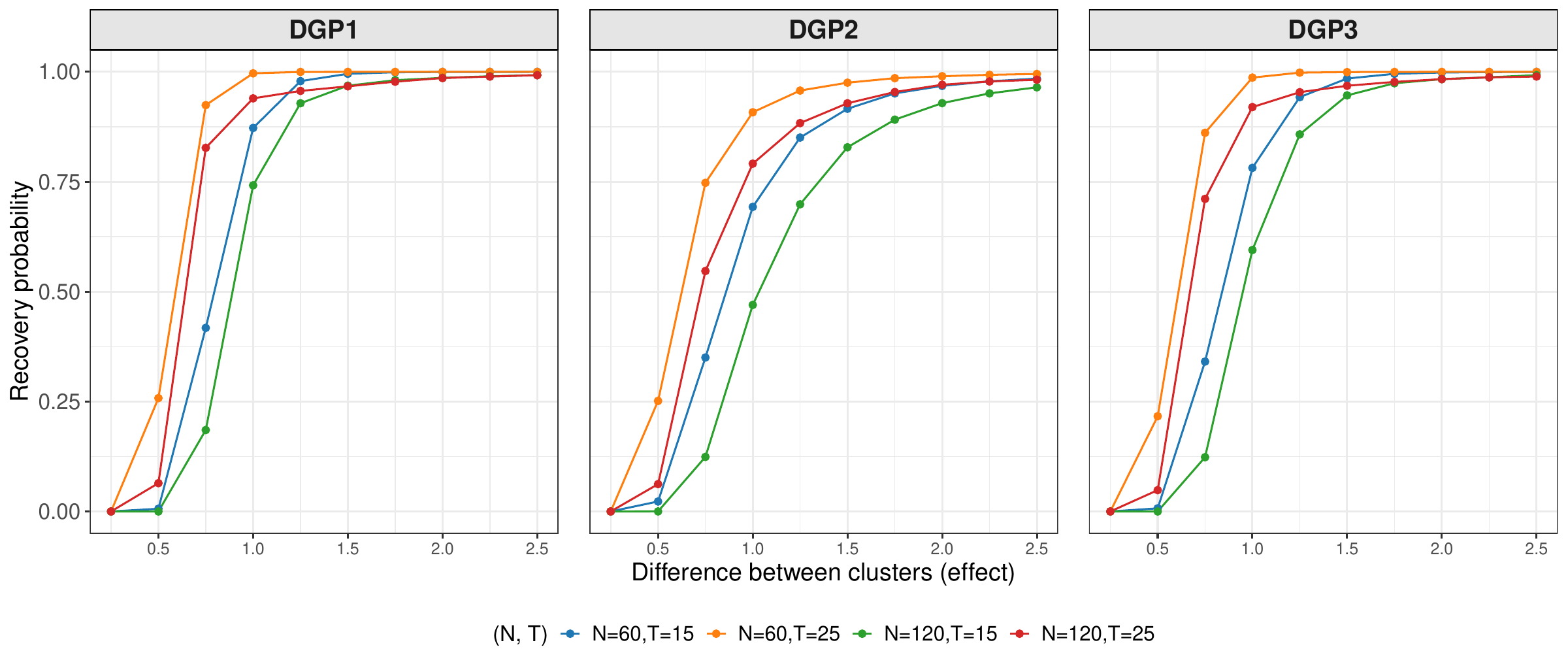}
	\caption{The recovery probability  across varying $\delta$'s under DGPs 1--3 with $K^0=3$.}
	\label{fig:pow123}
\end{figure}

 \begin{figure}[h]
	\centering
   \includegraphics[scale =0.36]{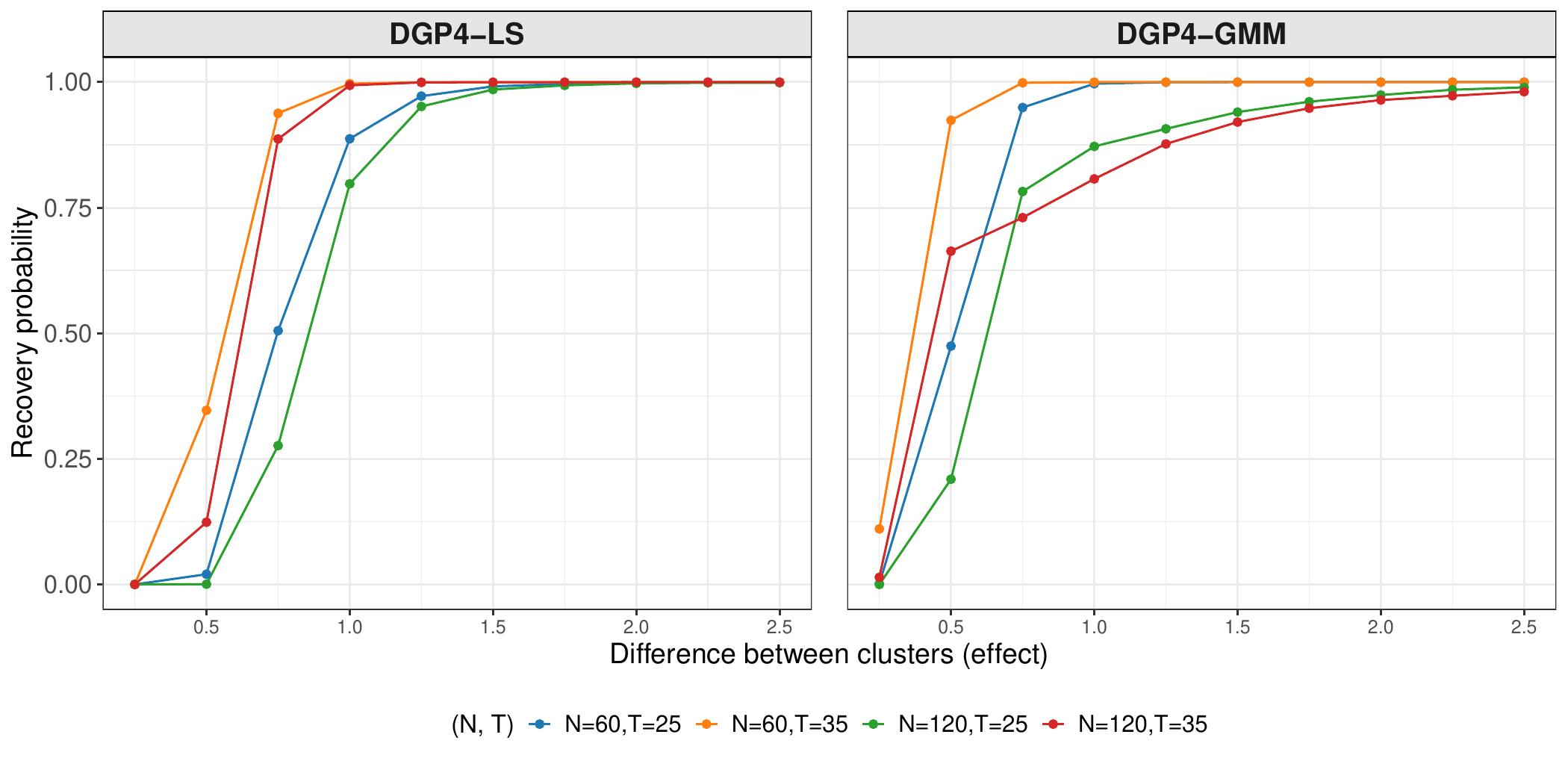}
	\caption{The recovery probability \eqref{eq:rp}  across varying $\delta$'s under DGP4 with $K^0=3$ and $\kappa=0.2$.}
	\label{fig:pow4}
\end{figure}

 \begin{figure}[h]
	\centering
   \includegraphics[scale =0.35]{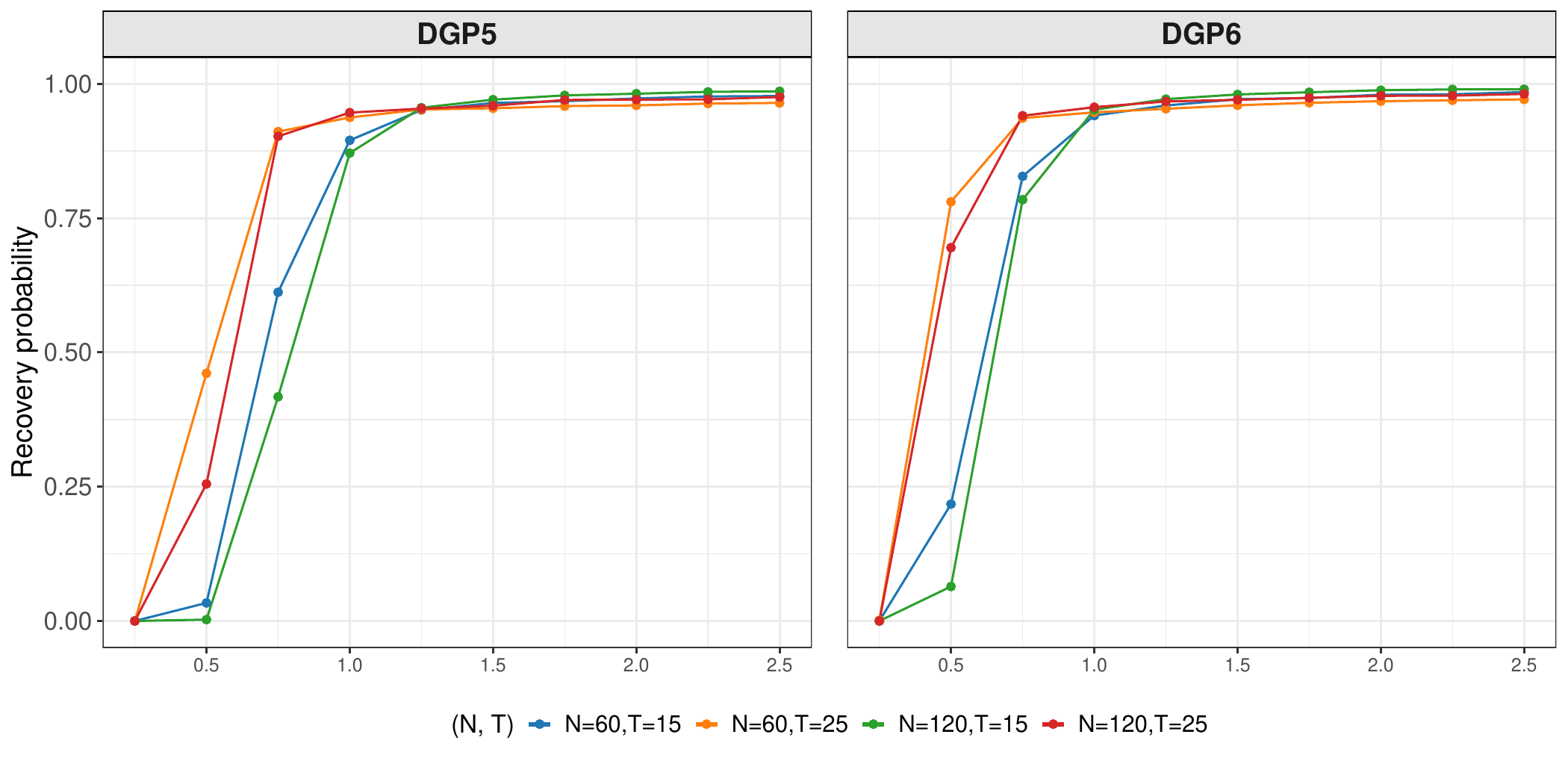}
	\caption{The recovery probability \eqref{eq:rp} across varying $\delta$'s under DGPs 5--6 with $K^0=3$.}
	\label{fig:pow56}
 \end{figure}

Figures \ref{fig:pow123}--\ref{fig:pow56} report the recovery probability for all simulation designs. Overall, the recovery probability increases as the signal strength $\delta$ becomes larger, indicating that stronger separation between the true group-specific parameters makes the latent group structure easier to recover. The improvement is also more pronounced when the time dimension $T$ increases, because longer panels provide more accurate individual initial estimators for clustering. For DGP 4, the GMM-based results are more reliable in the dynamic panel setting, whereas the LS-based results should be interpreted with caution due to the type I error distortion documented in the main text. For DGPs 5--6, the recovery probability follows the same increasing pattern and supports the use of the conditional power measure in Section \ref{sec:5}: once the relevant selected groups are correctly recovered, the proposed selective tests have substantial power.

\end{document}